\documentclass[useAMS,usenatbib,referee]{biom}

\newcommand{\add}[1]{\textcolor{red}{#1}}
\newcommand{\remove}[1]{\textcolor{blue}{$\langle$Remove: #1$\rangle$}}

\newcommand{\citeand}[2]{(\citeauthor{#1}, \citeyear{#1} and \citeauthor{#2}, \citeyear{#2})}

\def\bX{\mathbf{X}}
\def\bW{\mathbf{W}}
\def\bU{\mathbf{U}}
\def\bbeta{\boldsymbol{\beta}}

\def\bgamma{\boldsymbol{\gamma}}
\def\bdelta{\boldsymbol{\delta}}
\def\balpha{\boldsymbol{\alpha}}
\def\bSigma{\boldsymbol{\Sigma}}
\def\bOmega{\boldsymbol{\Omega}}
\def\bxi{\boldsymbol{\xi}}
\def\bmu{\boldsymbol{\mu}}
\def\bQ{\mathbf{Q}}

\def\bw{\mathbf{w}}
\def\bq{\mathbf{q}}
\def\bI{\mathbf{I}}
\def\bu{\mathbf{u}}
\def\bm{\mathbf{m}}
\def\be{\mathbf{e}}
\def\bp{\mathbf{p}}

\title[]{Semi-supervised linear regression: enhancing efficiency and robustness in high dimensions}

\author{Kai Chen$^{*}$\email{kaichen@ruc.edu.cn} and
	Yuqian Zhang$^{**}$\email{yuqianzhang@ruc.edu.cn} \\
	Institute of Statistics and Big Data, Renmin University of China, Beijing, China}

\def\T{{ \top }}
\usepackage{xcolor}
\usepackage{amssymb}
\usepackage{amsmath}
\usepackage{graphicx}
\usepackage{arydshln}
\usepackage{subfig}
\usepackage{booktabs}
\usepackage{amssymb}
\usepackage{amsmath}
\usepackage{graphicx}
\usepackage{arydshln}
\usepackage{subfig}
\usepackage{xcolor}
\newcommand{\BlackBox}{\rule{1.5ex}{1.5ex}} 

\begin{document}
	
	
	
	

	
	
	
	
	\label{firstpage}
	
	
	\begin{abstract}
		In semi-supervised learning, the prevailing understanding suggests that observing additional unlabeled samples improves estimation accuracy for linear parameters only in the case of model misspecification. In this work, we challenge such a claim and show that additional unlabeled samples are beneficial in high-dimensional settings. Initially focusing on a dense scenario, we introduce robust semi-supervised estimators for the regression coefficient without relying on sparse structures in the population slope. Even when the true underlying model is linear, we show that leveraging information from large-scale unlabeled data helps reduce estimation bias, thereby improving both estimation accuracy and inference robustness. Moreover, we propose semi-supervised methods with further enhanced efficiency in scenarios with a sparse linear slope. The performance of the proposed methods is demonstrated through extensive numerical studies.
		\vspace{1em}
	\end{abstract}

	%
	
	
	
	\maketitle
	
	
	%
	
		\section{Introduction}\label{sec:introduction}
	In many real-world applications, collecting the outcome of interest can be challenging due to accessibility constraints or high costs, whereas covariate information is typically abundant and inexpensive. Examples include speech recognition and image and document classification in machine learning \citep{zhu2005semi, chapelle2006semi, van2020survey}, as well as electronic health records (EHR) and genome-wide association studies (GWAS) in biomedical research \citeand{chakrabortty2018efficient}{tony2020semisupervised}. Beyond a labeled data set, we often have access to a large volume of unlabeled data, which leads to the problem of \emph{semi-supervised learning}. Let $Y_i \in \mathbb{R}$ represent the outcome variable and $\bX_i \in \mathbb{R}^{d+1}$ the covariates. A semi-supervised data set consists of independent and identically distributed (i.i.d.) labeled samples $\mathcal{L} = (Y_i, \bX_i)_{i \in \mathcal{I}}$ and unlabeled samples $\mathcal{U} = (\bX_j)_{j \in \mathcal{J}}$, where $\mathcal{I}$ and $\mathcal{J}$ are the index sets for labeled and unlabeled samples, respectively. Let $n = |\mathcal{I}|$, $m = |\mathcal{J}|$, and $N = n + m$. In semi-supervised learning, the focus is typically on scenarios where $N \gg n$.
	
	Semi-supervised learning has a longstanding presence in machine learning research, primarily focusing on classification tasks under assumptions like clustering, manifold, or smoothness conditions \citep{zhu2005semi, chapelle2006semi, van2020survey}. More recently, this area has gained substantial interest from the statistical community, which has shifted the focus toward statistical inference without relying on these additional conditions. Let $f_{Y|\bX}$ denote the conditional distribution of $Y$ given $\bX$, and $f_{\bX}$ the marginal distribution of $\bX$. When estimating a parameter of interest $\theta = \theta(f_{Y|\bX}, f_{\bX})$ that depends on both $f_{Y|\bX}$ and $f_{\bX}$, incorporating additional unlabeled samples that provide information about $f_{\bX}$ can enhance estimation efficiency. Examples can be found in the estimation of mean response \citeand{zhang2019semi}{zhang2022high}, explained variance  \citep{tony2020semisupervised}, M-estimation \citep{angelopoulos2023prediction}, causal parameters \citeand{cheng2021robust}{chakrabortty2022general}, and receiver operating characteristic (ROC) parameters \citetext{\citealp{gao2024semi} and \citealp{wang2024semisupervised}}. Semi-supervised learning is also closely related to fields such as data fusion \citep{li2024adaptive} and transfer learning \citetext{\citealp{cai2024semi} and \citealp{lu2024enhancing}}.

	The semi-supervised linear regression problem has recently been investigated by \cite{azriel2022semi, chakrabortty2018efficient, song2023general}, primarily in low-dimensional settings. \cite{deng2023optimal} extended this research to high-dimensional covariates and developed semi-supervised estimators for the linear slope. These approaches hinge on the availability of well-behaved non-parametric or non-linear estimates that outperform the linear ones -- a restrictive requirement in high dimensions. Even when such estimates are available, their semi-supervised methods improve efficiency over standard supervised approaches only when the linear model deviates from the truth; the greater the deviation, the more substantial the efficiency gain. However, when the linear model poorly approximates the true data-generating process, alternative approaches may be more appropriate. For prediction, non-parametric or non-linear models yield greater accuracy than relying on a misspecified linear model. For inference, interpreting coefficients from a misspecified model may offer limited practical relevance; targeting quantities such as the average treatment effect or coefficients in a partially linear model may be more appropriate.

	In this work, we aim to address the following question: \emph{Can the performance of linear regression be improved using additional unlabeled data when the linear model accurately approximates the truth?} Traditionally, there has been a widespread belief that unlabeled data offer no advantage when a parametric model is correctly specified and the parameter of interest depends solely on the conditional distribution $f_{Y|\bX}$ \citep{zhang2000value,seeger2001learning,chakrabortty2018efficient,azriel2022semi,deng2023optimal,song2023general}. This is because the supervised maximum likelihood estimator (MLE) remains the MLE in the semi-supervised setting. As the MLE is known to be the most asymptotically efficient estimator under regularity conditions, the supervised MLE retains this efficiency even in the presence of additional unlabeled samples. However, this statement may no longer applies in high dimensions as the MLE becomes ill-defined due to identifiability issues. In high-dimensional settings, estimates often exhibit substantial bias. Unlike most existing semi-supervised methods that focus on reducing variance using additional unlabeled samples -- an approach that is typically ineffective when the linear model is correctly specified -- we investigate whether unlabeled data can instead be used to reduce bias.
	
	For any $i \in \mathcal{I}$, define the population slope \citep{buja2019models} as $\bgamma^* = \mathop{\arg\min}_{\bgamma \in \mathbb{R}^{d+1}} {E}(Y_i - \bX_i^\T\bgamma)^2$. Let the residual be $\varepsilon_i = Y_i - \bX_i^\T\bgamma^*$, and possibly, $E(\varepsilon_i | \bX_i) \neq 0$. This allows for model misspecification, although our primary interest lies in assessing the role of unlabeled samples under correct model specification. Specifically, we focus on the statistical inference of a single linear coefficient. Without loss of generality, we take $\theta = \mathbf{e}_1^\T \bgamma^*$ as the parameter of interest, where $\mathbf{e}_1$ is the vector with $1$ in the first position and $0$ elsewhere.
	
	In fact, recent advances in high-dimensional statistics already highlight the potential value of unlabeled data. For instance, \cite{javanmard2014hypothesis, javanmard2018debiasing, zhu2018linear, bradic2022testability} established statistical inference for single linear coefficients under supervised settings, assuming different sparse structures with correct linear models. These works also consider the oracle case where $\boldsymbol{\Sigma} = E(\bX_i \bX_i^\T)$ is known, corresponding to having infinitely many unlabeled samples. In this setting, inference requires weaker sparsity conditions. These insights indicate that, at least in the oracle semi-supervised situation where $m = \infty$, unlabeled samples can be beneficial in high dimensions. In the following, we formally investigate the utility of additional unlabeled samples in high-dimensional linear regression.

	\section{Semi-supervised inference under dense loadings}\label{sec:dense}
	
	We first consider dense loadings without imposing sparse structures on the population slope \(\bgamma^*\), a scenario that has been explored by \cite{bradic2022testability} and \cite{zhu2018linear} in supervised settings. We introduce the following models:
	\begin{align}
		Y_i &= \bX_i^\T\bgamma^* + \varepsilon_i = Z_i\theta + \bW_i^\T\balpha^* + \varepsilon_i\;\;\text{for}\;\;i\in\mathcal I,\label{eq:linear1} \\
		Z_i &= \bW_i^\T\bbeta^* + v_i\;\;\text{for}\;\;i\in\mathcal I\cup\mathcal J,\label{eq:linear2}
	\end{align}
	where $\bgamma^* = (\theta, {\balpha^*}^\T)^\T$ is defined as in Section \ref{sec:introduction}, $\bX_i=(Z_i,\bW_i^\T)^\T$, $\bbeta^* = \mathop{\arg\min}_{\bbeta \in \mathbb{R}^d} E(Z_i - \bW_i^\T\bbeta)^2$, and $v_i = Z_i - \bW_i^\T\bbeta^*$. 
	Model \eqref{eq:linear1} is the primary model, where the parameter of interest, $\theta$, is the linear coefficient associated with the primary predictor $Z_i=\mathbf{e}_1^\T\bX_i$. Model \eqref{eq:linear2} is an auxiliary model that captures the association between $Z_i$ and the remaining control variables $\bW_i$. Incorporating the auxiliary model helps reduce regularization bias in the estimation of Model \eqref{eq:linear1}. This approach aligns with strategies used in the supervised literature \citep{zhang2014confidence, chernozhukov2018double, javanmard2018debiasing, bradic2022testability}.
	
We assume that the marginal distribution of \(\bX_i\) remains consistent across labeled and unlabeled groups, a standard assumption in the semi-supervised literature \citep{zhang2019semi,zhang2022high,tony2020semisupervised,cheng2021robust,chakrabortty2018efficient,azriel2022semi,deng2023optimal,song2023general}. In other words, the outcomes are missing completely at random (MCAR). For simplicity, we assume normally distributed covariates and residuals, with independent \(\bX_i \sim N(0, \boldsymbol{\Sigma})\) and \(\varepsilon_i \sim N(0, \sigma_\varepsilon^2)\). Similar assumptions can be found in \cite{javanmard2018debiasing, bradic2022testability, bellec2022biasing}. Extensions addressing covariate shifts and generalization to sub-Gaussian distributions will be discussed in Sections S2-S3 of the Supplementary Materials.

\subsection{The semi-supervised sparsity-robust estimator}\label{sec:SS-SR}

Based on the construction of \(\bgamma^*\) and \(\bbeta^*\), we observe that $E(v_iY_i)=\theta E(v_iZ_i) + E\{v_i(\bW_i^\T\balpha^*+\varepsilon_i)\}=\theta E(v_i^2) + E\{v_i\bW_i^\T(\balpha^* + \theta\bbeta^* + \varepsilon_i)\} = \theta E(v_i^2)$.
Therefore, the parameter of interest can be represented as $\theta = E(v_iY_i)/E(v_i^2)$.
A straightforward plug-in approach would involve constructing a Lasso estimate \(\widehat{\bbeta}\) for \(\bbeta^*\), followed by constructing \(\hat{\theta}_{\rm PI} = \sum_{i\in\mathcal{I}} \hat{v}_i Y_i / \sum_{i\in\mathcal{I}} \hat{v}_i^2\), where \(\hat{v}_i = Z_i - \bW_i^\T \widehat{\bbeta}\). This method avoids the need to estimate the entire population slope \(\bgamma^*\), thus accommodating cases where \(\bgamma^*\) is completely dense. However, in high-dimensional settings, regularized estimates are known to suffer from relatively large biases, leading to slow convergence rates for the plug-in estimator. Specifically, the estimation error can be decomposed as $\hat{\theta}_{\rm PI} - \theta = t_1 + t_2 + t_3$, where $t_1 = O_p(1/\sqrt{n})$ is asymptotically normal under regularity conditions, and $t_2 = O_p(\|\widehat{\bbeta} - \bbeta^*\|_2^2)$ depends quadratically on the estimation error from the auxiliary model \eqref{eq:linear2}. The primary source of bias is $t_3 = -(\sum_{i \in \mathcal{I}} \bW_i Y_i)^\T(\widehat{\bbeta} - \bbeta^*)/\sum_{i \in \mathcal{I}} \hat{v}_i^2$. In the following, we present a novel debiasing technique to correct for this bias.

Let $\boldsymbol{\xi} = -\sum_{i \in \mathcal{I}} \bW_i Y_i/\sum_{i \in \mathcal{I}} \hat{v}_i^2$. The bias term $t_3 = \boldsymbol{\xi}^\T(\widehat{\bbeta} - \bbeta^*)$ can be interpreted as the error in estimating $\boldsymbol{\xi}^\T \bbeta^*$ using the plug-in estimator $\boldsymbol{\xi}^\T \widehat{\bbeta}$. Prior work \citep{zhang2014confidence, van2014asymptotically, javanmard2014confidence, cai2017confidence} has proposed debiasing terms of the form $-|\mathcal{I}|^{-1} \sum_{i \in \mathcal{I}} \widehat{\mathbf{u}}^\T \bX_i(Y_i - \bX_i^\T \widehat{\bgamma})$ to correct the bias $\boldsymbol{\xi}^\T(\widehat{\bgamma} - \bgamma^*)$ when estimating linear combinations of coefficients from the primary model \eqref{eq:linear1}. Although the target parameter $\theta = \mathbf{e}_1^\T \bgamma^*$ is a linear combination of $\bgamma^*$, the density of the primary model prevents the construction of a reliable initial estimator $\widehat{\bgamma}$. Instead, we estimate $\theta$ using a plug-in estimator $\hat{\theta}_{\rm PI}$ derived from the auxiliary model, where the leading bias is $t_3$. Since $t_3$ is a linear function of $\widehat{\bbeta} - \bbeta^*$, the estimation error from the auxiliary model, we replace $(\bX_i, Y_i)$ with $(\bW_i, Z_i)$ and introduce a debiasing term of the form $-|\mathcal{I}|^{-1} \sum_{i \in \mathcal{I}} \widehat{\mathbf{u}}^\T \bW_i(Z_i - \bW_i^\T \widehat{\bbeta})$. A key distinction from prior work is that the vector $\boldsymbol{\xi}$ is data-dependent, which induces dependence between $\boldsymbol{\xi}$ and $\widehat{\bbeta}$. To address this, we introduce a sample splitting procedure that separates the data used to estimate $\boldsymbol{\xi}$ from that used to estimate $\widehat{\bbeta}$, ensuring valid inference. Furthermore, unlike in supervised settings, we can improve estimation of $\bbeta^*$ by leveraging additional unlabeled data.

We formally introduce our method below. Divide the labeled samples into two sets, indexed by \(\mathcal{I}_1\) and \(\mathcal{I}_2\) with \(|\mathcal{I}_1| \asymp |\mathcal{I}_2| \asymp n\). Let \(\bar{\mathcal{J}} = \mathcal{I}_1 \cup \mathcal{J}\). Leveraging the unlabeled samples, we first construct a Lasso estimate of \(\bbeta^*\) using the training samples \((\bX_i)_{i \in \bar{\mathcal{J}}}\), as the following:
$$
\widehat{\bbeta}_{\rm SR} = \mathop{\arg\min}_{\bbeta\in\mathbb{R}^{d}}\bigg\{|\bar{\mathcal{J}}|^{-1}\sum_{i\in \bar{\mathcal{J}}}(Z_i-\bW_i^\T\bbeta)^2+ \lambda_{\bbeta}\|\bbeta\|_1
\bigg\},
$$
where $\lambda_{\bbeta}\asymp(\log d /N)^{1/2}$ is a tuning parameter. Define $\widehat{\boldsymbol{\xi}}=|\mathcal{I}_2|^{-1}\sum_{i\in \mathcal{I}_2}\bW_iY_i$, $\widehat{\boldsymbol{\Sigma}}_{\bW}=|\bar{\mathcal{J}}|^{-1}\sum_{i\in \bar{\mathcal{J}}} \bW_i\bW_i^\T$, and \(\hat{v}_i = Z_i - \bW_i^\T \widehat{\bbeta}_{\rm SR}\). Choose some $\lambda_{\mathbf{u}}\asymp (M+\sigma_{\varepsilon})(\log d /n)^{1/2}$ with $M=\|\bgamma^*\|_2$, we solve the following optimization problem:
\begin{equation*}
\widehat{\mathbf{u}}_{\rm SR} = \mathop{\arg\min}_{\mathbf{u}\in\mathbb{R}^{d}}\mathbf{u}^{\T}\widehat{\boldsymbol{\Sigma}}_{\bW}\mathbf{u}\ \ {\rm s.t.}\ \ \|\widehat{\boldsymbol{\xi}}-\widehat{\boldsymbol{\Sigma}}_{\bW} \mathbf{u}\|_\infty\le \lambda_{\mathbf{u}}.
\end{equation*}
The \emph{semi-supervised sparsity-robust (SS-SR) estimator} is proposed as
\begin{equation*}\label{def:theta_d}
\hat{\theta}_{\rm SR}=\dfrac{|\mathcal{I}_1|^{-1}\sum_{i\in \mathcal{I}_1}\hat{v}_iY_i-|\bar{\mathcal{J}}|^{-1}\sum_{i\in \bar{\mathcal{J}}}\widehat{\mathbf{u}}_{\rm SR}^{\T}\bW_i\hat{v}_i}{|\mathcal{I}_1|^{-1}\sum_{i\in \mathcal{I}_1}\hat{v}_i^2}.
\end{equation*}

Notably, the proposed SS-SR estimator extends its applicability to degenerate supervised settings, representing a novel addition to the supervised high-dimensional linear regression literature. In previous work, \cite{bradic2022testability} studied the same dense scenario under supervised settings. Compared to their approach, our method offers several advantages, including a simpler sample splitting design and the avoidance of matrix and non-convex optimization. A detailed comparison with their estimator, as well as with other supervised methods, is provided in Section S1 of the Supplementary Materials.

\subsection{Theoretical properties}\label{sec:theory-dense}	
The following results characterize the convergence rate of the proposed estimator $\hat\theta_{\rm SR}$.

\begin{assumption}[Bounded eigenvalues]\label{ass:bound_eigenvalues}
Suppose that there exists a constant $C_{\Sigma}>1$ such that the eigenvalues of $\boldsymbol{\Sigma}=E(\bX_i\bX_i^\T)$ satisfy $1/C_{\boldsymbol{\Sigma}}\le\lambda_{\min}(\boldsymbol{\Sigma})\le\lambda_{\max}(\boldsymbol{\Sigma})\le C_{\boldsymbol{\Sigma}}$.
\end{assumption}

\begin{theorem}\label{thm:SR_consist_rate}
Let Assumption \ref{ass:bound_eigenvalues} hold and $s=\|\bbeta^*\|_0\ll(Nn)^{1/2}/\log d$. Then, as $n,d\to\infty$, 
$\hat{\theta}_{\rm SR}-\theta=O_p[(M+\sigma_{\varepsilon})\{n^{-1/2}+s\log d/(Nn)^{1/2}\}]$.
\end{theorem}

\begin{remark}[Semi-supervised minimax lower bound]\label{remark:minimax}
In the semi-supervised setting, the minimax expected length of confidence intervals for $\theta$ is bounded from below by $C(M + \sigma_{\varepsilon})n^{-1/2}$ for some constant $C > 0$. This lower bound holds uniformly over all values of $s$ and $N$, including the idealized case where $s = 0$ and $N = \infty$. Importantly, the bound is governed by the labeled sample size $n$, regardless of how large $N$ is. This is because, for parameters that depend on the outcome's distribution, the stochastic variation introduced by the noise in the outcome model cannot be reduced by incorporating unlabeled data. Further explanations and theoretical support are provided in Section S7 of the Supplementary Materials.
\end{remark}

Theorem \ref{thm:SR_consist_rate} imposes sparsity conditions solely on the vector \(\bbeta^*\), allowing \(\bgamma^*\) to be fully dense. Importantly, the sparsity requirement on \(\bbeta^*\) depends on the total sample size, making it easily achievable when the unlabeled sample size is sufficiently large. The convergence rate in Theorem \ref{thm:SR_consist_rate} is influenced not only by the unexplained variance \(\sigma_\varepsilon^2 = E(\varepsilon_i^2)\) but also by the linear signal level \(M^2 = \|\bgamma^*\|_2^2\). This dependence arises because, in the absence of any sparse constraints on the vector \(\bgamma^*\), consistent estimates in high dimensions are not expected. As a result, we cannot distinguish the linear signal \(\bX_i^\T\bgamma^*\) from the linear residual variable \(\varepsilon_i\) in \eqref{eq:linear1}. As noted in Remark \ref{remark:minimax}, the presence of $M$ in the convergence rate is unavoidable, even when $s$ is extremely small and $N$ is extremely large.

In supervised settings, Corollary 6 of \cite{bradic2022testability} shows that the minimax estimation error for \(\theta\) is of the order \(n^{-1/2} + s\log d / n\) when \(M\) and \(\sigma_{\varepsilon}\) are constants. In comparison, our semi-supervised estimator achieves the same variance order \(n^{-1/2}\), but with an improved bias rate of \(s\log d/(Nn)^{1/2}\). Thus, \(\hat{\theta}_{\rm SR}\) achieves a strictly faster convergence rate when \(N \gg n\) and \(n\ll s^2\log^2 d\), while matches the supervised rate otherwise. The condition \(N \gg n\) ensures a sufficient amount of unlabeled data, while \(n\ll s^2\log^2 d\) indicates that unlabeled samples help only when the model is not easily estimable. Besides, when \(\boldsymbol{\Sigma} = E(\bX_i\bX_i^\T)\) is known (as in the oracle case with \(N = \infty\)), Theorem 1 of \cite{bradic2022testability} shows root-\(n\) estimation is possible. Our findings indicate that such a rate can be attained as long as \(N \gg s^2\log^2 d\). 

The following theorem establishes the asymptotic normality of the proposed estimator.


\begin{theorem}\label{thm:SR_asy_normal}
Let Assumption \ref{ass:bound_eigenvalues} hold, $s=\|\bbeta^*\|_0\ll N^{1/2}/\log d$, and $M/\sigma_{\varepsilon}=O(1)$. Then, as $n,d\to\infty$, 
$\hat{\theta}_{\rm SR}-\theta=O_p\{(M+\sigma_\varepsilon)n^{-1/2}\}\;\;\text{and}\;\;(\hat{\theta}_{\rm SR}-\theta)/\hat\sigma_{\rm SR}\xrightarrow{\rm d} N(0,1)$, where $\hat\sigma_{v,1}^2=|\mathcal{I}_1|^{-1}\sum_{i\in \mathcal{I}_1}\hat{v}_i^2$ and $\hat\sigma_{\rm SR}^2=(\hat\sigma_{v,1}^2)^{-1}\sum_{i\in\mathcal{I}_1}\{(Y_i-\hat{\theta}_{\rm SR} \hat{v}_i)/|\mathcal{I}_1|-\widehat{\mathbf{u}}_{\rm SR}^\T \bW_i/|\bar{\mathcal{J}}|\}^2+(\hat\sigma_{v,1}^2)^{-1}|\bar{\mathcal{J}}|^{-2}\sum_{i\in\mathcal{J} }(\widehat{\mathbf{u}}_{\rm SR}^\T \bW_i)^2$.
\end{theorem}

Theorem \ref{thm:SR_asy_normal} indicates that the proposed estimator $\hat\theta_{\rm SR}$ is asymptotically normal, enabling the construction of root-$n$ inference as long as the total sample size $N$ is sufficiently large. In comparison, Corollary 6 of \cite{bradic2022testability} highlights that under supervised settings, root-$n$ inference for $\theta$ is achievable only when $n\gg s^2\log^2d$. Leveraging unlabeled samples in our semi-supervised approach relaxes this requirement to $N=n+m\gg s^2\log^2d$. Given the typically lower cost of collecting unlabeled data, the semi-supervised method provides a more cost-effective route to valid inference. When the labeled sample size is insufficient, a large number of unlabeled samples can be gathered to bridge the gap, ultimately leading to asymptotically normal results. The proposed estimator demonstrates superior robustness to sparsity constraints, where the population slope $\bgamma^*$ in the primary model can be entirely dense, and the sparsity requirement on $\bbeta^*$ in the auxiliary model is solely dependent on the total sample size $N$, not the labeled size $n$. Even if both the linear slopes $\bgamma^*$ and $\bbeta^*$ in \eqref{eq:linear1}-\eqref{eq:linear2} are completely dense, the normal results still hold if $N$ is large.

The asymptotic normality results in Theorem \ref{thm:SR_asy_normal} hold under the condition that the signal-to-noise ratio is bounded, \(M/\sigma_{\varepsilon} = O(1)\). In the Supplementary Materials, we demonstrate that both this constraint and the Gaussian assumptions can be relaxed through a modified sample splitting procedure involving additional splits; see Section S3.


\section{Improved efficiency under sparse structures}\label{sec:sparse}

In this section, we extend our analysis to scenarios where \(\bgamma^*\) in the primary model is sparse.

\subsection{The semi-supervised degree-of-freedom adjusted estimator}\label{sec:SS-DFA}

When \(\bgamma^*\) is sparse, unlike the dense situation discussed in Section \ref{sec:dense}, we can obtain a consistent estimate using the labeled data \((\bX_i, Y_i)_{i \in \mathcal{I}}\) through \(\ell_1\)-regularization. Specifically, with \(\lambda_{\bgamma} \asymp \sigma_{\varepsilon} (\log d / n)^{1/2}\), we derive the Lasso estimate: $\widehat{\bgamma}_{\rm DFA} = \mathop{\arg\min}_{\bgamma\in\mathbb{R}^{d+1}}\{n^{-1}\sum_{i\in \mathcal{I}}(Y_i-\bX_i^\T\bgamma)^2+\lambda_{\bgamma}\|\bgamma\|_1\}.$
Further, with \(\lambda_{\bbeta} \asymp (\log d / N)^{1/2}\), we estimate \(\bbeta^*\) using the Dantzig selector, incorporating both labeled and unlabeled samples \((\bX_i)_{i \in \mathcal{I} \cup \mathcal{J}}\), as follows:
\begin{equation}\label{def:beta_hat3}
\begin{split}
	\widehat{\bbeta}_{\rm DFA}=\mathop{\arg\min}_{\bbeta\in\mathbb{R}^{d}}\|\bbeta\|_1,\quad \mbox{s.t.}\quad &\Big\|n^{-1}\sum_{i\in\mathcal{I}}\bW_i(Z_i-\bW_i^\T\bbeta)\Big\|_\infty\le (N/n)^{1/2}\cdot\lambda_{\bbeta},\\
	&\Big\|N^{-1}\!\sum_{i\in \mathcal{I}\cup\mathcal{J}}\bW_i(Z_i-\bW_i^\T\bbeta)\Big\|_\infty\le \lambda_{\bbeta}.
\end{split}
\end{equation}
The first constraint controls the error from the labeled samples, ensuring the effectiveness of the debiasing procedure in \eqref{def:theta_df}, while the second constraint boosts estimation efficiency by leveraging the entire set of observations. With a slight abuse of notation, let \(\hat{v}_i = Z_i - \bW_i^\T \widehat{\bbeta}_{\rm DFA}\). Denote \(\hat{q} = \|\widehat{\bgamma}_{\rm DFA}\|_0\), which represents the degrees of freedom for the Lasso estimator \citep{zou2007degrees,Tibshirani2012degrees}. Drawing inspiration from the degrees-of-freedom adjustment technique used in supervised settings \citetext{\citealp{javanmard2014hypothesis} and \citealp{bellec2022biasing}}, we introduce the \emph{semi-supervised degrees-of-freedom adjusted} (SS-DFA) estimator:
\begin{equation}\label{def:theta_df}
\hat{\theta}_{\rm DFA} = \mathbf{e}_1^\T \widehat{\bgamma}_{\rm DFA} + \frac{n^{-1}\sum_{i\in \mathcal{I}} \hat{v}_i(Y_i - \bX_i^\T \widehat{\bgamma}_{\rm DFA})}{(1 - \hat{q}/n)n^{-1}\sum_{i\in \mathcal{I}} \hat{v}_i Z_i}.
\end{equation}
The proposed estimator serves as a semi-supervised generalization of the supervised degrees-of-freedom adjusted estimator from \cite{bellec2022biasing}. The distinction lies in the estimation of $\bbeta^*$. While \cite{bellec2022biasing} estimate $\bbeta^*$ using only labeled data, our approach constructs an estimator that incorporates both labeled and unlabeled samples, as defined in \eqref{def:beta_hat3}. To ensure the validity of the debiasing procedure while improving estimation accuracy, we solve a constrained optimization problem with two separate $\ell_\infty$-norm constraints, in contrast to the Lasso estimator employed in \cite{bellec2022biasing}.

\subsection{Theoretical properties}\label{sec:theory-DFA}

The following theorem describes the asymptotic behavior of \(\hat{\theta}_{\rm DFA}\) under Gaussian designs; extensions to sub-Gaussian designs are given in Section S4 of the Supplementary Materials.

\begin{theorem}\label{thm:DF_consist_rate}
Let Assumption \ref{ass:bound_eigenvalues} hold, $s=\|\bbeta^*\|_0\ll(Nn)^{1/2}/\log d$, $k=\|\bgamma^*\|_0\ll n/\log d$, $k\ll d$, and $\Sigma_{jj}\le 1$ for all $j\in [d+1]$. Then, as $n,d\to\infty$, 
$\hat{\theta}_{\rm DFA}-\theta=O_p[\sigma_{\varepsilon}n^{-1/2}+\sigma_{\varepsilon}\min\{k\log d/n,s\log d/(Nn)^{1/2}\}]$.

Moreover, assume at least one of the following sparsity conditions holds: (a) $k\ll n^{1/2}/\log d$ and $s\ll(Nn)^{1/2}/\log d$ or (b) $k\ll n/\log d$ and $s\ll N^{1/2}/\log d$. Let $\sigma ^2={E}(v_i^2\varepsilon_i^2)/\{E(v_i^2)\}^2$. Then, as $n,d\to\infty$, $\hat\theta_{\rm DFA}-\theta=O_p(\sigma_\varepsilon/n^{1/2})$, $n^{1/2}(1-\hat{q}/n)(\hat\theta_{\rm DFA}-\theta)/\sigma\xrightarrow{\rm d} N(0,1)$, and $\hat{\sigma}_{\rm DFA}^2/\sigma ^2=1+o_p(1)$, where $\hat\sigma_{\rm DFA}^2={\sum_{i\in \mathcal{I}}(Y_i-\bX_i^\T\widehat{\bgamma}_{\rm DFA})^2}/{\sum_{i\in \mathcal{I}}(Z_i-\bW_i^\T\widehat{\bbeta}_{\rm DFA})^2}$.
\end{theorem}

In supervised settings, \cite{javanmard2018debiasing} indicates that the minimax lower bound for estimating \(\theta\) is of the order \(n^{-1/2} + \min(k, s) \log d / n\) when \(\sigma_\varepsilon\) is treated as a constant. Theorem \ref{thm:DF_consist_rate} demonstrates that the estimator \(\hat{\theta}_{\rm DFA}\) achieves a convergence rate that is strictly faster than that of supervised methods if \(N \gg (1+s/k)n\) and \(n \ll (s \land k)^2 \log^2 d\). In scenarios where the unlabeled size is insufficient or the labeled size is already large enough, the convergence rate aligns with the minimax lower bound for supervised estimation.

Comparing with the results in Theorem \ref{thm:SR_consist_rate}, \(\hat{\theta}_{\rm DFA}\) also offers a convergence rate that is either faster or equal to that of \(\hat{\theta}_{\rm SR}\). The strict improvement occurs under two conditions: (a) when the signal-to-noise ratio is high, such that \(M \gg \sigma_\varepsilon\), or (b) when the sparsity level of \(\beta^*\) is relatively large, specifically \(s \gg N^{1/2}/\log d + k(N/n)^{1/2}\). However, unlike \(\hat{\theta}_{\rm SR}\), the consistency of \(\hat{\theta}_{\rm DFA}\) necessitates additional sparsity constraints on \(\bgamma^*\), requiring that \(k = \|\bgamma^*\|_0 \ll n / \log d\). This makes \(\hat{\theta}_{\rm DFA}\) less robust to non-sparse structures. In summary, while \(\hat{\theta}_{\rm DFA}\) is more efficient, it demands stronger conditions; conversely, \(\hat{\theta}_{\rm SR}\) is more conservative, being less efficient but applicable under broader circumstances.

In ultra-sparse settings, the supervised estimator of \cite{bellec2022biasing} achieves root-$n$ consistency and asymptotic normality with the same variance as Theorem \ref{thm:DF_consist_rate}. While their method requires $\min(k,s) \ll n^{1/2}/\log d$ and $\max(k,s) \ll n/\log d$, our use of unlabeled data relaxes these sparsity constraints. Specifically, the condition on $s = \|\bbeta^*\|_0$ depends on $N$ and is easily met when $N$ is large. Thus, additional unlabeled samples reduce bias without increasing variance, enabling weaker sparsity assumptions.

\section{Numerical experiments}\label{sec:num}
Section \ref{sec:sim} presents simulation studies under various settings, followed by a real data application using the AIDS Clinical Trials Group Study 175 (ACTG175) dataset \citep{hammer1996trial} in Section \ref{sec:real}. Additional simulations and a second real data example based on the National Health and Nutrition Examination Survey Data I Epidemiologic Follow-up Study (NHEFS) dataset \citep{hernan2020causal} are provided in Sections S5 and S6 of the Supplementary Materials.
\subsection{Simulation results}\label{sec:sim}

In the following, outcomes are generated exclusively in the labeled groups \(i \in \mathcal{I}\). We denote the \(d \times d\) identity matrix as \(\mathbf{I}_d\) and the \(d\)-dimensional vector of all ones as \(\mathbf{1}_d\).

Model 1: Linear models with sparse $\bbeta^*$ and $\bgamma^*$. For all $i\in\mathcal I\cup\mathcal J$, $\bW_i\sim N(0,\boldsymbol{\Sigma}_{\bW})$, $Z_i=-\sum_{j=1}^{s}W_{ij}/s^{1/2}+v_i$ with $s=25$, and $Y_i=0.4Z_i+\sum_{j=1}^{9}W_{ij}/(10)^{1/2}+\varepsilon_i$, where $\boldsymbol{\Sigma}_{\bW}=(1-\mu)\mathbf{I}_d+\mu \mathbf{1}_d\mathbf{1}_d^\T$, $\mu=1/(d+1)$, $v_i\sim N(0,1)$, and $\varepsilon_i\sim N(0,0.4^2)$.

Model 2: Linear models with sparse $\bbeta^*$ and dense $\bgamma^*$. Generate $Z_i$, $\bW_i$, $v_i$, and $\varepsilon_i$ as in Model 1. Let $Y_i=0.4Z_i+\sum_{j=1}^{d}W_{ij}/(d+1)^{1/2}+\varepsilon_i$. 

Model 3: Linear models with dense $\bbeta^*$ and $\bgamma^*$. Generate $\bW_i$, $v_i$, and $\varepsilon_i$ as in Model 1. Let $Z_i=-\sum_{j=1}^{s}W_{ij}/s^{1/2}+v_i$ with $s=d=499$ and $Y_i=0.8Z_i+\sum_{j=1}^{d}W_{ij}/(d+1)^{1/2}+\varepsilon_i$.

\begin{figure}[t]
\centering
\includegraphics[width=\textwidth]{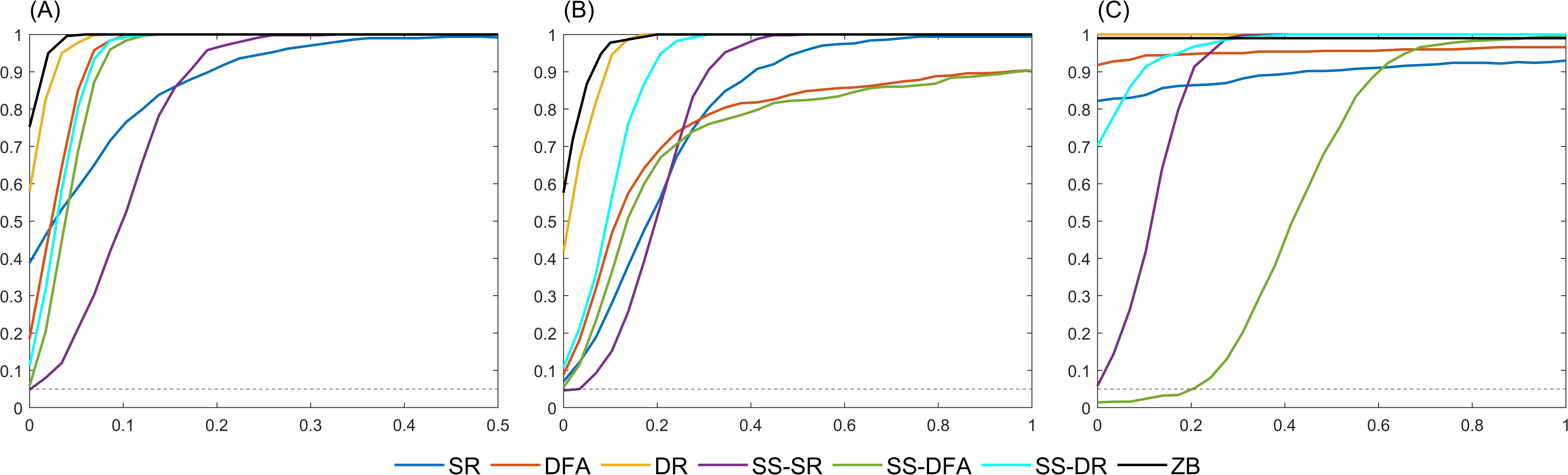}
\caption{Panel (A): Model 1 ($m=1000$) with $\|\bbeta^*\|_0=25$ and $\|\bgamma^*\|_0=10$; 
	Panel (B): Model 2 with $\|\bbeta^*\|_0=25$ and $\|\bgamma^*\|_0=500$; 
	Panel (C): Model 3 with $\|\bbeta^*\|_0=499$ and $\|\bgamma^*\|_0=500$. 
	Each panel shows power curves of competing methods corresponding to the hypothesis $H_0:\theta=\theta_0+h$, where $\theta_0$ is the true regression coefficient, and the significance level is $\alpha=0.05$. The x-axis represents the deviation $h$, and the y-axis denotes the average rejection rate over 500 repetitions. Hence, $h=0$ corresponds to the Type I error, and the remaining values correspond to the power.}
\label{fig:power}
\end{figure}

We evaluate the numerical performance of the SS-SR and SS-DFA estimators, \(\hat{\theta}_{\rm SR}\) and \(\hat{\theta}_{\rm DFA}\), introduced in Sections \ref{sec:dense} and \ref{sec:sparse}, respectively. Additionally, we incorporate a semi-supervised doubly robust (SS-DR) estimator, as presented in Section S4 of the Supplementary Materials, which extends the doubly robust approach \citep{chernozhukov2018double} to semi-supervised settings. For comparison, we include several supervised estimators: the degenerate supervised version of the sparsity-robust (SR) estimator, the supervised degrees-of-freedom adjusted (DFA) estimator \citep{bellec2022biasing}, and the supervised doubly robust (DR) estimator \citep{chernozhukov2018double} using Lasso for nuisance parameter estimation. Tuning parameters are selected via additional cross-validation. We also consider the single coefficient estimators derived from the supervised Lasso and the safe semi-supervised learning (S-SSL) method of \cite{deng2023optimal}, although their focus was on estimation rather than inference. The S-SSL method is implemented using random forests to estimate the conditional mean function. For hypothesis testing, we further consider the method of \cite{zhu2018linear}, referred to as ZB. The simulations are repeated 500 times, and the results are summarized in Table \ref{model1}. We report the empirical bias (Bias), root mean square error (RMSE), average confidence interval length (Length), and the average coverage probability of $95\%$ confidence intervals (Coverage). Power curves for hypothesis tests are displayed in Figure \ref{fig:power}.

\begin{table}[h]
\caption{Simulation results under Models 1-3}
\vspace{1em}
{\resizebox{1\linewidth}{!}{\begin{tabular}{lcccclcccc}
			\hline
			\hline\vspace{-2.6em}\\
			Estimator  & Bias  & RMSE  & Length   & Coverage &Estimator  & Bias  & RMSE  & Length   & Coverage\\
			\hline\vspace{-2.6em}\\
			\multicolumn{10}{c}{(A) Model 1 with $d=499$, $n=300$, $\|\bbeta^*\|_0=25$, $\|\bgamma^*\|_0=10$, and varied $m$}\\
			\hdashline\vspace{-2.6em}\\
			&\multicolumn{4}{c}{$m=0$ (supervised)}&& \multicolumn{4}{c}{$m=500$}\\[-0.5em]
			Lasso& $-0.0861$ & $0.0891$ & / & / &S-SSL  & $-0.1435$ & $0.1466$ & / & / \\
			SR&$-0.0864$ & $0.1376$ & $0.2758$ & $61.2\%$& SS-SR&$-0.0283$ & $0.0644$ & $0.2418$ & $93.6\%$ \\
			DFA&$-0.0350$ & $0.0421$ & $0.1191$ & $81.6\%$&SS-DFA&$-0.0227$ & $0.0333$ & $0.1249$ & $94.2\%$ \\
			DR&$-0.0468$ & $0.0524$ & $0.0858$ & $42.0\%$&SS-DR&$-0.0276$ & $0.0375$ & $0.1020$ & $82.4\%$\\[-0.5em]
			&\multicolumn{4}{c}{$m=250$}&&\multicolumn{4}{c}{$m=1000$}\\[-0.5em]
			S-SSL&$-0.1308$ & $0.1340$ & / &/ &S-SSL& $-0.1560$ & $0.1594$ & / & /\\
			SS-SR&$-0.0374$ & $0.0711$ & $0.2534$ & $92.6\%$&SS-SR&$-0.0190$ & $0.0586$ & $0.2345$ & $95.2\%$ \\
			SS-DFA&$-0.0268$ & $0.0360$ & $0.1202$ & $90.8\%$&SS-DFA&$-0.0220$ & $0.0334$ & $0.1228$ & $94.2\%$\\
			SS-DR&$-0.0330$ & $0.0416$ & $0.0969$ & $71.6\%$ &SS-DR&$-0.0228$ & $0.0337$ & $0.1056$ & $88.8\%$\\
			\hline\vspace{-2.6em}\\
			\multicolumn{10}{c}{(B) Model 2 with $d=499$, $n=300$, $\|\bbeta^*\|_0=25$, $\|\bgamma^*\|_0=500$, and $m=1000$}\\
			\hdashline\vspace{-2.6em}\\
			Lasso  &$-0.1804$ & $0.1885$ & / &/ &S-SSL& $-0.2037$ & $0.2085$ & / & /\\
			SR   &$-0.0073$ & $0.1405$ & $0.4327$ & $93.0\%$&SS-SR  & $0.0102$ & $0.0913$ & $0.3696$ & $95.4\%$\\
			DFA   & $0.0391$ & $0.3069$ & $0.6221$ & $91.2\%$&SS-DFA  & $0.0823$ & $0.3454$ & $0.6051$ & $94.6\%$\\
			DR   &$-0.1000$ & $0.1149$ & $0.2230$ & $58.4\%$ &SS-DR  &$-0.0460$ & $0.0810$ & $0.2748$ & $89.2\%$\\
			\hline\vspace{-2.6em}\\
			\multicolumn{10}{c}{(C) Model 3 with $d=499$, $n=150$, $\|\bbeta^*\|_0=499$, $\|\bgamma^*\|_0=500$, and $m=1500$}\\
			\hdashline\vspace{-2.6em}\\
			Lasso   &$-0.7466$ & $0.7486$ & / &/ &S-SSL& $-0.7040$ & $0.7064$ & / & /\\
			SR    &$-0.1280$ & $1.2703$ & $0.7990$ & $17.8\%$ &SS-SR   &$-0.0365$ & $0.0798$ & $0.3011$ & $94.2\%$ \\
			DFA    &$-0.6107$ & $0.6140$ & $0.5676$ &  $8.2\%$ & SS-DFA   &$-0.0663$ & $0.2935$ & $1.0376$ & $98.6\%$ \\
			DR    &$-0.6308$ & $0.6328$ & $0.1843$ &    $0\%$ & SS-DR   &$-0.2070$ & $0.2332$ & $0.3210$ & $29.8\%$ \\
			\hline
			\hline
\end{tabular}}}
\label{model1}
\end{table}

We first observe that both Lasso and S-SSL estimators exhibit relatively large bias and RMSE, as they target the full coefficient vector without debiasing for individual inference. Additionally, due to the correct linear model and slow convergence of random forests in high dimensions, S-SSL sometimes yield larger estimation error than the supervised Lasso.

Table \ref{model1} (A) shows the performance of the estimators as the unlabeled sample size $m$ increases, when both the primary and auxiliary models are sparse. All proposed methods exhibit improvements in RMSE and coverage as $m$ increases, except for the SS-DFA, whose performance remains similar at $m=500$ and $m=1000$. This aligns with our theory that once the unlabeled sample size is large enough and root-$n$ inference is achieved, additional unlabeled data does not improve efficiency. Given the sparse structure of the primary model, SS-DFA and SS-DR yield lower RMSEs than SS-SR. Moreover, SS-DFA also provides shorter intervals than SS-SR and better coverage than SS-DR.	

The SS-DR estimator exhibits poor coverage when the primary model is dense, as shown in Table \ref{model1} (B)-(C). SS-SR consistently achieves accurate coverage and relatively low RMSEs, particularly when both models are dense. SS-DFA also attains coverage close to the nominal $95\%$, but with substantially longer intervals. This is due to the degree-of-freedom adjustment factor $(1 - \|\widehat{\bgamma}_{\rm DFA}\|_0/n)^{-1}$, which becomes unstable when $\|\bgamma^*\|_0$ is large relative to $n$. The supervised SR method also performs relatively well when $\bbeta^*$ is sparse, as seen in Table \ref{model1} (B). Table \ref{model1} (C) further shows that all supervised methods fail to provide valid coverage, underscoring the benefit of incorporating unlabeled data.

Figure \ref{fig:power} shows power curves for testing $H_0: \theta = \theta_0 + h$ under varying $h > 0$. Across all considered models, only SS-DFA and SS-SR exhibit Type I errors smaller than or close to $\alpha=0.05$. Specifically, under Model 1, SS-DFA demonstrates higher powers than SS-SR for any $h>0$. For Model 2, SS-SR surpasses SS-DFA in power when $h$ is large, but shows lower powers when $h$ is small. The supervised SR method maintains a Type I error rate near $\alpha=0.05$; in comparison, the method proposed by \cite{zhu2018linear} displays a significantly larger Type I error. In Model 3, SS-DFA exhibits a Type I error significantly smaller than $\alpha=0.05$, while SS-SR demonstrates higher powers for arbitrary $h>0$. Additionally, all other methods perform poorly with large Type I errors.

\subsection{Application to ACTG175}\label{sec:real}

We evaluate the proposed methods using data from ACTG175, a clinical trial assessing the efficacy and safety of antiretroviral regimens for HIV-infected individuals. Patients with baseline CD4 counts between \(200\) and \(500\) cells/mm\(^3\) were randomly assigned to one of four treatment groups: ZDV only, ZDV + ddI, ZDV + Zal, and ddI only. The corresponding group sizes are $532$, $522$, $524$, and $561$, and we denote the treatment variable as $T = 0, 1, 2, 3$, respectively. Prior studies have analyzed the causal effects of these treatments \citep{tsiatis2008covariate, liang2020empirical}. Since treatment outcomes are known to depend on age -- older individuals often exhibit slower immune recovery and are more susceptible to adverse effects \citep{li2022treatment, negin2012prevalence} -- we focus on the association between age and the change in CD4 count within each group, using CD4 change as a measure of immune recovery. Although analyzing each treatment subgroup separately is a natural approach, it may lead to inefficient use of available data. Instead, we treat the problem as a semi-supervised learning task, using information from other subgroups to improve the association estimate within one subgroup. We believe this strategy can be broadly applied to other subgroup analysis problems.

We adopt the potential outcome framework. For each treatment \(t \in \{0, 1, 2, 3\}\), let $Y_i(t)$ denote the potential outcome a patient would receive if assigned to treatment group $t$. Our goal is to estimate the linear association between the primary predictor $Z_i$ (age) and the potential outcome $Y_i(t)$ within each treatment group. For any fixed $t$, the value of $Y_i(t)$ is observed only when $T_i = t$, yielding a labeled set \(\mathcal{L}_t = \{(Y_i(t), \bX_i): i \leq N, T_i = t\}\), along with an unlabeled set \(\mathcal{U}_t = \{\bX_i: i \leq N, T_i \neq t\}\). This setup allows us to restructure the dataset into four semi-supervised datasets \((\mathcal{L}_t, \mathcal{U}_t)_{t=0}^3\). Since treatments were randomly assigned in the trial, there is no covariate shift between the labeled and unlabeled samples, allowing for the reliable application of semi-supervised methods. The control variables $\bW_i$ are derived from additional baseline covariates, resulting in a transformed feature dimension of $d = 434$. Further details are provided in Section S6 of the Supplementary Materials.

\begin{figure}[t]
\centering
\includegraphics[height=0.53\linewidth,keepaspectratio]{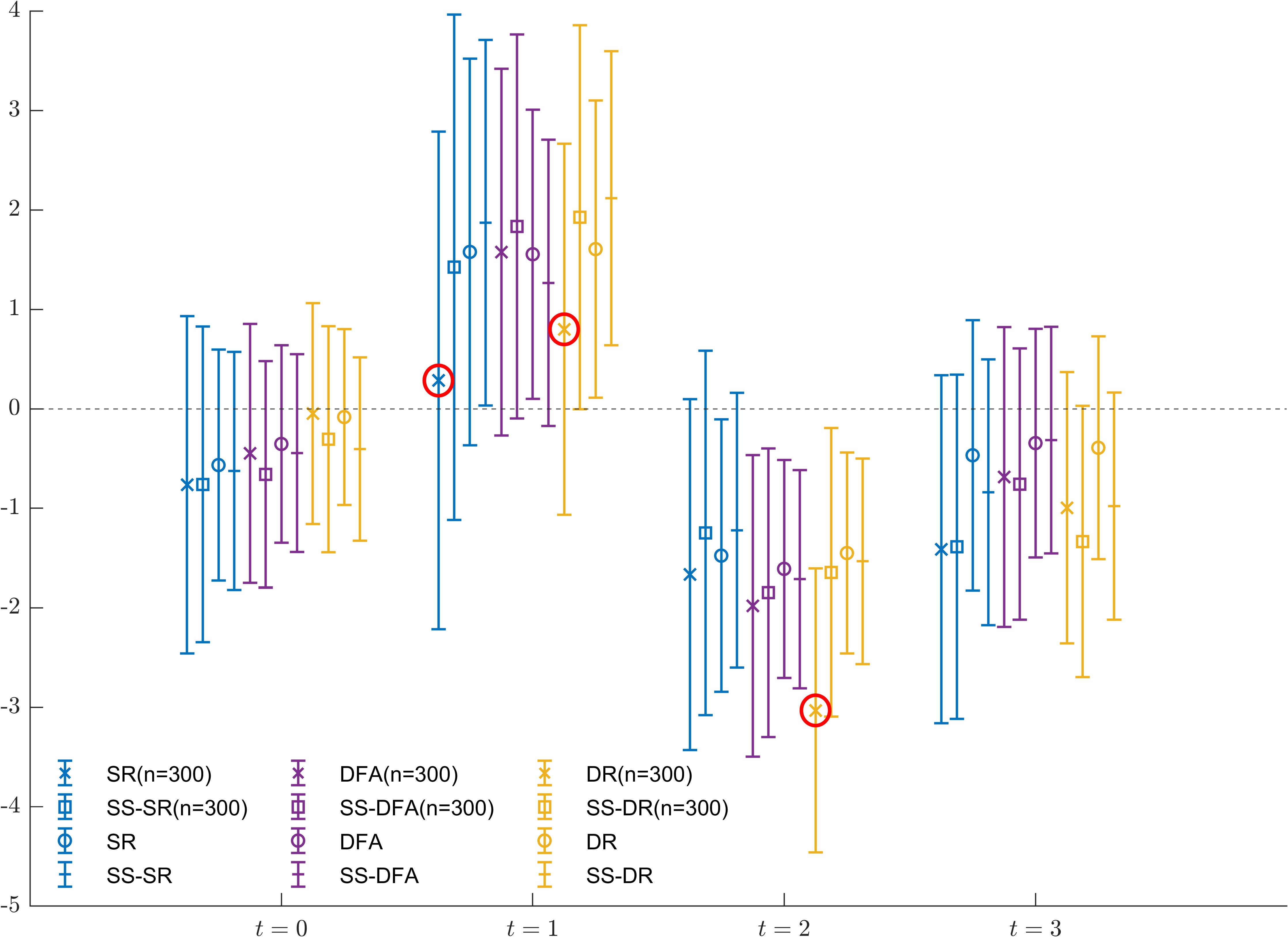}
\caption{The $95\%$ confidence intervals of the regression coefficient correspond to age-related changes in CD4 cell count within each treatment group. The x-axis, denoted by $t=0,1,2,3$, represents the treatment groups: ZDV only, ZDV + ddI, ZDV + Zal, and ddI only, respectively. The marker shapes ``x'' and ``o'' represent the results of supervised methods based on 300 labeled samples and all labeled samples, respectively. The marker shape ``square'' represents the results of semi-supervised methods using 300 labeled samples and all unlabeled samples, while the marker shape ``-'' represents the results obtained using all labeled samples and unlabeled samples.}
\label{fig:real_data}
\end{figure}

For each treatment group, Figure \ref{fig:real_data} shows the 95\% confidence intervals of the regression coefficients based on the full semi-supervised dataset and on the labeled samples only. To assess the impact of labeled and unlabeled sample sizes, we also report results based on a randomly selected subsample of 300 labeled samples combined with all unlabeled samples, as well as results using only 300 labeled samples. For $t=1$ (ZDV+ddI), most results from full labeled samples suggest a positive association between age and changes in CD4 count. For $t=2$ (ZDV+Zal), most methods show a negative association. These findings indicate that age may influence the effectiveness of antiretroviral therapies, with different regimens potentially yielding varying responses in older patients. Our results align with previous research on individualized treatment regimens \citep{lu2013variable, fan2017concordance, cui2023estimating}.

Comparing results across varying sample sizes, most point estimates remain consistent, except for a few cases (highlighted in red in Figure \ref{fig:real_data}) 
from supervised methods using only $n = 300$ labeled samples, indicating potential bias from the limited sample size. In contrast, semi-supervised methods that use the same $300$ labeled samples along with additional unlabeled data produce more stable estimates and confidence intervals comparable to those from larger labeled datasets. These findings demonstrate the advantage of semi-supervised methods in reducing bias and enhancing the reliability of inference when labeled data is scarce.

\section{Discussion}\label{sec:dis}

This paper delves into the high-dimensional linear regression problem in the presence of additional unlabeled samples. Our primary focus is on the estimation and inference of a linear coefficient $\theta=\mathbf{e}_1^\T\bgamma^*$. When the linear model is correctly specified, the parameter of interest $\theta$ depends solely on the conditional distribution $P_{Y|\bX}$ and is independent of the marginal distribution $P_{\bX}$. Unlike existing semi-supervised literature, we show that unlabeled data improve estimation accuracy and inference robustness for the linear coefficient $\theta$ in high dimensions, even when the linear model is correct. We believe this observation contributes to a deeper understanding of the usefulness of additional unlabeled samples. 

While we considered the linear regression problem as a starting point of the analysis, we believe that the proposed framework can be extended to other parametric models, such as general M-estimation problems, which often involve additional analytical challenges due to the potential non-linearity of the estimating equations. It is also of interest to study the use of non-parametric or machine learning methods in semi-supervised settings. For example, it is valuable to investigate whether and how the performance of non-parametric models can be improved by incorporating additional unlabeled data under correct model specifications. Moreover, it is worthwhile to develop semi-supervised methods that are more robust to the labeling mechanisms and to explore the use of transfer learning techniques in such scenarios.

	\newpage
	\renewcommand\thesection{S\arabic{section}}
	\setcounter{section}{0} 
	\renewcommand\thesubsection{\thesection.\arabic{subsection}}
	\renewcommand{\theequation}{S\arabic{equation}}
	\setcounter{equation}{0}
	\renewcommand{\theassumption}{S\arabic{assumption}}
	\setcounter{assumption}{0}
	\renewcommand\thetheorem{S\arabic{theorem}}
	\setcounter{theorem}{0}
	\renewcommand\thetable{S\arabic{table}}
	\renewcommand\thelemma{S\arabic{lemma}}
	\renewcommand{\thesubfigure}{\Alph{subfigure}}
	\renewcommand\thefigure{S\arabic{figure}}
	\begin{center}
	\textbf{Supplementary Materials for “Semi-supervised linear
		regression: enhancing efficiency and robustness in high
		dimensions”}
	\end{center}
	
	\section{Comparison of the SS-SR estimator with existing supervised methods}
		The SS-SR estimator proposed in Section 2.1 offers a novel contribution to the literature, even in degenerate supervised settings. Below, we provide comparisons with existing methods under the supervised setup.
		
				\paragraph*{\textbf{Comparison with \cite{javanmard2014confidence} and \cite{ cai2017confidence}.}}\ In scenarios involving a sparse primary model (1), \cite{javanmard2014confidence} introduced a debiased estimator for \(\theta\) that accommodates a dense auxiliary model (2). This method was further examined by \cite{cai2017confidence}. 
				
				Their approach begins by constructing a Lasso estimate \(\widehat{\bgamma}\) for \(\bgamma^*\) and solving the optimization problem:  
		\begin{equation}\label{def:mhat}  
			\widehat{\bm} = \mathop{\arg\min}_{\bm \in \mathbb{R}^{d+1}} \bm^\top\widehat{\bSigma}_n \bm \ \ {\rm s.t.} \ \ \|\widehat{\bSigma}_n \bm - \be_1\|_\infty \le \lambda_{\bm},  
		\end{equation}  
		where \(\widehat{\bSigma}_n = n^{-1} \sum_{i\in\mathcal{I}} \bX_i \bX_i^\top\). The resulting debiased estimator is  
		\[
		\hat{\theta}_{\rm JM} = \be_1^\top \widehat{\bgamma} + n^{-1} \sum_{i\in\mathcal{I}} \widehat{\bm}^\top \bX_i \hat{\varepsilon}_i = \theta + n^{-1} \sum_{i\in\mathcal{I}} \widehat{\bm}^\top \bX_i \varepsilon_i - (\widehat{\bSigma}_n \widehat{\bm} - \be_1)^\top (\widehat{\bgamma} - \bgamma^*),
		\]
		where \(\hat{\varepsilon}_i := Y_i - \bX_i^\top \widehat{\bgamma}\). This procedure reduces the bias \(\be_1^\top(\widehat{\bgamma} - \bgamma^*)\) by incorporating residuals \(\hat{\varepsilon}_i = \varepsilon_i - \bX_i^\top(\widehat{\bgamma} - \bgamma^*)\). A debiasing term is constructed as a weighted average of residuals, with weights \(\widehat{\bm}^\top \bX_i\) determined by solving \eqref{def:mhat}. This reduces the bias to \((\widehat{\bSigma}_n \widehat{\bm} - \be_1)^\top (\widehat{\bgamma} - \bgamma^*)\) but introduces an additional variance term, \(n^{-1} \sum_{i\in\mathcal{I}} \widehat{\bm}^\top \bX_i \varepsilon_i\). Since \(\widehat{\bm}\) depends only on \((\bX_i)_{i\in\mathcal{I}}\), it is independent of the noise \(\varepsilon_i\) under Gaussian designs, enabling direct characterization of the variance term: $n^{-1} \sum_{i\in\mathcal{I}} \widehat{\bm}^\top \bX_i \varepsilon_i \mid (\bX_i)_{i\in\mathcal{I}} \sim N(0, \sigma_\varepsilon^2 \widehat{\bm}^\top \widehat{\bSigma}_n \widehat{\bm}/n)$. This result leverages the independence (or lack of correlation) between the noise \(\varepsilon_i\) from the sparse model (1) and the variables \((\bW_i, Z_i)\) from the dense model (2).  
		
		Our proposed debiasing procedure is motivated by this framework but addresses a different setup. Under supervised settings, we focus on a dense primary model (1) and a sparse auxiliary model (2), which is complementary to the setup in \cite{javanmard2014confidence} and \cite{cai2017confidence}. For sparse \(\bbeta^*\), we first construct a Lasso estimate \(\widehat{\bbeta}_{\rm SR}\) and then develop a debiasing term using residuals \(Z_i - \bW_i^\T \widehat{\bbeta}_{\rm SR} = v_i - \bW_i^\T (\widehat{\bbeta}_{\rm SR} - \bbeta^*)\). However, as the noise \(v_i\) from the sparse model is generally correlated with \(Y_i = v_i\theta + \bW_i^\T (\balpha^* + \theta \bbeta^*) + \varepsilon_i\) from the dense model, directly applying a debiasing procedure may result in a non-decaying variance term. To address this, we incorporate a sample-splitting technique to ensure independence between the weights \(\widehat{\bu}_{\rm SR}^\T \bW_i\) and \(v_i\). This step guarantees the asymptotic normality of the estimator, as shown in Theorem 2.  
		
		The considered sparse auxiliary model presents greater challenges compared to the sparse primary model due to the inherent asymmetry between the two models. The primary model \(Y \sim \bX\) encompasses all variables \((Z, \bW, Y)\), making it ``broader in scope'' than the auxiliary model \(Z \sim \bW\), which is restricted to \((Z, \bW)\). As a result, studying a dense structure for \(Y \sim \bX\) is inherently more complex than for \(Z \sim \bW\). This distinction underscores the unique contributions of our approach compared to the methods proposed by \cite{javanmard2014confidence} and \cite{cai2017confidence}.
		
		Furthermore, our method is particularly effective in semi-supervised settings. With access to additional unlabeled data, the sparsity assumptions for the auxiliary model depend on the total sample size \(N = n + m\) rather than the labeled sample size \(n\); see Theorems 1 and 2. As unlabeled data are often inexpensive and widely available, the required sparsity conditions are easier to meet. When $N$ is sufficiently large, our approach remains effective even when both the primary and auxiliary models are dense, demonstrating robustness to sparsity assumptions. This adaptability underpins the term ``sparsity-robust'' for the proposed method.

		\paragraph*{\textbf{Comparison with \cite{bradic2022testability}.}}\ \cite{bradic2022testability} investigated the dense setting where the linear coefficients of the primary model may exhibit no sparsity and offered useful insights into the testability of individual linear coefficients. However, their estimator is described as being ``of pure theoretical interest'' and is primarily presented to demonstrate the sharpness of their minimax lower bounds. We present their proposed estimator and offer detailed comparisons below.

		First, divide the samples into four equal-sized groups, indexed by \(\mathcal{I}_1, \dots, \mathcal{I}_4\). Let \(\lambda_1, \dots, \lambda_5\) be some positive constants depending on \(n\). Construct \(\widehat{\bbeta}_2\) as the Lasso estimate of \(\bbeta^*\) using samples from \(\mathcal{I}_2\) with tuning parameter \(\lambda_1\). Define a set \(A = \{j \in [d] : ||\mathcal{I}_1|^{-1} \sum_{i \in \mathcal{I}_1} \bW_i Y_i| > \lambda_2\}\), representing the set of ``important'' features after feature screening. Construct a truncated vector \(\widehat{\bxi}_A\) such that \((\widehat{\bxi}_A)_j = 0\) for \(j \notin A\), and \((\widehat{\bxi}_A)_j = |\mathcal{I}_3|^{-1} \sum_{i \in \mathcal{I}_3} W_{ij} Y_i\) otherwise. Define \(\widehat{\bSigma}_4 = |\mathcal{I}_4|^{-1} \sum_{i \in \mathcal{I}_4} \bW_i \bW_i^\T\) and construct an estimate of $\bOmega=\bSigma_{\bW}^{-1}$ as below:
		\begin{equation}\label{def:hat-Omega}
			\widehat{\bOmega} = \mathop{\arg\min}_{\bQ \in \mathbb{R}^{d \times d}} \lambda_{\rm max}(\bQ) \ \ {\rm s.t.}\ \ \bQ = \bQ^\T, \ \|(\bI_d - \widehat{\bSigma}_4 \bQ) \widehat{\bxi}_A\|_\infty \leq \lambda_3, \ \widehat{\bxi}_A^\T \bQ \widehat{\bSigma}_4 \bQ \widehat{\bxi}_A \leq \lambda_4,
		\end{equation}
		where \(\bI_d\) is the \(d \times d\) identity matrix. A debiased Lasso estimate of $\bbeta^*$ is first constructed as \(\widehat{\bbeta}_4 = \widehat{\bbeta}_2 + |\mathcal{I}_4|^{-1} \sum_{i \in \mathcal{I}_4} \widehat{\bOmega} \bW_i (Z_i - \bW_i^\T \widehat{\bbeta}_2)\). Then, a projected debiased estimator is further introduced:
		\begin{align}
			\breve{\bbeta}=\mathop{\arg\min}_{\bq\in\mathbb{R}^d}\|\bq\|_1\ \ {\rm s.t.}\ \ &|\widehat{\bxi}_A^\T(\bq-\widehat{\bbeta}_4)|\le \lambda_5,\ \bigg\||\mathcal{I}_4|^{-1}\sum_{i\in\mathcal{I}_4}\bW_i(Z_i-\bW_i^\T q)\bigg\|_\infty\le \lambda_1/4,\nonumber\\
			&|\mathcal{I}_4|^{-1}\sum_{i\in\mathcal{I}_4}(Z_i-\bW_i^\T q)^2\ge C,\label{def:breve-beta}
		\end{align}
		where \(C > 0\) is some constant. Define \(\breve{v}_i = Z_i - \bW_i^\T \breve{\beta}\). \cite{bradic2022testability} proposed the following estimator of $\theta$: 
		\begin{align}\label{def:BFZ}
		\hat{\theta}_{\rm BFZ} =\sum_{i \in \mathcal{I}_4} \breve{v}_i Y_i/\sum_{i \in \mathcal{I}_4} \breve{v}_i^2.
		\end{align}
		
		As described in Section 2.1, a plug-in estimate of the form \eqref{def:BFZ} mainly suffers from a bias $\breve{\bxi}^\top(\breve{\bbeta} - \bbeta^*)$, where $\breve{\bxi} = \sum_{i \in \mathcal{I}_4} \bW_i Y_i / \sum_{i \in \mathcal{I}_4} \breve{v}_i^2$. The projected estimator $\breve{\bbeta}$ is introduced to reduce this bias. Since the set $A$ identifies the large components of the vector $\breve{\bxi}$, it ensures a small $\|\breve{\bxi}_{A^c}\|_\infty$, which leads to a controlled bound
$|\breve{\bxi}_{A^c}^\top(\breve{\bbeta} - \bbeta^*)| \leq \|\breve{\bxi}_{A^c}\|_\infty \|\breve{\bbeta} - \bbeta^*\|_1,$
as long as $\breve{\bbeta}$ converges sufficiently fast to the target $\bbeta^*$. On the other hand, as shown in \cite{cai2017confidence}, the debiasing term introduced in $\widehat{\bbeta}_4$ leads to a relatively small value of $|\breve{\bxi}_A^\top(\widehat{\bbeta}_4 - \bbeta^*)|$. Together with the first constraint in \eqref{def:breve-beta}, we also have $
|\breve{\bxi}_A^\top(\breve{\bbeta} - \bbeta^*)| \leq |\breve{\bxi}_A^\top(\widehat{\bbeta}_4 - \bbeta^*)| + |\breve{\bxi}_A^\top(\breve{\bbeta} - \widehat{\bbeta}_4)|$ being small enough.
Combining these results, the projected estimator $\breve{\bbeta}$ yields a small bias $\breve{\bxi}^\top(\breve{\bbeta} - \bbeta^*)$.	
	
		While \cite{bradic2022testability} reduce the plug-in bias by carefully designing a projected debiased estimate $\breve{\bbeta}$, they still construct their final estimate \eqref{def:BFZ} in a plug-in form. In contrast, we directly incorporate debiasing terms into the plug-in estimate to cancel out the bias; see our construction of the estimate $\hat\theta_\mathrm{SR}$. Compared to the existing proposal, the method we introduce leads to notable simplification. First, we observe that the truncation and feature screening steps are not necessary, since debiasing terms can be applied directly to the entire bias $t_3 = -(\sum_{i \in \mathcal{I}} \bW_i Y_i)^\top(\widehat{\bbeta} - \bbeta^*)/ \sum_{i \in \mathcal{I}} \hat{v}_i^2$. Consequently, we can also avoid the sample splittings that were introduced for the screening steps. From a computational perspective, the third constraint in \eqref{def:breve-beta}, which was added to control the denominator of the plug-in estimate \eqref{def:BFZ}, results in a non-convex optimization problem. We find that this constraint becomes unnecessary when the debiasing term is directly included in the plug-in estimator. Moreover, unlike the matrix optimization problem in \eqref{def:hat-Omega}, our construction only involves solving vector optimization problems.

		In summary, our method offers the following practical advantages: (a) The proposed procedure involves only two tuning parameters, \(\lambda_{\bbeta}\) and \(\lambda_{\bu}\), both of which can be selected using cross-validation. In comparison, \cite{bradic2022testability} rely on the selection of \(\lambda_1, \dots, \lambda_5\) and \(C\), some of which depend on unknown constants such as the eigenvalues of \(\bSigma = E(\bX_i\bX_i^\T)\). (b) Our method divides the sample into two parts instead of four, enhancing finite-sample efficiency. (c) The optimization problems are solved in \(\mathbb{R}^d\), whereas obtaining \(\hat{\Omega}\) in (7) involves solving for a \(d \times d\) matrix, which is computationally more demanding. (d) Both \(\widehat{\bbeta}_{\rm SR}\) and \(\widehat{\bu}_{\rm SR}\) can be efficiently computed using standard packages such as coordinate descent and quadratic programming. In contrast, solving \eqref{def:breve-beta} requires addressing a non-convex optimization problem.  
		
		In summary, our method represents the first practically feasible approach for point estimation in the presence of a dense primary model, even under traditional supervised settings. Moreover, with additional unlabeled samples, the bias of the point estimate can be further reduced, which relaxes the sparsity requirement on the auxiliary model while ensuring asymptotic normality and valid inference guarantees.
		
		\paragraph*{\textbf{Comparison with \cite{zhu2018linear}.}}\
		The scenario involving a dense primary model and a sparse auxiliary model was first introduced by \cite{zhu2018linear}, where the authors developed linear hypothesis tests with controlled Type I error but did not address point estimation for \(\theta\). While their framework ensures controlled Type I error even with dense \(\bgamma^*\), the test power is limited by the additional requirement of an ultra-sparse \(\bgamma^*\), \(k = \|\bgamma^*\|_0 \ll n^{1/2}/\log d\). In contrast, our asymptotic normality results in Theorem 2 accommodate dense \(\bgamma^*\) without imposing such restrictive sparsity assumptions.

	\section{Semi-supervised linear regression under covariate shift}\label{sec:CS}
	
	In this section, we extend our analysis to scenarios where covariate shift may occur. While the assumption of data being missing completely at random (MCAR) is commonly made in the semi-supervised literature \citep{zhang2019semi,zhang2022high,deng2023optimal,tony2020semisupervised}, it can be practically restrictive, as it assumes researchers have full control over the missing mechanism. To address this limitation, we take an initial step towards relaxing this assumption and introduce the following alternative condition.
	
	\begin{assumption}[Covariate shift]\label{ass:MAR}
		For any $i \in \mathcal{I}$, $j \in \mathcal{J}$, and $\bw\in\mathbb{R}^d$, let ${E}(Z_i \mid \bW_i = \bw) = {E}(Z_j \mid \bW_j = \bw)$.
	\end{assumption}
	
	Let \( R_i \in \{0,1\} \) denote the labeling indicator, where \( R_i = 1 \) if \( Y_i \) is labeled and \( R_i = 0 \) otherwise. When treating \( R_i \) as random, Assumption \ref{ass:MAR} follows from the missing at random (MAR) condition \( R_i \perp Z_i \mid \bW_i \), which states that the control variables \( \bW_i \) account for all the dependence between \( Z_i \), the primary predictor, and the labeling of \( Y_i \). This assumption is weaker than the more restrictive MCAR condition \( R_i \perp (Z_i, \bW_i) \).
	
	The MAR condition we require also differs from another type of MAR condition \( R_i \perp Y_i \mid \bX_i \), which is assumed in works such as  \cite{zhang2023decaying}, \cite{zhang2023double}, and \cite{zhang2023efficient}. Unlike these studies, we allow for arbitrary dependence between \( R_i \) and \( Y_i \), since our parameter of interest is defined based solely on the labeled group. Our goal is to improve the accuracy of estimating parameters within the labeled group using additional unlabeled data, rather than to generalize findings from the labeled group to a different population. In practice, the outcomes \( Y_i \) are typically observed after the covariates \( (Z_i, \bW_i) \), making it difficult to control the data generating process associated with \( Y_i \). In contrast, it is usually more feasible to influence the generation of the primary predictor \( Z_i \).
	
	For example, the MAR condition \( R_i \perp Z_i \mid \bW_i \) is satisfied when \( Z_i \) is generated based only on the control variables \( \bW_i \) and independent exogenous errors. This commonly occurs when \( Z_i \) represents a treatment variable assigned randomly according to \( \bW_i \). It also includes settings where treatments are assigned using stratified designs and the strata are fully captured by the control variables. Furthermore, the MAR condition also holds when the labeling indicator \( R_i \) is generated based solely on \( \bW_i \) and exogenous randomness.

	Unless otherwise stated, let \( i \in \mathcal{I} \) and \( j \in \mathcal{J} \) throughout. We assume a Gaussian design, where \( \bX_i \sim N(0, \bSigma) \), \( \bX_j \sim N(0, \bSigma') \), and \( \varepsilon_i \sim N(0, \sigma_\varepsilon^2) \). For further extensions to sub-Gaussian designs, refer to Sections \ref{sec:SS-SR-subG}-\ref{sec:sparse-subG}. Define \( \sigma_{v,1}^2 = E(v_i^2) \) and \( \sigma_{v,2}^2 = E(v_j^2) \). The following assumption is introduced for the covariance associated with labeled and unlabeled groups.
	
	\begin{assumption}[Bounded eigenvalues]\label{ass:bound_eigenvalues_s}
		With $i\in\mathcal I$ and $j\in\mathcal J$, suppose that there exists a constant $C_{\bSigma}>1$ such that the eigenvalues of $\bSigma$ and $\bSigma'$ satisfy $1/C_{\bSigma}\le\lambda_{\min}(\bSigma)\le\lambda_{\max}(\bSigma)\le C_{\bSigma}$ and $1/C_{\bSigma}\le\lambda_{\min}(\bSigma')\le\lambda_{\max}(\bSigma')\le C_{\bSigma}$.
	\end{assumption}
	
	We now characterize the asymptotic behavior of the semi-supervised sparsity-robust estimator \(\hat{\theta}_{\rm SR}\), introduced in Section 2, under potential covariate shift.

	\begin{theorem}\label{thm:SR_consist_rate_cs}
		Let Assumptions \ref{ass:MAR}-\ref{ass:bound_eigenvalues_s} hold and $s=\|\bbeta^*\|_0\ll(Nn)^{1/2}/\log d$. Then, as $n,d\to\infty$, 
		$\hat{\theta}_{\rm SR}-\theta=O_p[(M+\sigma_{\varepsilon})\{n^{-1/2}+s\log d/(Nn)^{1/2}\}].$
	\end{theorem}
	
	\begin{theorem}\label{thm:SR_asy_normal_cs}
		Let Assumptions \ref{ass:MAR}-\ref{ass:bound_eigenvalues_s} hold, $s=\|\bbeta^*\|_0\ll N^{1/2}/\log d$, and $M/\sigma_{\varepsilon}=O(1)$. Let $\hat\sigma_{v,2}^2=|\mathcal{J}|^{-1}\sum_{i\in \mathcal{J}}\hat{v}_i^2$, where \(\hat{v}_i = Z_i - \bW_i^\T \widehat{\bbeta}\).
		
		(a) If $\bSigma=\bSigma'$ (without covariate shift). Then, as $n,d\to\infty$, $\hat{\theta}_{\rm SR}-\theta=O_p\{(M+\sigma_\varepsilon)n^{-1/2}\}$ and 
		$(\hat{\theta}_{\rm SR}-\theta)/\hat\sigma_{\rm SR}\to N(0,1)$ in distribution, where 
		$$\hat\sigma_{\rm SR}^2=(\hat\sigma_{v,1}^2)^{-1}\sum_{i\in\mathcal{I}_1}\left(\dfrac{Y_i-\hat{\theta}_{\rm SR} \hat{v}_i}{|\mathcal{I}_1|}-\dfrac{\widehat{\bu}_{\rm SR}^\T \bW_i}{|\bar{\mathcal{J}}|}\right)^2+(\hat\sigma_{v,1}^2)^{-1}|\bar{\mathcal{J}}|^{-2}\sum_{i\in\mathcal{J} }(\widehat{\bu}_{\rm SR}^\T \bW_i)^2.$$
		
		(b) If $\bSigma\neq\bSigma'$ (with covariate shift). Then, as $n,m,d\to\infty$, $\hat{\theta}_{\rm SR}-\theta=O_p\{(M+\sigma_\varepsilon)/\sqrt n\}$ and $(\hat{\theta}_{\rm SR}-\theta)/\hat\sigma_{\rm SR}'\to\mathcal{N}(0,1)$ in distribution, where 
		\begin{align*}
			\hat\sigma_{\rm SR}'^2=(\hat\sigma_{v,1}^2)^{-1}\sum_{i\in\mathcal{I}_1}\left(\dfrac{Y_i-\hat{\theta}_{\rm SR} \hat{v}_i}{|\mathcal{I}_1|}-\dfrac{\widehat{\bu}_{\rm SR}^\T \bW_i}{|\bar{\mathcal{J}}|}\right)^2+\dfrac{\hat\sigma_{v,2}^2}{(\hat\sigma_{v,1}^2)^2}|\bar{\mathcal{J}}|^{-2}\sum_{i\in\mathcal{J} }(\widehat{\bu}_{\rm SR}^\T \bW_i)^2.
		\end{align*}
	\end{theorem}
	
	The following theorems outline the properties of the semi-supervised degrees-of-freedom adjusted estimator \(\hat{\theta}_{\rm DFA}\), as introduced in Section 3, when subject to potential covariate shift. 
	
	\begin{theorem}\label{thm:DF_consist_rate_cs}
		Let Assumption \ref{ass:MAR}-\ref{ass:bound_eigenvalues_s} hold, $s=\|\bbeta^*\|_0\ll(Nn)^{1/2}/\log d$, $k=\|\bgamma^*\|_0\ll n/\log d$, $k\ll d$, and $\Sigma_{jj}\le 1$ for all $j\in [d+1]$. Then, as $n,d\to\infty$, 
		$\hat{\theta}_{\rm DFA}-\theta=O_p[\sigma_{\varepsilon}n^{-1/2}+\sigma_{\varepsilon}\min\{k\log d/n,s\log d/(Nn)^{1/2}\}]$.
	\end{theorem}
	
	\begin{theorem}\label{thm:DF_asy_normal_cs}
		Let the assumptions of Theorem \ref{thm:DF_consist_rate_cs} hold. Further assume at least one of the following sparsity conditions holds: (a) $k\ll n^{1/2}/\log d$ and $s\ll(Nn)^{1/2}/\log d$ or (b) $k\ll n/\log d$ and $s\ll N^{1/2}/\log d$. Denote $\sigma_{v,1}^2=E(v_i^2)$ and $\sigma ^2={E}(v_i^2\varepsilon_i^2)/\sigma_{v,1}^4$. Then, as $n,d\to\infty$, 
		$\hat\theta_{\rm DFA}-\theta=O_p(\sigma_\varepsilon/n^{1/2})$, $n^{1/2}(1-\hat{q}/n)(\hat\theta_{\rm DFA}-\theta)/\sigma\to N(0,1)$ in distribution, and 
		$$\hat{\sigma}_{\rm DFA}^2/\sigma ^2=1+o_p(1),\ {\rm where}\ \hat\sigma_{\rm DFA}^2={\sum_{i\in \mathcal{I}}(Y_i-\bX_i^\T\widehat{\bgamma}_{\rm DFA})^2}/{\sum_{i\in \mathcal{I}}(Z_i-\bW_i^\T\widehat{\bbeta}_{\rm DFA})^2}.$$
	\end{theorem}
	
	Theorems 1-3 in the main document are special cases of the more general Theorems \ref{thm:SR_consist_rate_cs}-\ref{thm:DF_asy_normal_cs} above, which account for covariate shift.

	\section{Modified semi-supervised sparsity-robust estimator under sub-Gaussian designs}\label{sec:SS-SR-subG}
	We extend our analysis to sub-Gaussian designs and address the possibility of a misspecified linear model (1). In particular, we allow for \(E(\varepsilon_i \mid \bX_i) \neq 0\), while requiring only that \(E(v_i \mid \bW_i) = 0\). Accordingly, we propose a modified semi-supervised sparsity-robust estimator that incorporates an additional sample-splitting technique, designed to mitigate bias in more general scenarios.
	
	Divide the labeled samples into three parts, indexed by $(\mathcal{I}_k)_{k=1}^3$, with $|\mathcal I_k|\asymp n$ for each $k\le3$. Denote $\mathcal{J}_1=\mathcal{I}_1\cup\mathcal{J}$ and $\mathcal{J}_2=\mathcal{I}_2\cup\mathcal{J}$. With some $ \lambda_{\bbeta}\asymp(\log d /n)^{1/2}$, we first obtain a Lasso estimate of $\bbeta^*$ using training samples $(\bX_i)_{i\in\mathcal{J}_1}$:
	\begin{equation*}
		\widetilde{\bbeta}_{\rm SR} = \mathop{\arg\min}_{\bbeta\in\mathbb{R}^{d}}\bigg\{|\mathcal{J}_1|^{-1}\sum_{i\in \mathcal{J}_1}(Z_i-\bW_i^\T\bbeta)^2+  \lambda_{\bbeta}\|\bbeta\|_1
		\bigg\}.
	\end{equation*}
	Let $\tilde{v}_i=Z_i-\bW_i^\T\widetilde{\bbeta}_{\rm SR}$, $\widetilde{\bxi}=|\mathcal{I}_3|^{-1}\sum_{i\in \mathcal{I}_3}\bW_iY_i$, $\widetilde{\bSigma}_{\bW}=|\mathcal{J}_2|^{-1}\sum_{i\in \mathcal{J}_2} \bW_i\bW_i^{\T}$, $\bar{Y}=|\mathcal{I}_3|^{-1}\sum_{i\in\mathcal{I}_3}Y_i$, and $\hat{V}_Y=|\mathcal{I}_3|^{-1}\sum_{i\in\mathcal{I}_3}(Y_i-\bar{Y})^2$. With some $\lambda_{\bu}\asymp (M+\sigma_{\varepsilon})(\log d /n)^{1/2}$ and constant $q>0$, we obtain
	\begin{equation}\label{def:u_hat2}
		\widetilde{\bu}_{\rm SR}= \mathop{\arg\min}_{\bu\in\mathbb{R}^{d}}\bigg\{\bu^{\T}\widetilde{\bSigma}_{\bW}\bu\ \ {\rm s.t.}\ \|\widetilde{\bxi}-\widetilde{\bSigma}_{\bW} \bu\|_\infty\le \lambda_{\bu},\ \max_{i\in \mathcal{J}_2} | \bW_i^\T \bu|\le \hat{V}_Y^{1/2}|\mathcal{J}_2|^{q}\bigg\}.
	\end{equation}
	The \emph{modified semi-supervised sparsity-robust estimator} of $\theta$ is proposed as
	\begin{equation*}
		\tilde{\theta}_{\rm SR}=\dfrac{|\mathcal{I}_1|^{-1}\sum_{i\in \mathcal{I}_1}\tilde{v}_iY_i-|\mathcal{J}_2|^{-1}\sum_{i\in \mathcal{J}_2}\widetilde{\bu}_{\rm SR}^{\T}\bW_i(Z_i-\bW_i^{\T}\widetilde{\bbeta}_{\rm SR})}{|\mathcal{I}_1|^{-1}\sum_{i\in \mathcal{I}_1}\tilde{v}_i^2}.
	\end{equation*}
	
	To generalize our results to sub-Gaussian designs and eliminate additional constraints on the signal-to-noise ratio outlined in Theorem 2, we implement a modified sample-splitting approach, dividing the labeled data into three parts. The additional constraint on \(\max_{i\in \mathcal{J}_2} | \bW_i^\T \bu|\) in \eqref{def:u_hat2} is introduced to ensure the asymptotic normality of the proposed estimator; see similar designs in \cite{javanmard2014confidence} and \cite{athey2018approximate}.
	
	
	\begin{assumption}[Sub-Gaussian Design]\label{ass:sub_gau_design}
		Let $\sigma_u>0$ be a constant independent of $n$.
		For any $i\in\mathcal I\cup\mathcal J$, let $\bX_i$ be a mean-zero sub-Gaussian random vector satisfying $\|\bX_i^\T \mathbf{a}\|_{\psi_2}\le \sigma_u\|\mathbf{a}\|_2$ for all $\mathbf{a}\in \mathbb{R}^{d+1}$. For any $i\in\mathcal I$, let $\varepsilon_i$ be a mean-zero sub-Gaussian random variable satisfying $\|\varepsilon_i\|_{\psi_2}\le\sigma_u\sigma_{\varepsilon}$, where $\sigma_{\varepsilon}^2=E(\varepsilon_i^2)$ possibly depends on $n$.
	\end{assumption}
	
	
	In the following, we characterize the asymptotic behavior of the modified semi-supervised sparsity-robust estimator under sub-Gaussian designs and potential covariate shifts.
	
	\begin{theorem}\label{thm:SSR_consist_rate}
		Let Assumptions \ref{ass:MAR}-\ref{ass:sub_gau_design} hold, ${E}(v_i\mid \bW_i)=0$, and $s\ll(Nn)^{1/2}/\log d$. Then, as $n,d\to\infty$, 
		$\tilde{\theta}_{\rm SR}-\theta=O_p[(M+\sigma_{\varepsilon})\{n^{-1/2}+s\log d/(Nn)^{1/2}\}].$
	\end{theorem}

	\begin{theorem}\label{thm:SSR_asy_normal}
		Let the assumption of Theorem \ref{thm:SSR_consist_rate} hold and $s\ll N^{1/2}/\log d$. Additionally, with positive constants $c,c_t,c_v,C_v$, suppose that $\liminf \log d /\log (N)>0$, ${\rm var}(v_i\mid \bW_i)\ge c_v$, ${E}(|v_i|^{4+c}\mid \bW_i)\le C_v$, and ${E}[v_i^2\{\bW_i^\T(\theta\bbeta^*+\balpha^*)+\varepsilon_i\}^2]\ge c_t(M+\sigma_\varepsilon)^2$ for any $i\in\mathcal{I}\cup\mathcal{J}$. Let the constant $q$ in \eqref{def:u_hat2} satisfies $q<(2+c)/(8+2c)$. Denote $\tilde{\sigma}_{v,1}^2=|\mathcal I_1|^{-1}\sum_{i\in \mathcal{I}_1}\tilde{v}_i^2$. Then, as $n,d\to\infty$, $\tilde{\theta}_{\rm SR}-\theta=O_p\{(M+\sigma_{\varepsilon})n^{-1/2}\}$ and $(\tilde{\theta}_{\rm SR}-\theta)/\tilde\sigma_{\rm SR}\xrightarrow{\rm d}\mathcal{N}(0,1)$, where 
		$$\tilde{\sigma}_{\rm SR}^2=(\tilde{\sigma}_{v,1}^2)^{-2}|\mathcal I_1|^{-2}\sum_{i\in \mathcal{I}_1}(Y_i-\tilde{\theta}_{\rm SR}\tilde{v}_i)^2\tilde{v}_i^2+(\tilde{\sigma}_{v,1}^2)^{-2}|\mathcal J_2|^{-2}\sum_{i\in \mathcal{J}_2}(\widetilde{\bu}_{\rm SR}^\T \bW_i)^2\tilde{v}_i^2.$$
	\end{theorem}
	
	The convergence rate in Theorem \ref{thm:SSR_consist_rate} and the required sparsity condition in Theorem S6 align with those in Theorems 1 and 2, respectively. However, we now extend the framework to accommodate sub-Gaussian \(\bX_i\) and \(\varepsilon_i\), as well as a misspecified linear model (1). To the best of our knowledge, when \(\bgamma^*\) is a high-dimensional dense vector, Theorems \ref{thm:SSR_consist_rate}-\ref{thm:SSR_asy_normal} are the first to offer estimation and inference results under non-Gaussian designs and model misspecification, even in degenerate supervised settings.
	
	\section{Semi-supervised doubly robust estimator under sub-Gaussian designs}\label{sec:sparse-subG}
	
	We further explore scenarios where \(\bgamma^*\) is sparse under sub-Gaussian designs. In this section, we allow both linear models (1) and (2) to be misspecified, permitting conditions where \(E(\varepsilon_i \mid \bX_i) \neq 0\) and \(E(v_i \mid \bW_i) \neq 0\). In the following, we introduce a cross-fitting technique to the debiased estimator proposed in Section 3, which can also be interpreted as a semi-supervised adaptation of the doubly robust or double machine learning method of \cite{chernozhukov2018double}.
	
	For a fixed integer $K\ge2$, divide both the labeled and unlabeled samples into $K$ equal parts, indexed by $(\mathcal{I}_{k})_{k=1}^K$ and $(\mathcal{J}_{k})_{k=1}^K$, respectively. For each $k\in[K]$, denote $\mathcal{I}_{-k}=\mathcal{I}/\mathcal{I}_{k}$, $\mathcal{J}_{-k}=\mathcal{J}/\mathcal{J}_{k}$, and $\bar{\mathcal{J}}_{-k}=\mathcal{I}_{-k}\cup\mathcal{J}_{-k}$. With some $\lambda_{\gamma}\asymp\sigma_{\varepsilon}(\log d /n)^{1/2}$, we obtain a Lasso estimate of $\bgamma^*$ using training samples $(\bX_i,Y_i)_{i\in\mathcal{I}_{-k}}$, that is, the labeled samples out of the $k$-th fold:
	\begin{equation*}
		\widehat{\bgamma}_{{\rm DR},-k}= \mathop{\arg\min}_{\gamma\in\mathbb{R}^{d+1}}\bigg\{|\mathcal{I}_{-k}|^{-1}\sum_{i\in  \mathcal{I}_{-k}}(Y_i-\bX_i^\T\gamma)^2+\lambda_{\gamma}\|\gamma\|_1\bigg\}.
	\end{equation*}
	With some $ \lambda_{\bbeta}\asymp(\log d /n)^{1/2}$, we also get a Lasso estimate of $\bbeta^*$ using both labeled and unlabeled covariates $(\bX_i)_{i\in\bar{\mathcal{J}}_{-k}}$ as follows:
	\begin{equation*}
		\widehat{\bbeta}_{{\rm DR},-k} = \mathop{\arg\min}_{\bbeta\in\mathbb{R}^{d}}\bigg\{|\bar{\mathcal{J}}_{-k}|^{-1}\sum_{i\in  \bar{\mathcal{J}}_{-k}}(Z_i-\bW_i^\T\bbeta)^2+ \lambda_{\bbeta}\|\bbeta\|_1\bigg\}.
	\end{equation*}
	The \emph{semi-supervised doubly robust} estimator is proposed as:
	\begin{align*}
		\hat\theta_{{\rm DR}}=K^{-1}\sum_{k=1}^{K}\left\{\be_1^\T\widehat{\bgamma}_{{\rm DR},-k}+\dfrac{\sum_{i\in\mathcal{I}_{k}} (Z_i-\bW_i^\T\widehat{\bbeta}_{{\rm DR},-k})(Y_i-\bX_i^\T\widehat{\bgamma}_{{\rm DR},-k})}{\sum_{i\in \mathcal{I}_{k}} (Z_i-\bW_i^\T\widehat{\bbeta}_{{\rm DR},-k})Z_i}\right\}.
	\end{align*}


	The asymptotic properties of the semi-supervised doubly robust estimator are characterized below.
	
	\begin{theorem}\label{thm:DR_consist_rate}
		Let Assumption \ref{ass:MAR}-\ref{ass:sub_gau_design} hold, $s\ll N/\log d$ and $k\ll n/\log d$. Suppose that either (a) there is no covariate shift, that is, $\bSigma=\bSigma'$ or (b) the model (2) is correctly specified, that is, $E(v_i\mid \bW_i)=0$. Then, as $n,d\to\infty$, 
		$\hat{\theta}_{\rm DR}-\theta=O_p[\sigma_{\varepsilon}\{n^{-1/2}+(sk)^{1/2}\log d/(Nn)^{1/2}\}]$.
	\end{theorem}
	\begin{theorem}\label{thm:DR_asy_normal}
		Let the assumptions of Theorem \ref{thm:DR_consist_rate} hold. Further suppose that there exist a constant $c_{\varepsilon}>0$ such that ${E}(\varepsilon_i^2\mid \bX_i)\ge c_{\varepsilon}\sigma_{\varepsilon}^2$. Let $sk\log^2d\ll N$. Denote $\sigma_{v,1}^2=E(v_i^2)$ and $\sigma ^2={E}(v_i^2\varepsilon_i^2)/\sigma_{v,1}^4$. Then, as $n,d\to\infty$, 
		$\hat{\theta}_{\rm DR}-\theta=O_p(\sigma_\varepsilon n^{-1/2})$ and $n^{1/2}(\hat{\theta}_{\rm DR}-\theta)/\sigma\xrightarrow{\rm d}\mathcal{N}(0,1)$.
		In addition, we have 
		\begin{align*}
			\dfrac{\hat\sigma_{\rm DR}^2}{\sigma ^2}=1+o_p(1),\;\;\mbox{where}\;\;\hat\sigma_{\rm DR}^2=\dfrac{n^{-1}\sum_{k=1}^{K}\sum_{i\in\mathcal I_k}(Z_i-\bW_i^\T\widehat{\bbeta}_{{\rm DR},-k})^2(Y_i-\bX_i^\T\widehat{\bgamma}_{{\rm DR},-k})^2}{\left\{n^{-1}\sum_{k=1}^{K}\sum_{i\in \mathcal{I}_{k}}(Z_i-\bW_i^\T\widehat{\bbeta}_{{\rm DR},-k})^2\right\}^2}.
		\end{align*}
	\end{theorem}

	In contrast to the semi-supervised degrees-of-freedom adjusted method and its supervised versions \citep{bellec2022biasing,javanmard2018debiasing,celentano2023lasso}, the above results do not rely on Gaussian designs. In Theorems \ref{thm:DR_consist_rate}-\ref{thm:DR_asy_normal}, both (1) and (2) can be misspecified if $\bX_i$ has the same marginal distribution among the labeled and unlabeled groups; when covariate shift occurs, we still allow a misspecified model (1). The asymptotic variance in Theorem \ref{thm:DR_asy_normal} coincides with the semi-supervised degrees-of-freedom adjusted estimator, but the normal results require a different product sparsity condition $sk\log^2d\ll N$.
	
	The supervised doubly robust method \citep{chernozhukov2018double} is a degenerate case of the introduced semi-supervised doubly robust method, yielding root-$n$ inference under the conditions $s\ll n/\log d$, $k\ll n/\log d$, and $sk\log^2d\ll n$. As demonstrated in Theorem \ref{thm:DR_asy_normal}, the semi-supervised doubly robust method accommodates broader sparsity scenarios, requiring $s\ll N/\log d$, $k\ll n/\log d$, and $sk\log^2d\ll N$ -- with the additional flexibility of only needing $k=\|\bgamma^*\|_0\ll n/\log d$ when the total sample size $N$ is sufficiently large.

	\section{Additional simulation results}\label{sec:sim-add}
	
	We illustrate the performance of considered estimators in Section 4.1 under additional setups.
	
	Model 4: Non-linear $E(Y_i\mid \bX_i)$ with sparse models, $\|\bbeta^*\|_0=1$ and $\|\bgamma^*\|_0=5$ (Model 2 of \cite{deng2023optimal}). For $i\in \mathcal{I}\cup \mathcal{J}$, we first generate $(Z_i,\bW_i^\top)^\top\sim \mathcal{N}(0,\bSigma)$ with $\bSigma_{jk}=0.3^{|j-k|}$. Then, we replace the first component $W_{i1}$ of $\bW_i$ by its absolute value while keeping other components unchanged. Let $Y_i=0.6(W_{i1}+W_{i2})^2+0.4W_{i4}^3-W_{i5}+2Z_i+\varepsilon_i$ for $i\in\mathcal{I}$, where $\varepsilon_i\sim\mathcal{N}(0,1)$. We set $d=399$, $n=150$, and $m=1500$.
	
	Model 5: Non-linear $E(Y_i\mid \bX_i)$ with relatively sparse models, $\|\bbeta^*\|_0=25$ and $\|\bgamma^*\|_0=10$. Let $Y_i=0.4Z_i+0.4\sum_{j=1}^{4}W_{ij}+0.4\sum_{j=4}^{9}W_{ij}W_{i,j+1} + \varepsilon_i$, where $\varepsilon_i\sim\mathcal{N}(0,0.2^2)$. The choices of $Z_i$, $\bW_i$, and $v_i$ are the same as in Model 1. We set $d=499$, $n=300$, $s=25$, and $m=1000$.
	
	Model 6: Non-linear $E(Y_i\mid \bX_i)$ with a dense primary model and a sparse auxiliary model, $\|\bbeta^*\|_0=25$ and $\|\bgamma^*\|_0=500$. Let $Y_i=0.4Z_i+0.06\sum_{j=1}^{399}W_{ij}+0.03\sum_{j=399}^{498}W_{ij}W_{i,j+1}+\varepsilon_i$ where $\varepsilon_i\sim\mathcal{N}(0,0.2^2)$. The choices of $Z_i$, $\bW_i$, and $v_i$ are the same as in Model 1. We set $d=499$, $n=300$, $s=25$, and $m=1000$.
	
	Model 7: Covariate shift with sparse linear models, $\|\bbeta^*\|_0=9$ and $\|\bgamma^*\|_0=5$. For any $i\in\mathcal I$ and $j\in\mathcal J$, let $\bW_i\sim\mathcal{N}(0,\bSigma_{\bW})$, $\bW_j\sim \mathcal{N}(\mathbf{1}_d,\bSigma_{\bW}^\prime)$, $Z_i=-\sum_{l=1}^{s}0.4W_{il}+v_i$, $Z_j=-\sum_{l=1}^{s}0.4W_{jl}+v_j$, and $Y_i=0.4Z_i+0.4\sum_{l=1}^{4}W_{il}+\varepsilon_i$, where $\bSigma_{\bW}$ is defined in Model 1, $(\bSigma_{\bW}^\prime)_{il}=I_{\{i=l\}}+0.1I_{\{1\le |i-l|\le 5\}}$, $v_i\sim\mathcal{N}(0,0.8^2)$, $v_j\sim\mathcal{N}(0,1.2^2)$, and $\varepsilon_i\sim\mathcal{N}(0,0.4^2)$. We set $d=399$, $n=350$, $s=9$, and $m=1000$.

	Table \ref{models456} shows that the single linear coefficient estimate obtained from the supervised Lasso method has relatively large bias and RMSE. Although the S-SSL method of \cite{deng2023optimal} incorporates unlabeled samples, it is primarily developed for estimating the full linear coefficient vector rather than a single component. In addition, the S-SSL approach depends on a reliable non-linear estimate of the true primary model. However, random forests, which are used for this step, tend to produce substantial estimation errors in high-dimensional settings. Consequently, the single coefficient estimates from the S-SSL method also suffer from relatively large errors and can sometimes perform worse than those from the supervised Lasso.
	
		\begin{table}[t]
		\centering
		\renewcommand{\arraystretch}{0.8}
		\caption{Simulation results under Models 4-7}
		\vspace{0em}
		\hspace*{-0.84cm}
		{\resizebox{1.035\textwidth}{!}{\begin{tabular}{lcccclcccc}
					\hline
					\hline\vspace{-1.6em}\\
					Estimator  & Bias  & RMSE  & Length   & Coverage &Estimator  & Bias  & RMSE  & Length   & Coverage\\[0.3em]
					\hline\vspace{-1.6em}\\
					\multicolumn{10}{c}{(A) Model 4 with $d=399$, $n=150$, $\|\bbeta^*\|_0=1$, $\|\bgamma^*\|_0=5$, and $m=1500$}\\[0.3em]
					\hdashline\vspace{-1.6em}\\
					Lasso   &$-0.4463$ & $0.4881$ & / &/ &S-SSL& $-0.6498$ & $0.6859$ & / & /\\
					SR   &$0.1943$ & $0.7599$ & $1.3220$ & $89.0\%$ & SS-SR  &$0.0028$ & $0.3101$ & $1.1832$ & $94.6\%$\\
					DFA   &$-0.0837$ & $0.1980$ & $0.7880$ & $94.4\%$ & SS-DFA  &$-0.0722$ & $0.1950$ & $0.7501$ & $94.6\%$\\
					DR   &$-0.1051$ & $0.2257$ & $0.7504$ & $91.0\%$ & SS-DR  &$-0.0354$ & $0.2005$ & $0.8083$ & $95.4\%$\\
					\hline\vspace{-1.6em}\\
					\multicolumn{10}{c}{(B) Model 5 with $d=499$, $n=300$, $\|\bbeta^*\|_0=25$, $\|\bgamma^*\|_0=10$, and $m=1000$}\\[0.3em]
					\hdashline\vspace{-1.6em}\\
					Lasso   &$-0.1455$ & $0.1545$ & / &/ &S-SSL& $-0.1767$ & $0.1845$ & / & /\\
					SR    & $0.0403$ & $0.1419$ & $0.3873$ & $90.0\%$ & SS-SR   & $0.0262$ & $0.0872$ & $0.3286$ & $94.8\%$\\
					DFA    &$-0.0390$ & $0.0648$ & $0.2799$ & $95.6\%$ &SS-DFA   &$-0.0278$ & $0.0602$ & $0.2746$ & $97.6\%$  \\
					DR    &$-0.0504$ & $0.0735$ & $0.2053$ & $83.4\%$ & SS-DR   &$-0.0244$ & $0.0610$ & $0.2434$ & $95.8\%$\\
					\hline\vspace{-1.6em}\\
					\multicolumn{10}{c}{(C) Model 6 with $d=499$, $n=300$, $\|\bbeta^*\|_0=25$, $\|\bgamma^*\|_0=500$, and $m=1000$}\\[0.3em]
					\hdashline\vspace{-1.6em}\\
					Lasso   &$-0.2146$ & $0.2233$ & / &/ &S-SSL& $-0.2234$ & $0.2282$ & / & /\\
					SR    &$-0.0551$ & $0.1555$ & $0.4826$ & $90.2\%$ & SS-SR   &$-0.0066$ & $0.1016$ & $0.4137$ & $94.8\%$\\
					DFA    & $0.0241$ & $0.3133$ & $0.6270$ & $88.2\%$ & SS-DFA   & $0.0681$ & $0.3660$ & $0.5980$ & $94.8\%$ \\
					DR    &$-0.1244$ & $0.1380$ & $0.2460$ & $48.4\%$ & SS-DR   &$-0.0579$ & $0.0927$ & $0.3040$ & $89.8\%$\\
					\hline\vspace{-1.6em}\\
					\multicolumn{10}{c}{(D) Model 7 with $d=399$, $n=350$, $\|\bbeta^*\|_0=9$, $\|\bgamma^*\|_0=5$, and $m=1000$}\\[0.3em]
					\hdashline\vspace{-1.6em}\\
					Lasso   &$-0.0931$ & $0.0959$ & / &/ &S-SSL& $-0.1841$ & $0.1866$ & / & /\\
					SR   &$-0.0007$ & $0.0542$ & $0.2198$ & $95.2\%$ & SS-SR  &$-0.0121$ & $0.0604$ & $0.2166$ & $93.0\%$\\
					DFA   &$-0.0274$ & $0.0367$ & $0.1205$ & $91.0\%$ & SS-DFA  &$-0.0158$ & $0.0315$ & $0.1189$ & $94.2\%$\\
					DR   &$-0.0336$ & $0.0416$ & $0.1086$ & $78.6\%$ & SS-DR  &$-0.0161$ & $0.0329$ & $0.1160$ & $92.4\%$\\
					\hline
					\hline
		\end{tabular}}}
		\label{models456}
		\vspace{-1em}
	\end{table}

		\begin{figure}[t]
	\renewcommand{\thefigure}{S1} 
	\centering
	\subfloat[Model 4 with $\|\bbeta^*\|_0=1$ and $\|\bgamma^*\|_0=5$]{\includegraphics[width=.48\linewidth,trim={4cm 9.5cm 4cm 10cm},clip]{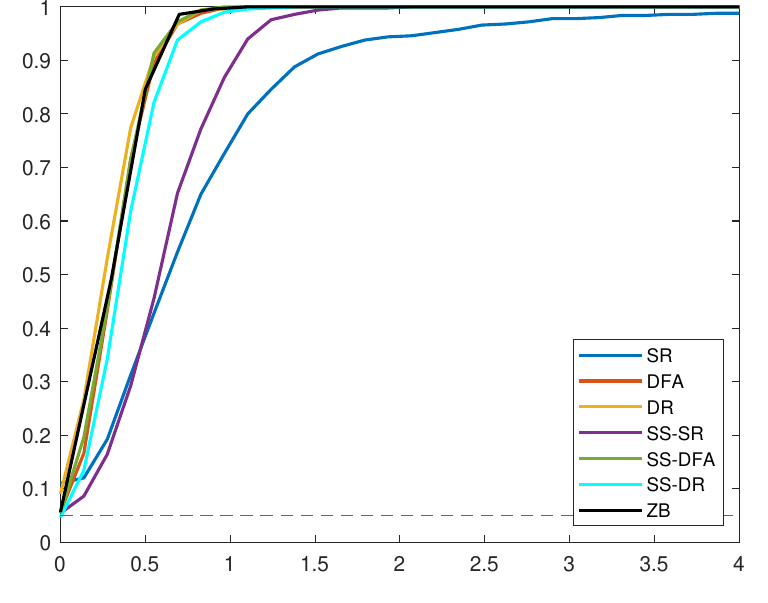}}\hfill
	\subfloat[Model 5 with $\|\bbeta^*\|_0=25$ and $\|\bgamma^*\|_0=10$]{\includegraphics[width=.48\linewidth,trim={4cm 9.5cm 4cm 10cm},clip]{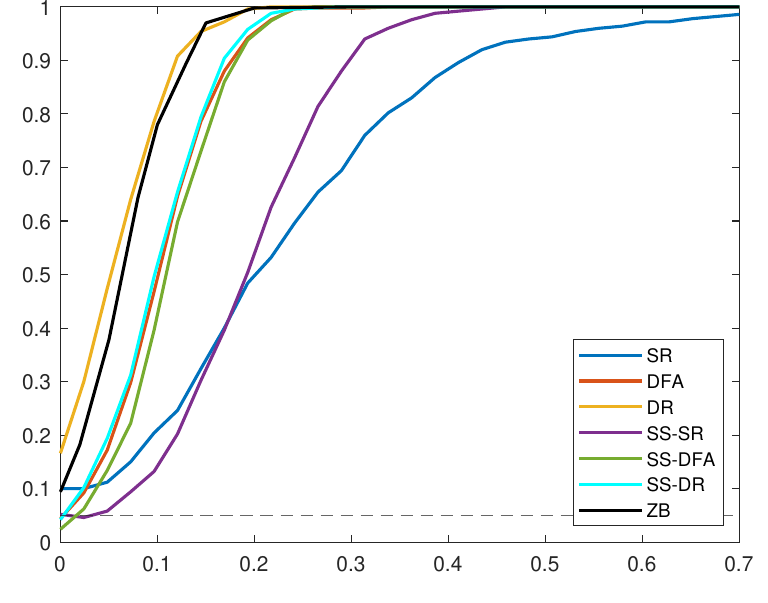}}\\[1ex]
	\subfloat[Model 6 with $\|\bbeta^*\|_0=25$ and $\|\bgamma^*\|_0=500$]{\includegraphics[width=.48\linewidth,trim={4cm 9.5cm 4cm 10cm},clip]{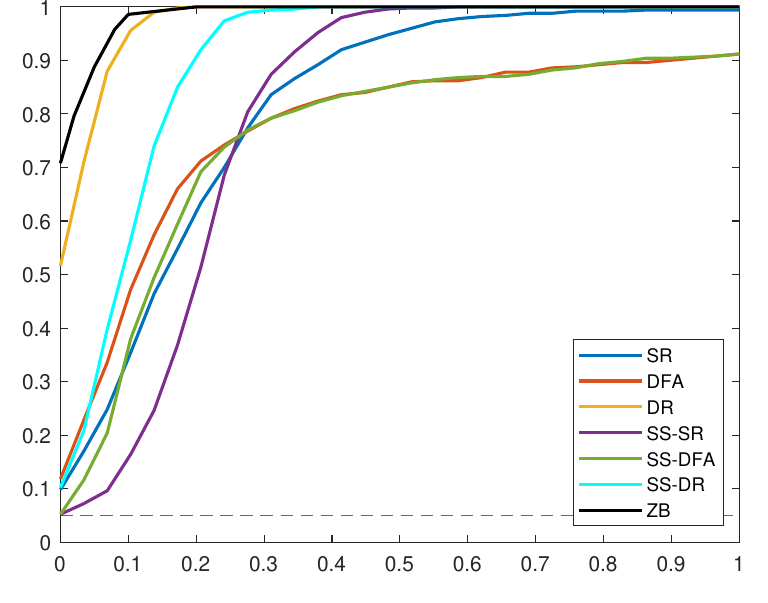}}\hfill
	\subfloat[Model 7 with $\|\bbeta^*\|_0=9$ and $\|\bgamma^*\|_0=5$]{\includegraphics[width=.48\linewidth,trim={4cm 9.5cm 4cm 10cm},clip]{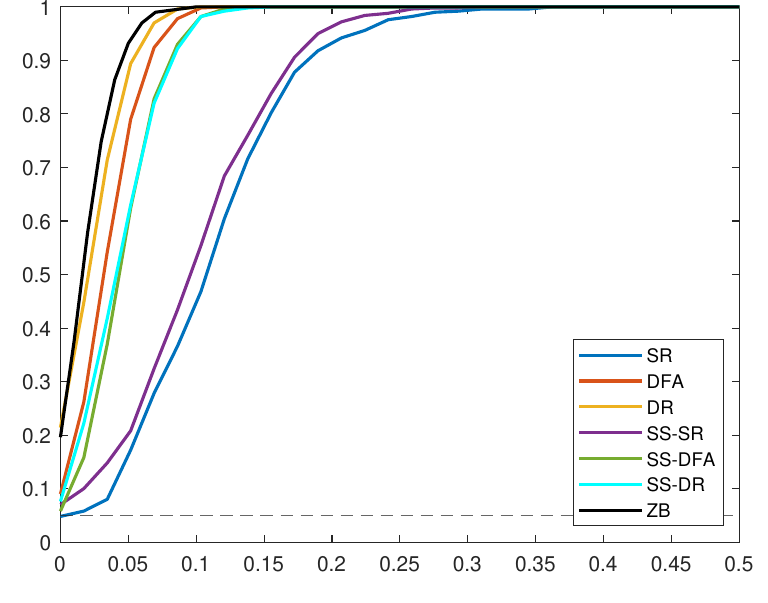}}
	\caption{Power curves of competing methods corresponding to the hypothesis $H_0:\theta=\theta_0+h$, where $\theta_0$ is the true regression coefficient, and the significance level is $\alpha=0.05$. The x-axis represents the deviation $h$, and the y-axis denotes the average rejection rate over 500 repetitions. Hence, $h=0$ corresponds to the Type I error, and the remaining values correspond to the power.}
	\label{fig:power2}
	\vspace{-2em}
\end{figure}
	
	Table \ref{models456} (A)-(C) illustrates the results when the linear model (1) is misspecified and the residual variable $\varepsilon_i$ is non-Gaussian. When the outcome model $E(Y_i\mid \bX_i)$ depends only on a small number of covariates, as in Table \ref{models456} (A)-(B), we can still observe good performance from the supervised degrees-of-freedom adjusted (DFA) and semi-supervised degrees-of-freedom adjusted (SS-DFA) methods, where the semi-supervised version provides a smaller bias and root mean squared error (RMSE). Additionally, the semi-supervised doubly robust (SS-DR) method has a similar performance compared with the SS-DFA method. The semi-supervised sparsity robust (SS-SR) method also provides robust inference results but with lower efficiency. On the other hand, if the outcome model has a dense structure, as in Table \ref{models456} (C), the RMSEs of the DFA and SS-DFA methods are relatively large. The SS-DR method provides a small RMSE, but the confidence interval's coverage is lower than the desired $95\%$. The supervised sparsity robust (SR) and SS-SR methods provide relatively small RMSEs, slightly larger than the SS-DR due to lower efficiency, along with robust inference results.

	We also consider the appearance of covariate shift. As shown in Table \ref{models456} (D), when both $\bgamma^*$ and $\bbeta^*$ are sparse, all the considered semi-supervised methods provide coverages close to $95\%$; due to the lower efficiency, the SS-SR method provides a wider confidence interval and a larger RMSE. For the supervised methods, we can see that the supervised SR method also provides a good coverage, although the RMSE is larger than the other supervised methods.

	The power curves are illustrated in Figure \ref{fig:power2}. Under Model 4, the SR and SS-SR methods yield relatively low power, while the other methods exhibit comparable performance. This is because both the primary and auxiliary models involve only a small number of covariates, resulting in relatively small estimation bias. In contrast, the SR and SS-SR methods do not account for the sparse structure of the primary model, which limits their effectiveness in this setting. Under Model 5, all three considered semi-supervised methods and the supervised DFA method exhibit Type I errors smaller than or close to $\alpha=0.05$. The SS-SR method also shows smaller powers compared to the others, attributed to its lower efficiency. For Model 6, only the SS-SR and SS-DFA methods yield acceptable Type I errors. The SS-SR method displays higher powers for large $h$ and lower powers for small $h$. Under Model 7, the SR, SS-SR, and SS-DFA methods maintain Type I errors close to $\alpha=0.05$, with the SS-DFA method demonstrating higher powers for arbitrary $h>0$.

	\section{Supplementary results from the real data application}\label{sec:real_data2}
	
	\subsection{Additional details on the application to the ACTG175 dataset}
	Now we introduce how the covariates are constructed. Pre-treatment baseline covariates of the ACTG175 dataset include $5$ continuous variables: age, weight, Karnofsky score (karnof), CD4 cell count (CD40), and CD8 cell count (CD80). Additionally, there are $7$ binary variables: hemophilia (hemo), homosexual (homo), history of intravenous drug use (drugs), race, gender, antiretroviral history (str2), and a symptomatic indicator (symptom).  We let $Z_i$ represent the individual's age. Following \cite{liang2020empirical}, the vector $\bW_i$ comprises linear and quadratic terms of the continuous baseline covariates, the linear and interaction terms of binary covariates, along with interactions between the two sets generated, i.e.,
	\begin{align*}
		&\left({\rm weight}+{\rm karnof}+{\rm CD40}+{\rm CD80}+1\right)^2\times\big\{({\rm hemo}+{\rm homo}+{\rm drugs}+{\rm race}+{\rm gender}+{\rm str2} \\
		&\qquad+{\rm symptom}+1)^2-{\rm hemo}^2-{\rm homo}^2-{\rm drugs}^2-{\rm race}^2-{\rm gender}^2-{\rm str}^2-{\rm symptom}^2\big\}.
	\end{align*}
	The quadratic terms of binary variables are excluded since they are the same as the linear ones. Additionally, we exclude all the terms involving the age variable $Z_i$ from $\bW_i$. This procedure leads to $d=434$ covariates. Since treatments are assigned randomly, there is no covariate shift between the labeled and unlabeled groups.
	
	\subsection{Analysis based on the NHEFS dataset}
	We further apply the proposed method to the data from the National Health and Nutrition Examination Survey Data I Epidemiologic Follow-up Study (NHEFS). The NHEFS was launched through a collaboration between the National Center for Health Statistics and the National Institute on Aging, along with several other agencies within the U.S. Public Health Service. Its primary goal was to examine how various clinical, nutritional, and behavioral factors influence health outcomes such as disease incidence and mortality. Previous studies have examined the impact of smoking cessation on weight gain \citeand{hernan2020causal}{ertefaie2020nonparametric}. In this section, we shift the focus to study the effect of individual age on weight gain and assess the performance of the proposed semi-supervised methods.
	
	The subset of the NHEFS data we analyze includes $1566$ individuals who participated in the study in 1971, with their age at that time used as the primary predictor. These participants were followed up in 1982, and the weight change between 1971 and 1982 serves as the outcome variable. The control covariates include $54$ original features measured in 1971, such as participants' sex, blood pressure, race, height, weight, education level, smoking status, medical history, and alcohol consumption. All categorical variables were converted into dummy variables, resulting in a total of $d=86$ covariates.
	
	\begin{table}[t]
		\centering
		\caption{Results from real-data analysis using the NHEFS dataset}
		\renewcommand{\arraystretch}{0.8}
		\vspace{0em}
		\begin{tabular}{lcccccc}
			\toprule
			Estimator  & SR & SS-SR & DFA & SS-DFA & DR & SS-DR\\
			\midrule
			\multicolumn{7}{c}{$r=1/3$}\\[0.3em]
			\hdashline\vspace{-1.6em}\\
			Estimate      & -0.2832 & -0.2770 & -0.2655 & -0.2692 & -0.2439 & -0.2427 \\
			Length        & 0.3422  & 0.3684  & 0.3024  & 0.3048  & 0.2950  & 0.3056 \\
			Bias          & -0.0165 & -0.0103 & -0.0034 & -0.0071 & 0.0052  & 0.0064 \\
			RMSE          & 0.1214  & 0.1146  & 0.0581  & 0.0562  & 0.0502  & 0.0491 \\
			\midrule
			\multicolumn{7}{c}{$r=1/2$}\\[0.3em]
			\hdashline\vspace{-1.6em}\\
			Estimate      & -0.2777 & -0.2647 & -0.2746 & -0.2560 & -0.2541 & -0.2443 \\
			Length        & 0.2920  & 0.3022  & 0.2566  & 0.2248  & 0.2626  & 0.2468 \\
			Bias          & -0.0110 & 0.0020  & -0.0125 & 0.0061  & -0.0050 & 0.0048 \\
			RMSE          & 0.1015  & 0.0994  & 0.0466  & 0.0323  & 0.0442  & 0.0316 \\
			\midrule
			\multicolumn{7}{c}{$r=2/3$}\\[0.3em]
			\hdashline\vspace{-1.6em}\\
			Estimate      & -0.2740 & -0.2653 & -0.2756 & -0.2614 & -0.2602 & -0.2455 \\
			Length        & 0.2598  & 0.2624  & 0.2244  & 0.2046  & 0.2342  & 0.2110 \\
			Bias          & -0.0073 & 0.0014  & -0.0135 & 0.0007  & -0.0111 & 0.0036 \\
			RMSE          & 0.0900  & 0.0825  & 0.0379  & 0.0252  & 0.0418  & 0.0225 \\
			\midrule
			\multicolumn{7}{c}{$r=1$}\\[0.3em]
			\hdashline\vspace{-1.6em}\\
			Estimate           & -0.2667 &   /      & -0.2621 &   /      & -0.2491 &   /     \\
			Length        & 0.2068  &   /      & 0.1732  &   /      & 0.1696  &   /     \\
			\toprule
		\end{tabular}
		\label{table:rd2}
	\end{table}

	In Table \ref{table:rd2}, we report the point estimates and $95\%$ confidence interval lengths for the supervised estimators SR, DFA, and DR described in Section 4.1, based on the full dataset. All results suggest a negative association between age and weight gain. This may be because aging is associated with a decline in metabolic rate, which could make it more difficult to gain weight.

	To evaluate the performance of the semi-supervised methods, we randomly draw a fraction $r \in \{1/3, 1/2, 2/3\}$ of samples from the dataset as labeled data, with the remaining samples treated as unlabeled data. We then apply the SS-SR, SS-DFA, and SS-DR estimators to the resulting semi-supervised datasets, and the SR, DFA, and DR estimators to the corresponding labeled datasets. The point estimates from the full-data supervised estimators are treated as reference values. We report the point estimate (Estimate), $95\%$ confidence interval length (Length), empirical bias (Bias), and root mean square error (RMSE). Table \ref{table:rd2} summarizes the results based on 500 repetitions. In all settings, the semi-supervised methods consistently outperform their supervised counterparts in terms of RMSE, demonstrating the benefit of incorporating unlabeled data.

	\section{Justifications of the minimax lower bounds in Remark 1}\label{S4:minimax}
	Let $Y=(Y_1,\cdots,Y_n)$, $X=(X_1^\T,\cdots,X_n^\T)^\T$, and $\tilde{X}=(X_1^\T,\cdots,X_{N}^\T)^\T$. We first show that for any $Q,M_0,\sigma_0>0$,
	\begin{equation}\label{half}
		Q\cdot L_\alpha^*\{\mathcal{T}(s,M_0,\sigma_0^2)\}\ge L_\alpha^*\{\mathcal{T}(s,QM_0,Q\sigma_0^2)\}.
	\end{equation}
	To this end, we first show that for any $\varepsilon > 0$, it holds that $Q\cdot[L_\alpha^*\{\mathcal{T}(s,M_0,\sigma_0^2)\} + \varepsilon] \ge L_\alpha^*\{\mathcal{T}(s,QM_0,Q\sigma_0^2)\}$. For the sake of simplicity, we use $(Y,\tilde{X}) \sim \tau$ to indicate that $(Y,\tilde{X})$ follows the joint distribution induced by parameter $\tau$. For any $\varepsilon>0$, by the definition of $L_\alpha^*\{\mathcal{T}(s,M_0,\sigma_0^2)\}$, there exist a $\tilde{J}_\alpha \in C_\alpha\{\mathcal{T}(s,M_0,\sigma_0^2)\}$ such that
	\begin{equation}\label{infL}
		\begin{split}
			L_\alpha^*\{\mathcal{T}(s,M_0,\sigma_0^2)\}+\varepsilon&\ge \sup_{\tau\in\mathcal{T}(s,M_0,\sigma_0^2)}{E}_{(Y,\tilde{X})\sim \tau}[L\{\tilde{J}_\alpha(Y,\tilde{X})\}]\\
			&=\sup_{\tau\in\mathcal{T}(s,M_0,\sigma_0^2)}{E}_{(Q^{-1}Y,\tilde{X})\sim \tau}[L\{\tilde{J}_\alpha(Q^{-1}Y,\tilde{X})\}].
		\end{split}
	\end{equation}
	For any $\tau=(\gamma^*,\Sigma,\sigma_{\varepsilon})$, denote $Q\odot\tau =(Q\gamma^*,\Sigma,Q\sigma_{\varepsilon})$, then
	\begin{equation}\label{infQ_1}
		\begin{split}
			Q[L_\alpha^*\{\mathcal{T}(s,M_0,\sigma_0^2)\}+\varepsilon]&\ge Q\sup_{\tau\in\mathcal{T}(s,M_0,\sigma_0^2)}{E}_{(Q^{-1}Y,\tilde{X})\sim \tau}[L\{\tilde{J}_\alpha(Q^{-1}Y,\tilde{X})\}]\\
			&=Q\sup_{\tau\in\mathcal{T}(s,M_0,\sigma_0^2)}{E}_{(Y,\tilde{X})\sim Q\odot\tau}[L\{\tilde{J}_\alpha(Q^{-1}Y,\tilde{X})\}]\\
			&=Q\sup_{\tau\in\mathcal{T}(s,QM_0,Q\sigma_0^2)}{E}_{(Y,\tilde{X})\sim \tau}[L\{\tilde{J}_\alpha(Q^{-1}Y,\tilde{X})\}]\\
			&=\sup_{\tau\in\mathcal{T}(s,QM_0,Q\sigma_0^2)}{E}_{(Y,\tilde{X})\sim \tau}[L\{Q\tilde{J}_\alpha(Q^{-1}Y,\tilde{X})\}],
		\end{split}
	\end{equation}
	where the first inequality is by \eqref{infL}, the first equality holds since $(Q^{-1}Y,\tilde{X})\sim \tau$ implies $(Y,\tilde{X})\sim Q\odot \tau$, the second equality is by $\{Q\odot\tau\,|\,\tau\in\mathcal{T}(s,M_0,\sigma_0^2)\}=\{\tau\,|\,\tau\in\mathcal{T}(s,QM_0,Q\sigma_0^2)\}$,
	and the third equality is by $QL\{\tilde{J}_\alpha(Q^{-1}Y,\tilde{X})\}=L\{Q\tilde{J}_\alpha(Q^{-1}Y,\tilde{X})\}$.
	We next show that $QJ_\alpha^*(Q^{-1}Y,\tilde{X})\in\mathcal{T}(s,QM_0,Q\sigma_0^2)$. For any $\tau_0\in \mathcal{T}(s,QM_0,Q\sigma_0^2)$,
	\begin{align*}
		{P}_{(Y,\tilde{X})\sim \tau_0}\{\theta\in QJ_\alpha^*(Q^{-1}Y,\tilde{X})\}
		&={P}_{(Y,\tilde{X})\sim \tau_0}\{Q^{-1}\theta\in \tilde{J}_\alpha(Q^{-1}Y,\tilde{X})\}\\
		&={P}_{(Q^{-1}Y,\tilde{X})\sim Q^{-1}\odot\tau_0}\{Q^{-1}\theta\in \tilde{J}_\alpha(Q^{-1}Y,\tilde{X})\}
		\ge 1-\alpha,
	\end{align*}
	where the second equality holds as $(Y,\tilde{X})\sim \tau$ implies $(Q^{-1}Y,\tilde{X})\sim Q^{-1}\odot \tau$, the last inequality is by $Q^{-1}\odot \tau\in\mathcal{T}(s,M_0,\sigma_0^2)$ and $\tilde{J}_\alpha\in C_\alpha\{\mathcal{T}(s,M_0,\sigma_0^2)\}$. Thus $Q\tilde{J}_\alpha(Q^{-1}Y,\tilde{X})\in\mathcal{T}(s,QM_0,Q\sigma_0^2)$ and 
	\begin{equation}\label{infQ_2}
		\sup_{\tau\in\mathcal{T}(s,QM_0,Q\sigma_0^2)}{E}_{(Y,\tilde{X})\sim \tau}[L\{Q\tilde{J}_\alpha(Q^{-1}Y,\tilde{X})\}]\ge L_\alpha^*\{\mathcal{T}(s,QM_0,Q\sigma_0^2)\}.
	\end{equation}
	Combining \eqref{infQ_1} and \eqref{infQ_2}, $QL^*\{\mathcal{T}(s,M_0,\sigma_0^2)\}\ge L^*\{\mathcal{T}(s,QM_0,Q\sigma_0^2)\}$.
	Similarly, we also have $Q^{-1}L_\alpha^*\{\mathcal{T}(s,QM_0,Q\sigma_0^2)\}\ge L_\alpha^*\{\mathcal{T}(s,M_0,\sigma_0^2)\}$ by setting $(Q,M_0,s)=(Q^{-1},QM_0,Q\sigma_0^2)$ in \eqref{half}. 
	Thus 
	\begin{equation}\label{QL}
		QL_\alpha^*\{\mathcal{T}(s,M_0,\sigma_0^2)\}= L_\alpha^*\{\mathcal{T}(s,QM_0,Q\sigma_0^2)\}.
	\end{equation}
	We next prove that there exist a constant $C$ such that 
	\begin{equation}\label{L110}
		L^*\{\mathcal{T}(1,1,0)\}\ge \dfrac{C}{n^{1/2}}.
	\end{equation}
	Combining \eqref{QL} and \eqref{L110}, 
	\begin{equation}\label{M_sqrtn}
		L_\alpha^*\{\mathcal{T}(s,M_0,\sigma_0^2)\}\ge L_\alpha^*\{\mathcal{T}(1,M_0,0)\}\ge M_0L_\alpha^*\{\mathcal{T}(1,1,0)\}\ge \dfrac{CM_0}{n^{1/2}},
	\end{equation}
	since $\mathcal{T}(1,M_0,0)\subseteq\mathcal{T}(s,M_0,\sigma_0^2)$ as $s\ge 1$ and $\sigma_0^2\ge 0$. We define a null space $\mathcal{H}_0 = \{((0,{\alpha^*}^\T)^\T, I_{d+1}, 0)\}$ and an alternative space $\mathcal{H}_1 = \{((r_n, {\alpha^*}^\T)^\T, I_{d+1}, 0)\}$, where $r_n$ is a sequence to be determined. Define $\mathcal{B} = \{\alpha \,|\, \|\alpha\|_2 \le 1\}$. Let $\alpha^*$ follow a Gaussian prior with mean zero and variance $C_1^2I_{d}/d$, truncated by $\mathcal{B}$, where $C_1$ is a constant to be determined, $I_d$ denotes the $d\times d$ identity matrix. Let $\mu(\cdot)$ denotes the density function of the Gaussian distribution $\mathcal{N}(0, C_1^2I_{d}/d)$. Let $f_i(t), i = 0,1$ denote the density of $T=(Y,\tilde{X})\sim\tau$ when $\tau$ lies in $\mathcal H_0$ and $\mathcal H_1$ defined above, that is,
	$f_0(t)=\{1/(\int_{\mathcal{B}}\mu(\alpha)d\alpha)\}\int_{\mathcal{B}}g_{\{(0,{\alpha}^\T)^\T,I_{d+1},0\}}(t)\mu(\alpha)d\alpha$ and $f_1(t)=\{1/(\int_{\mathcal{B}}\mu(\alpha)d\alpha)\}\int_{\mathcal{B}}g_{\{(r_n,{\alpha}^\T)^\T,I_{d+1},0\}}(t)\mu(\alpha)d\alpha$,
	where $g_\tau(t)$ denotes the density function of $T=(Y,\tilde{X})$ induced by a deterministic parameter $\tau$. By Lemma 1 of \cite{cai2017confidence}, 
	\begin{equation}\label{lower_TV}
		{E}[L\{J_\alpha(Y,\tilde{X})\}]\ge r_n\{1-2\alpha-{\rm TV}(f_1,f_0)\}_{+},
	\end{equation}
	where ${\rm TV}(f_1,f_0)=\frac{1}{2}\int |f_1(t)-f_0(t)|dt$ denotes the total variation (TV) distance. To prove \eqref{L110}, we choose appropriate $r_n$ and $C_1$ to control ${\rm TV}(f_1,f_0)$. Further define spaces $\mathcal{H}_0^\prime=\{((0,{\alpha^*}^\T)^\T,I_{d+1},0)\}$ and $\mathcal{H}_1^\prime=\{((r_n,{\alpha^*}^\T)^\T,I_{d+1},0)\}$, where $\alpha^*$ follows a Gaussian prior $\mathcal{N}(0,C_1^2I_{d}/d)$ (without truncation). Let
	$f_0^\prime(t)=\int_{\mathbb{R}^d} g_{\{(0,{\alpha}^\T)^\T,I_{d+1},0\}}(t)\mu(\alpha)d\alpha$ and $f_1^\prime(t)=\int_{\mathbb{R}^d} g_{\{(r_n,{\alpha}^\T)^\T,I_{d+1},0\}}(t)\mu(\alpha)d\alpha$.
	Notice that $${\rm TV}(f_1,f_0)\le {\rm TV}(f_1,f_1^\prime)+{\rm TV}(f_1^\prime,f_0^\prime)+{\rm TV}(f_0^\prime,f_0).$$ We next control the three terms on the right-hand-side above. For any arbitrary constant $\alpha\in(0,1/2)$, choose $\delta$ satisfying $0<\delta<(1/2-\alpha)/5$. Without loss of generality, we consider the case that $d=2n$. If $d > 2n$, we can simply apply the distribution to the first $2n$ elements of $\alpha^*$ and leave the other elements to be zero.
	
	\textbf{Step (1)}: Control ${\rm TV}(f_1,f_1^\prime)$ and ${\rm TV}(f_0^\prime,f_0)$. 
	We use $g_{\alpha}(t)$ to represent $g_{((0,{\alpha}^\T)^\T,I_{d+1},0)}(t)$ for the sake of simplifying notation. Then
	\begin{align*}
		{\rm TV}(f_1,f_1^\prime)&=\dfrac{1}{2}\int \left|\dfrac{1}{\int_{\mathcal{B}}\mu(\alpha)d\alpha}\int_{\mathcal{B}} g_{\alpha}(t)\mu(\alpha)d\alpha-\int_{\mathbb{R}^d} g_{\alpha}(t)\mu(\alpha)d\alpha\right|dt\\
		&=\dfrac{1}{2}\int\left|\left\{\dfrac{1}{\int_{\mathcal{B}}\mu(\alpha)d\alpha}-1\right\}\int_{\mathcal{B}} g_{\alpha}(t)\mu(\alpha)d\alpha-\int_{\mathcal{B}^c} g_{\alpha}(t)\mu(\alpha)d\alpha\right|dt \\
		&\le \dfrac{1}{2}\left|\left\{\dfrac{1}{\int_{\mathcal{B}}\mu(\alpha)d\alpha}-1\right\}\right|\int\int_{\mathcal{B}}g_{\alpha}(t)\mu(\alpha)d\alpha dt+\dfrac{1}{2}\int\int_{\mathcal{B}^c}g_{\alpha}(t)\mu(\alpha)d\alpha dt\\
		&\le\dfrac{1}{2} \left|\left\{\dfrac{1}{\int_{\mathcal{B}}\mu(\alpha)d\alpha}-1\right\}\right|+\dfrac{\int_{\mathcal{B}^c}\mu(\alpha)d\alpha}{2}
		=\dfrac{\int_{\mathcal{B}^c}\mu(\alpha)d\alpha}{2\{1-\int_{\mathcal{B}^c}\mu(\alpha)d\alpha\}}+\dfrac{\int_{\mathcal{B}^c}\mu(\alpha)d\alpha}{2},
	\end{align*}
	where the second inequality is by Fubini theorem and $\int g_{\alpha}(t)dt=1$. By Markov's inequality,
	\begin{align*}
		\int_{\mathcal{B}^c}\mu(\alpha)d\alpha&={P}(\|\alpha\|_2^2\ge 1)\le {E}(\|\alpha\|_2^2)={E}({\alpha}^\T\alpha)={E}\{{\rm tr}(\alpha{\alpha}^\T)\}={\rm tr}\{{E}(\alpha{\alpha}^\T)\}=C_1^2.
	\end{align*}
	Choose $C_1=(2\delta/3)^{1/2},$ then $\int_{\mathcal{B}^c}\mu(\alpha)d\alpha\le 2\delta/3< 1/2$ and ${\rm TV}(f_1,f_1^\prime)\le 3\mu_n(\mathcal{B}^c)/2< \delta$.
	Similarly, we also have ${\rm TV}(f_0^\prime,f_0)< \delta$.
	
	\textbf{Step (2)}: Control ${\rm TV}(f_1^\prime,f_0^\prime)$. Let $X=(Z,W)$, where $Z=(Z_1,\cdots,Z_n)^\T$ and $W=(W_1^\T,\cdots,W_n^\T)^\T$. Define $\mathcal{A}=\{\tilde{X}\,|\,Z^\T(WW^\T)^{-1}Z\le C_2\}$.
	We next prove that ${P}(\tilde{X}\notin\mathcal{A})\le 2\delta$ when $C_2= 1/(0.09\delta)$. Define
	\begin{align*}
		\mathcal{M}_4=\Big\{(2n)^{1/2}-n^{1/2}-0.1n^{1/2}\le &\{\lambda_{\rm min}(WW^\T)\}^{1/2}\\
		\le &\{\lambda_{\rm max}(WW^\T)\}^{1/2}\le (2n)^{1/2}+n^{1/2}+0.1n^{1/2}\Big\}.
	\end{align*}
	By Corollary 5.35 of \cite{vershynin2010introduction}, $\mathcal{P}(\mathcal{M}_4)\ge 1-2\exp(-0.005n)$. On the event $\mathcal{M}_4$, $WW^\T$ is non-singular, as $2^{1/2}-1-0.1\ge 0.3$. In addition,
	\begin{align*}
		&{P}\{Z^\T(WW^\T)^{-1}Z\le C_2\mid\mathcal{M}_4\}\ge {P}[\lambda_{\rm max}\{(WW^\T)^{-1}\}\|Z\|_2^2\le C_2\mid\mathcal{M}_4]\\
		&\qquad={P}(\|Z\|_2^2\le \lambda_{\rm min}(WW^\T)C_2\mid\mathcal{M}_4)\ge {P}(\|Z\|_2^2\le 0.09nC_2\mid \mathcal{M}_4)\\
		&\qquad\ge 1-{E}(\|Z\|_2^2)/(0.09nC_2)=1-1/(0.09C_2),
	\end{align*}
	where the last equality is by ${E}(\|Z\|_2^2)=n{E}(Z_i^2)=n$. When $n\ge -200\log(\delta/2)$, ${P}(\mathcal{M}_4)\ge 1-\delta$. Let $C_2= 1/(0.09\delta)$, then 
	${P}(\tilde{X}\in \mathcal{A})\le 2\delta$. Additionally, when $\tau$ lies in $\mathcal{H}_0^\prime$ and $\mathcal{H}_1^\prime$, conditional on $\tilde{X}$, $Y$ follows the distributions $\mathcal{N}(0,C_1^2WW^\T/d)$ and $\mathcal{N}(r_nZ,C_1^2WW^\T/d)$, respectively. We denote $f_{\tilde{X},0}^\prime$ and $f_{\tilde{X},1}^\prime$ as the corresponding density functions. Conditional on $\tilde{X}$, when $WW^\T$ is non-singular, by Lemma \ref{KL}, the KL divergence of the two conditional distribution is 
	\begin{equation}\label{KLx}
		{\rm KL}\left(f_{\tilde{X},1}^\prime,f_{\tilde{X},0}^\prime\right)=\dfrac{dr_n^2}{2C_1^2}Z^\T(WW^\T)^{-1}Z=\dfrac{nr_n^2}{C_1^2}Z^\T(WW^\T)^{-1}Z.
	\end{equation}
	We have
	\begin{align*}
		&{\rm TV}(f_1^\prime,f_0^\prime)=\dfrac{1}{2}\int |f_1^\prime(t)-f_0^\prime(t)|dt={E}_{\tilde{X}}\left\{\dfrac{1}{2}\int|f_{\tilde{X},1}^\prime(y)-f_{\tilde{X},0}^\prime(y)|dy\right\}\\
		&\qquad={E}_{\tilde{X}}\left\{\dfrac{I_{\{\tilde{X}\in\mathcal{A}\}}}{2}\!\int|f_{\tilde{X},1}^\prime(y)-f_{\tilde{X},0}^\prime(y)|dy\right\}+{E}_{\tilde{X}}\left\{\dfrac{I_{\{\tilde{X}\notin\mathcal{A}\}}}{2}\!\int|f_{\tilde{X},1}^\prime(y)-f_{\tilde{X},0}^\prime(y)|dy\right\}\\
		&\qquad=V_1+V_2.
	\end{align*}
	We next bound $V_1$ and $V_2$. For $V_1$, let $r_n=2^{1/2}C_1\delta/(C_2n)^{1/2}$, then
	\begin{align*}
		&\dfrac{I_{\{\tilde{X}\in\mathcal{A}\}}}{2}\int|f_{\tilde{X},1}^\prime(y)-f_{\tilde{X},0}^\prime(y)|dy\le I_{\{\tilde{X}\in\mathcal{A}\}}\left\{\dfrac{1}{2}{\rm KL}(f_{\tilde{X},1}^\prime,f_{\tilde{X},0}^\prime)\right\}^{1/2}\\
		&\qquad\le I_{\{\tilde{X}\in\mathcal{A}\}}\left\{\dfrac{nr_n^2}{2C_1^2}Z^\T(WW^\T)^{-1}Z\right\}^{1/2}\le \dfrac{r_n(C_2n)^{1/2}}{2^{1/2}C_1}=\delta,
	\end{align*}
	where the first inequality is by Pinsker's inequality, the second inequality is by \eqref{KLx}, the last inequality holds as $V_1\le \delta$ by the definition of $\mathcal{A}$. For $V_2$, we have $V_2\le {P}(\tilde{X}\notin\mathcal{A})\le 2\delta$. Thus ${\rm TV}(f_1^\prime,f_0^\prime)\le 3\delta$. Therefore, it holds that ${\rm TV}(f_1,f_0)\le 5\delta<1/2-\alpha$. Since $r_n\asymp 1/n^{1/2}$, by \eqref{lower_TV}, it holds that $L_\alpha^*\{\mathcal{T}(1,1,0)\}\ge C/n^{1/2}$ for some constant $C$.
	
	We next prove that there exist a constant $C^\prime$ such that
	\begin{equation}\label{sigma_sqrtn}
		L_\alpha^*\{\mathcal{T}(s,M_0,\sigma_0^2)\}\ge \dfrac{C^\prime\sigma_0}{n^{1/2}}.
	\end{equation}
	Define $\mathcal{H}_0=\{(0_{d+1}^\T,I_{d+1},\sigma_0)\}$ and $\mathcal{H}_1=\{((\theta,0_d^\T)^\T,I_{d+1},\sigma_0)\}$, both with the point mass prior, where $\theta=C_3\sigma_0/n^{1/2}$ for some constant $C_3$ to be determined, $0_{d+1}$ and $0_d$ denote $d+1$-dimensional and a $d$-dimensional zero vector. For $i\in\mathcal{I}$, let $f_{1,i}$, $f_{0,i}$ be the density functions of $(y_i,X_i)$ induced by parameters $(0_{d+1}^\T,I_{d+1},0)$ and $((\theta,0_d^\T)^\T,I_{d+1},0)$, respectively. We use ${\rm KL}(f_{1,i},f_{0,i})$ to denote their KL divergence. By Lemma \ref{KL}, ${\rm KL}(f_{1,i},f_{0,i})=\theta^2/\sigma_0^2$. Similarly, for $j\in\mathcal{J}$, let $f_{1,j}$, $f_{0,j}$ be the marginal density functions of $X_j$ induced by parameters $((0_{d+1}^\T,I_{d+1},0)$ and $(\theta,0_d^\T)^\T,I_{d+1},0)$. Let ${\rm KL}(f_{1,j},f_{0,j})$ denote their KL divergence, we have ${\rm KL}(f_{1,j},f_{0,j})=0$. It holds that
	\begin{align*}
		{\rm TV}(f_1,f_0)&\le \sqrt{\dfrac{1}{2}{\rm KL}(f_1,f_0)}=\sqrt{\dfrac{n}{2}{\rm KL}(f_{1,i},f_{0,i})+\dfrac{m}{2}{\rm KL}(f_{1,j},f_{0,j})}\le \sqrt{\dfrac{n\theta^2}{2\sigma_0^2}}=\dfrac{C_3}{2^{1/2}}.
	\end{align*}
	When we choose $C_3$ sufficiently small, ${\rm TV}(f_1,f_0)<1/2-\alpha$. Notice that $\theta\asymp \sigma_0/n^{1/2}$, \eqref{sigma_sqrtn} is proved. Combining \eqref{M_sqrtn} and \eqref{sigma_sqrtn}, we can get the result.

	\section{Auxiliary lemmas}\label{sec:lemma} 
	
	The following lemmas will be helpful in our proofs.
	
	\begin{lemma}\label{sub_gaussian_bound}
		Denote $\Gamma(a)=\int_{0}^{\infty}x^{a-1}\exp(-x)dx$ for any $a>0$ as the Gamma function. If a random variable $X$ satisfies $\|X\|_{\psi_2}\le \sigma$, then ${P}(|X|>t)\le 2\exp(-t^2/\sigma^2)$ for any $t\ge 0$ and ${E}(|X|^m)\le 2\sigma^m\Gamma(m/2+1)$.
		Let $\{\bX_i\}_{i=1}^n$ be independent random variables with means $\{\mu_i\}_{i=1}^n$ such that $\|\bX_i-\mu_i\|_{\psi_2}\le \sigma_i$ for some $\{\sigma_i\}_{i=1}^n\ge 0$. Then for any set of real numbers $\{a_i\}_{i=1}^n$ and any $t\ge 0$, we have
		${P}\{|\sum_{i=1}^{n}a_i(\bX_i-\mu_i)|\ge t\}\le 2\exp\{-t^2/(8\sum_{i=1}^{n}\sigma_i^2a_i^2)\}$.
	\end{lemma}
	
	\begin{lemma}\label{Bernstein_inequality}
		Let $\{Z_i\}_{i=1}^n$ be i.i.d random variables
		with ${E}(Z_i)=\mu$ and $\|Z_i\|_{\psi_1}\le \sigma_Z$. Then, $\|Z_i-\mu\|_{\psi_1}\le 2\sigma_Z$, $E(|Z_i-\mu|^m)\le 2^m(m!/2)\sigma_z^m$ for $m\ge 2$, and ${P}\{|n^{-1}\sum_{i\in\mathcal{I}}(Z_i-\mu_i)|\ge 2\sigma_Zt(2+t)\}\le 2\exp(-nt^2)$.
		Additionally, if $Z=XY$, where $X$ and $Y$ are sub-Gaussian, then $\|Z\|_{\psi_1}\le \|X\|_{\psi_2}\|Y\|_{\psi_2}$.
	\end{lemma}
	
	Lemmas \ref{sub_gaussian_bound} and \ref{Bernstein_inequality} are from Lemmas D.2, D.4, and D.5 of \cite{chakrabortty2019high}.
	
	\begin{lemma}\label{best_linear_slope}
		Define $\varepsilon_i$ as in Section 1. Let $\bX_i$ and $\varepsilon_i$ be of mean zero. Then
		\begin{equation}\label{y_best_linear_slope}
			{E}(\bX_i\varepsilon_i)=0,\ \ i\in\mathcal{I}.
		\end{equation}
		Furthermore, assume at least one of the following holds: (a) Assumption \ref{ass:MAR} holds with Gaussian designs, (b) Assumption \ref{ass:MAR} holds and the model (2) is correctly specified, that is, $E(v_i\mid \bW_i)=0$, or (c) there is no covariate shift, that is, $\Sigma=\Sigma^\prime$. Then
		\begin{equation}\label{z_best_linear_slope}
			{E}(v_i)=0\ \ {\rm and}\ \ {E}(\bW_iv_i)=0,\ \ i\in\mathcal{I}\cup\mathcal{J}.
		\end{equation}
	\end{lemma}
	
	\begin{lemma}\label{psi}
		Let Assumptions \ref{ass:MAR}-\ref{ass:sub_gau_design} hold. For $i\in\mathcal{I}$ and for $j\in\mathcal{J}$, denote $\bSigma_{\bW}=E(\bW_i\bW_i^\T)$, $\Sigma'=E(\bW_j\bW_j^\T)$, and $r=|\mathcal{I}_1|/(|\mathcal{I}_1|+|\mathcal{J}|)$. Denote
		\begin{align}\label{def:sigma_mix}
			\bSigma_{\bW}^{\rm (mix)}=r\bSigma_{\bW}+(1-r)\bSigma_{\bW}^\prime,\ \  \bOmega_{\bW}^{\rm (mix)}=\big(\bSigma_{\bW}^{\rm (mix)}\big)^{-1}.
		\end{align}
		It holds that $1/C_{\bSigma}\le \sigma_{v,1}^2\le C_{\bSigma}$, $1/C_{\bSigma}\le \sigma_{v,2}^2\le C_{\bSigma}$, and
		\begin{align*}
			&\|\bW_i^\T(\theta\bbeta^*+\balpha^*)\|_{\psi_2}\le \sigma_u M(C_{\bSigma}^2+1),\ i\in \mathcal{I},\quad
			\|Y_i\|_{\psi_2}\le \sigma_u(M+\sigma_\varepsilon),\  i\in \mathcal{I},\\
			&\|v_i\|_{\psi_2}\le \sigma_u(1+C_{\bSigma}^4)^{1/2}, \  i\in \mathcal{I}\cup\mathcal{J},\quad
			\|\bW_{i}^\T\bOmega_{\bW}^{\rm (mix)}\xi\|_{\psi_2}\le \sigma_u MC_{\bSigma}^2(C_{\bSigma}^2+1),\  i\in \mathcal{I}\cup\mathcal{J}.
		\end{align*}
	\end{lemma}

	\begin{lemma}\label{pi_constraint}
		Let Assumptions \ref{ass:MAR}-\ref{ass:sub_gau_design} hold and $E(v_i\mid \bW_i)=0$. Then, as $n,d\to\infty$,
		$$
		\bigg\||\bar{\mathcal{J}}|^{-1}\sum_{i\in \bar{\mathcal{J}}}\bW_iv_i\bigg\|_\infty=O_p\left\{(\log d/N)^{1/2}\right\}.
		$$
	\end{lemma}
	
	\begin{lemma}\label{delta}
		Let Assumptions \ref{ass:MAR}-\ref{ass:sub_gau_design} hold, $E(v_i\mid \bW_i)=0$, and $s\ll(Nn)^{1/2}/\log d$. For any arbitrary constant $\alpha\in(0,1)$, there exist constants $\kappa>0$ and $c>0$ such that $|\bar{\mathcal{J}}|^{-1}\sum_{i\in \bar{\mathcal{J}}}(\bW_i^\T \bp)^2\ge \kappa\|\bp\|_2^2$
		for all $\bp\in\mathbb{R}^{d}$ such that $\|\bp_{S^c}\|_1\le 3\|\bp_S\|_1$ with probability at least $1-\alpha$, if $S$ satisfies $|S|\le s$ and $|\bar{\mathcal{J}}|\ge cs\log d/\kappa$. Furthermore, as $n,d\to\infty$, $
		\|\widehat{\bbeta}_{\rm SR}-\bbeta^*\|_2=O_p\left\{(s\log d /N)^{1/2}\right\}$, $\|\widehat{\bbeta}_{\rm SR}-\bbeta^*\|_1=O_p\left\{s(\log d/N)^{1/2}\right\}$.
	\end{lemma}
	
	\begin{lemma}\label{T_11}
		Let Assumptions \ref{ass:MAR}-\ref{ass:sub_gau_design} hold. Then, as $n,d\to\infty$,
		$$
		|\mathcal{I}_1|^{-1}\sum_{i\in \mathcal{I}_1}v_i\{\bW_i^\T(\theta\bbeta^*+\balpha^*)+\varepsilon_i\}=O_p\left\{(M+\sigma_{\varepsilon})n^{-1/2}\right\}.
		$$
	\end{lemma}
	
	\begin{lemma}\label{xi_infty}
		Let Assumptions \ref{ass:MAR}-\ref{ass:sub_gau_design} hold. Then, as $n,d\to\infty$,
		\begin{align*}
			\|\widehat{\bxi}-\bxi\|_\infty&=O_p\left\{(M+\sigma_{\varepsilon})(\log d/n)^{1/2}\right\},\\ 
			\bigg\||\mathcal{I}_1|^{-1}\sum_{i\in \mathcal{I}_1}\bW_iY_i-\widehat{\bxi}\,\bigg\|_\infty&=O_p\left\{(M+\sigma_{\varepsilon})(\log d/n)^{1/2}\right\},
		\end{align*}
		where $\bxi={E}(\bW_iY_i)$ and $\widehat{\bxi}=|\mathcal{I}_2|^{-1}\sum_{i\in \mathcal{I}_2}\bW_iY_i$.
	\end{lemma}
	
	\begin{lemma}\label{OmegaW_xi_constraint_cov_shift}
		Let Assumptions \ref{ass:MAR}-\ref{ass:sub_gau_design} hold. 
		For any arbitrary constant $\alpha\in (0,1)$, let 
		$
		\mathcal{M}_1=\left\{\left\|\widehat{\bxi}-\widehat{\bSigma}_{\bW} \bOmega_{\bW}^{\rm (mix)}\bxi\right\|_\infty\le \lambda_{\bu}\right\},
		$
		where $\bOmega_{\bW}^{\rm (mix)}$ is defined as in Lemma \ref{psi}. Choose
		$$\lambda_{\bu}=6(\sigma_u^2M+\sigma\sigma_\varepsilon)\{\log (2d/\alpha)/|\mathcal{I}_2|\}^{1/2}+6\sigma_u^2MC_{\bSigma}^2(C_{\bSigma}^2+1)\{\log (2d/\alpha)/|\bar{\mathcal{J}}|\}^{1/2}.$$
		When $|\mathcal{I}_2|\ge36\log(2d/\alpha)$ and $|\bar{\mathcal{J}}|\ge \log(2d/\alpha)$, we have ${P}(\mathcal{M}_1)\ge1-2\alpha$. Furthermore,
		\begin{equation}\label{OmegaW_xi_2}
			{P}\left\{\bxi^\T\bOmega_{\bW}^{\rm (mix)}\widehat{\bSigma}_{\bW} \bOmega_{\bW}^{\rm (mix)}\bxi\le M^2(6\sigma_u^2 C_{\bSigma}^4+C_{\bSigma}^3)(C_{\bSigma}^2+1)^2 \right\}\ge 1-\alpha.
		\end{equation}
		On the event $\mathcal{M}_1$, the feasible set of the problem is non-empty. Thus, the solution $\widehat{\bu}_{\rm SR}$ exists and satisfies the constraint $\|\widehat{\bxi}-\widehat{\bSigma}_{\bW} \widehat{\bu}_{\rm SR}\|_\infty\le \lambda_{\bu}$. Define
		\begin{equation}\label{def:M_2}
			\mathcal M_2=\left\{\widehat{\bu}_{\rm SR}\ {\rm exists\ and\ }\widehat{\bu}_{\rm SR}^\T\widehat{\bSigma}_{\bW} \widehat{\bu}_{\rm SR}\le  M^2(6\sigma_u^2 C_{\bSigma}^4+C_{\bSigma}^3)(C_{\bSigma}^2+1)^2\right\}.
		\end{equation}
		Then it follows that ${P}(\mathcal{M}_2)\ge 1-3\alpha$.
	\end{lemma}
	
	\begin{lemma}\label{T_2_2_cov_shift}
		Let Assumptions \ref{ass:MAR}-\ref{ass:sub_gau_design} hold and $E(v_i\mid \bW_i)=0$. Then, as $n,d\to\infty$,
		$$
		|\bar{\mathcal{J}}|^{-1}\sum_{i\in \bar{\mathcal{J}}}\widehat{\bu}_{\rm SR}^\T \bW_iv_i=O_p\left(MN^{-1/2}\right).
		$$
	\end{lemma}
	
	\begin{lemma}\label{B_6}
		Let Assumptions \ref{ass:MAR}-\ref{ass:sub_gau_design} hold, $E(v_i\mid \bW_i)=0$, and $s\ll (Nn)^{1/2}/\log d$. Then, as $n,d\to\infty$,
		$$
		|\mathcal{I}_1|^{-1}\sum_{i\in \mathcal{I}_1}\{\bW_i^\T(\widehat{\bbeta}_{\rm SR}-\bbeta^*)\}^2=O_p\left\{s\log d /(Nn)^{1/2}\right\}.
		$$
	\end{lemma}
	
	\begin{lemma}\label{numerator}
		Let Assumptions \ref{ass:MAR}-\ref{ass:sub_gau_design} hold, $E(v_i\mid \bW_i)=0$, and $s\ll(Nn)^{1/2}/\log d$. Then, as $n,d\to\infty$,
		$$
		|\mathcal{I}_1|^{-1}\sum_{i\in\mathcal{I}_1}\hat{v}_i^2/\sigma_{v,1}^2=1+o_p(1),\ \  1/\{|\mathcal{I}_1|^{-1}\sum_{i\in \mathcal{I}_1}\hat{v}_i^2\}=O_p(1).
		$$
	\end{lemma}
	
	\begin{lemma}\label{v_i}
		Let Assumptions \ref{ass:MAR}-\ref{ass:sub_gau_design} hold, $E(v_i|\bW_i)=0$, and $s\ll N^{1/2}/\log d$. Then, as $n,d\to\infty$,
		$$
		|\mathcal{I}_1|^{-1}\sum_{i\in \mathcal{I}_1}(v_i-\hat{v}_i)^4=o_p(1),\ \  |\mathcal{I}_1|^{-1}\sum_{i\in \mathcal{I}_1}\hat{v}_i^4=O_p(1).$$
	\end{lemma}
	
	\begin{lemma}\label{df_delta}
		Let the assumptions of Theorem \ref{thm:DF_consist_rate_cs} hold. Then, as $n,d\to\infty$,
		$$
		\|\widehat{\bbeta}_{\rm DFA}-\bbeta^*\|_2=O_p\left\{(s\log d /N)^{1/2}\right\},\ \ \|\widehat{\bbeta}_{\rm DFA}-\bbeta^*\|_1=O_p\left\{s(\log d /N)^{1/2}\right\}.
		$$
	\end{lemma}
	
	\begin{lemma}\label{vizi}
		Let the assumptions of Theorem \ref{thm:DF_consist_rate_cs} hold. Then, as $n,d\to\infty$,
		\begin{align*}
			n^{-1}\sum_{i\in \mathcal{I}}\hat{v}_i^2-\sigma_{v,1}^2&=O_p\left\{n^{-1/2}+s\log d /(Nn)^{1/2}\right\},\\
			n^{-1}\sum_{i\in \mathcal{I}}\hat{v}_iZ_i-\sigma_{v,1}^2&=O_p\left[n^{-1/2}+\{s\log d /(Nn)^{1/2}\}^{1/2}\right].
		\end{align*}
	\end{lemma}
	
	\begin{lemma}\label{Sigma_W_inf}
		Let the assumptions of Theorem \ref{thm:DF_consist_rate_cs} hold. Then, as $n,d\to\infty$,
		$$
		\Big\|\Big(n^{-1}\sum_{i\in \mathcal{I}} \bW_i\bW_i^\T-\bSigma_{\bW}\Big)\bbeta^*\Big\|_\infty = O_p\left\{(\log d /n)^{1/2}\right\}.
		$$
	\end{lemma}
	
	\begin{lemma}\label{xeps}
		Let the assumptions of Theorem \ref{thm:DF_consist_rate_cs} hold. Then, as $n,d\to\infty$,
		$$
		\Big\|n^{-1}\sum_{i\in\mathcal{I}}\bX_i\varepsilon_i\Big\|_\infty = O_p\left\{\sigma_\varepsilon(\log d /n)^{1/2}\right\}.
		$$
	\end{lemma}

	\begin{lemma}\label{KL}
		Let $f_0$ and $f_1$ denote the density functions of the distributions $\mathcal{N}(\bmu_0,\bSigma_0)$ and $\mathcal{N}(\bmu_1,\bSigma_1)$ respectively. Denote ${\rm KL}(f_0,f_1)$ as their Kullback-Leibler divergence. Then
		$$
		{\rm KL}(f_0,f_1)=\dfrac{1}{2}\left\{{\rm tr}\{\bSigma_1^{-1}(\bSigma_0-\bSigma_1)\}+\log\left(\dfrac{{\rm det}(\bSigma_1)}{{\rm det}(\bSigma_0)}\right)+(\bmu_1-\bmu_0)^\T\bSigma_1^{-1}(\bmu_1-\bmu_0)\right\}.
		$$
	\end{lemma}
	
	Lemma \ref{KL} is from Lemma 7 of \cite{bradic2022testability}.
	
	\section{Proof of auxiliary lemmas}\label{sec:proof_lemma}
	
	{\bf Proof of Lemma Lemma \ref{best_linear_slope}}.
	By Lemma 1 of \cite{zhang2019semi}, \eqref{y_best_linear_slope} holds. Additionally, since $\bX_i=(Z_i,\bW_i^\T)^\T$ has mean zero, it follows that $v_i=Z_i-\bW_i^\T\bbeta^*$ satisfies $E(v_i)=0$ for all $i\in\mathcal I\cup\mathcal J$. We prove the remaining part of \eqref{z_best_linear_slope} below. 
	
	\textbf{Case (b)}. Let Assumption 2 hold and $E(v_i\mid \bW_i)=0$ for $i\in\mathcal{I}$. Recall that $\bbeta^*$ is defined as the best linear slope for the labeled samples. By Lemma 1 of \cite{zhang2019semi}, $
	{E}(\bW_iv_i)=0$ for $i\in\mathcal{I}$.
	By Assumption 2 and $E(v_i\mid \bW_i)=0$, ${E}(Z_j\mid \bW_j=\bw)={E}(Z_i\mid \bW_i=\bw)=\bw^\T\bbeta^*$ for $ i\in\mathcal{I},\ j\in\mathcal{J}$
	Hence, ${E}(v_j\mid \bW_j)={E}(Z_j-\bW_j^\T\bbeta^*\mid \bW_j)=0$ for $j\in\mathcal{J}$. It follows that
	${E}(\bW_jv_j)={E}\{\bW_j{E}(v_j\mid \bW_j)\}=0$ for $j\in\mathcal{J}$.
	
	\textbf{Case (a)}. Let Assumption 2 hold with Gaussian design. Notice that $v_i$ is independent of $\bW_i$, ${E}(v_i\mid \bW_i)={E}(v_i)=0$. Hence, Case (a) is a special case of Case (b), and \eqref{z_best_linear_slope} holds.
	
	\textbf{Case (c)}. Suppose that there is no covariate shift. By Lemma 1 of \cite{zhang2019semi},
	$
	{E}(\bW_iv_i)=0$ for $i\in\mathcal{I}$.
	Note that $E(\bX_i\bX_i^\T)=\bSigma=\bSigma^\prime=E(\bX_j\bX_j^\T)$, where $\bX_i=(Z_i,\bW_i^\T)^\T$ and $\bX_j=(Z_j,\bW_j^\T)^\T$ for any $i\in\mathcal I$ and $j\in\mathcal{J}$. Hence,
	$
	{E}(\bW_jv_j)={E}\{\bW_j(Z_j-\bW_j^\T\bbeta^*)\}={E}\{\bW_i(Z_i-\bW_i^\T\bbeta^*)\}={E}(\bW_iv_i)=0$ for $j\in\mathcal{J}$.
	\hfill\BlackBox
	\\ \hspace*{\fill} \\
	{\bf Proof of Lemma \ref{psi}}.
	Let Assumptions \ref{ass:MAR}-\ref{ass:sub_gau_design} hold. We obverse that the first row of $\bOmega=\bSigma^{-1}$ can be expressed as $(1,-{\bbeta^*}^\T)/\sigma_{v,1}^2$. Additionally, since $1/\sigma_{v,1}^2=\be_1^\T\bOmega \be_1$, we have
	$$
	1/C_{\bSigma}\le 1/\lambda_{\rm max}(\bSigma)=\lambda_{\rm min}(\bOmega)\le 1/\sigma_{v,1}^2\le \lambda_{\rm max}(\bOmega)=1/\lambda_{\rm min}(\bSigma)\le C_{\bSigma}.
	$$
	It follows that $1/C_{\bSigma}\le \sigma_{v,1}^2\le C_{\bSigma}$. Similarly, we also have $1/C_{\bSigma}\le \sigma_{v,2}^2\le C_{\bSigma}$. Since $1/\sigma_{v,1}^2+\|\bbeta^*\|_2^2/\sigma_{v,1}^2=\be_1^\T\bOmega^\T\bOmega \be_1\le C_{\bSigma}^2$, together with $\sigma_{v,1}^2\le C_{\bSigma}$, we have $\|\bbeta^*\|_2\le C_{\bSigma}^2$.
	Notice that $\bSigma_{\bW}$ is a principle sub-matrix of $\bSigma$ and $\bSigma_{\bW}^\prime$ is a principle sub-matrix of $\bSigma^\prime$, we have
	\begin{gather*}
		1/C_{\bSigma}\le \lambda_{\rm min}(\bSigma)\le \lambda_{\rm min}(\bSigma_{\bW})\le \lambda_{\rm max}(\bSigma_{\bW})\le\lambda_{\rm max}(\bSigma) \le  C_{\bSigma},\\
		1/C_{\bSigma}\le \lambda_{\rm min}(\bSigma^\prime)\le \lambda_{\rm min}(\bSigma_{\bW}^\prime)\le \lambda_{\rm max}(\bSigma_{\bW}^\prime)\le\lambda_{\rm max}(\bSigma^\prime) \le  C_{\bSigma}.
	\end{gather*}
	By Weyl's inequality,
	\begin{align}
		1/C_{\bSigma}\le \lambda_{\rm min}(\bSigma_{\bW}^{\rm (mix)})&\le \lambda_{\rm max}(\bSigma_{\bW}^{\rm (mix)})\le C_{\bSigma},\label{bound:Sigma-mix}\\
		1/C_{\bSigma}\le \lambda_{\rm min}(\bOmega_{\bW}^{\rm (mix)})=1/\lambda_{\rm max}(\bSigma_{\bW}^{\rm (mix)})&\le 1/\lambda_{\rm min}(\bSigma_{\bW}^{\rm (mix)})=\lambda_{\rm max}(\bOmega_{\bW}^{\rm (mix)})\le C_{\bSigma}.\label{bound:Omega-mix}
	\end{align}
	Denote $\bxi={E}(\bW_iY_i)$. Since $Y_i=\theta Z_i+\bW_i^\T\balpha^*+\varepsilon_i = \bW_i^\T(\theta\bbeta^*+\balpha^*)+\theta v_i+\varepsilon_i$, by Lemma \ref{best_linear_slope}, we have $\bxi=\bSigma_{\bW}(\theta\bbeta^*+\balpha^*)$. Furthermore, as $\max(|\theta|,\|\balpha^*\|_2)\le\|\bgamma^*\|_2=M$,
	$$
	\|\bxi\|_2\le \lambda_{\rm max}(\bSigma_{\bW})\|\theta\bbeta^*+\balpha^*\|_2\le \lambda_{\rm max}(\bSigma_{\bW})(|\theta|\cdot\|\bbeta^*\|_2+\|\balpha^*\|_2)\le MC_{\bSigma}(C_{\bSigma}^2+1).
	$$
	Under Assumption 3, it holds that
	\begin{align*}
		\|\bW_i^\T(\theta\bbeta^*+\balpha^*)\|_{\psi_2}&\le \sigma_u\|\theta\bbeta^*+\balpha^*\|_2\le \sigma_u M(C_{\bSigma}^2+1),\ \  i\in \mathcal{I},\\
		\|Y_i\|_{\psi_2}&\le\|\bX_i^\T\bgamma^*\|_{\psi_2}+\|\varepsilon_i\|_{\psi_2}\le \sigma_u\|\bgamma^*\|_2+\sigma_u\sigma_\varepsilon= \sigma_u (M+\sigma_\varepsilon),\ \  i\in\mathcal{I},\\
		\|v_i\|_{\psi_2}&=\|\bX_i^\T(1,-{\bbeta^*}^\T)^\T\|_{\psi_2}\le \sigma_u (1+\|\bbeta^*\|_2^2)^{1/2}\le \sigma_u(1+C_{\bSigma}^4)^{1/2},\ \  i\in \mathcal{I}\cup\mathcal{J},\\
		\|\bW_{i}^\T\bOmega_{\bW}^{\rm (mix)}\bxi\|_{\psi_2}&\le \sigma_u\|\bOmega_{\bW}^{\rm (mix)}\bxi\|_2\le\sigma_u\lambda_{\rm max}(\bOmega_{\bW}^{\rm (mix)})\|\bxi\|_2\le \sigma_u MC_{\bSigma}^2(C_{\bSigma}^2+1),\ \  i\in \mathcal{I}\cup\mathcal{J}.
	\end{align*}
	\hfill\BlackBox
	\\ \hspace*{\fill} \\
	{\bf Proof of Lemma \ref{pi_constraint}}.
	By Lemma \ref{best_linear_slope}, we have ${E}(\bW_iv_i)=0$ for $i\in \mathcal{I}\cup\mathcal{J}$. By Lemmas \ref{Bernstein_inequality} and \ref{psi}, for any $j\in [d]$, $\|W_{ij}v_i\|_{\psi_1}\le \|W_{ij}\|_{\psi_2}\|v_i\|_{\psi_2}\le\sigma_u^2(1+C_{\bSigma}^4)^{1/2}$. For any $t>0$, by Lemma \ref{Bernstein_inequality} and Boole's inequality, 
	$$
	{P}\bigg\{\bigg\||\bar{\mathcal{J}}|^{-1}\sum_{i\in \bar{\mathcal{J}}}\bW_iv_i\bigg\|_\infty\ge 2\sigma_u^2t(2+t)(1+C_{\bSigma}^4)^{1/2}\bigg\}\le 2d\exp(-|\bar{\mathcal{J}}|t^2).
	$$
	For any arbitrary constant $\alpha\in (0,1)$, let $t=\{\log(2d/\alpha)/|\bar{\mathcal{J}}|\}^{1/2}$. When $|\bar{\mathcal{J}}|\ge \log(2d/\alpha)$, $t\le1$ and hence
	$$
	{P}\bigg[\bigg\||\bar{\mathcal{J}}|^{-1}\sum_{i\in \bar{\mathcal{J}}}\bW_iv_i\bigg\|_\infty\ge 6\sigma_u^2\{(1+C_{\bSigma}^4)\log(2d/\alpha)/|\bar{\mathcal{J}}|\}^{1/2}\bigg]\le \alpha.
	$$
	The conclusion then follows by $|\bar{\mathcal{J}}|\asymp N$.
	\hfill\BlackBox
	\\ \hspace*{\fill} \\
	{\bf Proof of Lemma \ref{delta}}.
	Note that $(\bW_i)_{i\in\mathcal I\cup\mathcal J}$ are independent but not necessarily identical (if covariate shift occurs) random vectors. Let $(\epsilon_i)_{i=1}^N$ be i.i.d. Rademacher random variables. By Lemma D.1 (ii) of \cite{chakrabortty2019high}, $\|\epsilon_i\|_{\psi_2}\le(\log 2)^{-1/2}<2$. Repeating the proof of Lemma \ref{pi_constraint}, for any $t>0$,
	$$
	{P}\bigg\{\bigg\||\bar{\mathcal{J}}|^{-1}\sum_{i\in \bar{\mathcal{J}}}\bW_i\epsilon_i\bigg\|_\infty\ge 4\sigma_u^2t(2+t)\bigg\}\le 2d\exp(-|\bar{\mathcal{J}}|t^2).
	$$
	Let $|\bar{\mathcal{J}}|\ge\log(2d)$ and $t=[\{\log(2d)+u^2\}/|\bar{\mathcal{J}}|]^{1/2}$ with some $u>0$, then we have $t\le\{\log(2d)/|\bar{\mathcal{J}}|\}^{1/2}+u/|\bar{\mathcal{J}}|^{1/2}$, $t^2=\{\log(2d)+u^2\}/|\bar{\mathcal{J}}|\le u^2/|\bar{\mathcal{J}}|+\{\log(2d)/|\bar{\mathcal{J}}|\}^{1/2}$, and
	$$
	{P}\bigg[\bigg\||\bar{\mathcal{J}}|^{-1}\sum_{i\in \bar{\mathcal{J}}}\bW_i\epsilon_i\bigg\|_\infty\ge 4\sigma_u^2u^2/|\bar{\mathcal{J}}|+8\sigma_u^2u/|\bar{\mathcal{J}}|^{1/2}+12\sigma_u^2\{\log(2d)/|\bar{\mathcal{J}}|\}^{1/2}\bigg]\le\exp(-u^2).
	$$
	Together with Lemma G.3 of \cite{zhang2023decaying}, as $n,d\to\infty$,
	\begin{align*}
		E\bigg(\bigg\||\bar{\mathcal{J}}|^{-1}\sum_{i\in \bar{\mathcal{J}}}\bW_i\epsilon_i\bigg\|_\infty\bigg)&\le 12\sigma_u^2\left\{\dfrac{\log(2d)}{|\bar{\mathcal{J}}|}\right\}^{1/2}+\dfrac{16\sigma_u^2}{|\bar{\mathcal{J}}|}+\dfrac{8\pi^{1/2}\sigma_u^2}{|\bar{\mathcal{J}}|^{1/2}}=O\left\{(\log d/N)^{1/2}\right\}.
	\end{align*}
	Repeating the proof of Theorem 9.36 for independent but non-identical $\bW_i$ using \eqref{bound:Sigma-mix}-\eqref{bound:Omega-mix}, we also obtain that for any arbitrary constant $\alpha\in (0,1)$, $|\bar{\mathcal{J}}|^{-1}\sum_{i\in \bar{\mathcal{J}}}(\bW_i^\T \bp)^2
	\ge 2\kappa\|\bp\|_2^2-c\log d\|\bp\|_1^2/|\bar{\mathcal{J}}|$
	with probability at least $1-\alpha$ and some constants $\kappa,c>0$. When $\|\bp_{S^c}\|_1\le 3\|\bp_S\|_1$ with $|S|\le s$, we have $\|\bp\|_1=\|\bp_{S^c}\|_1+\|\bp_S\|_1\le4\|\bp_S\|_1\le4 s^{1/2}\|\bp_S\|_2\le s^{1/2}\|\bp\|_2$. Therefore, $
	|\bar{\mathcal{J}}|^{-1}\sum_{i\in \bar{\mathcal{J}}}(\bW_i^\T \bp)^2
	\ge (2\kappa-cs\log d/|\bar{\mathcal{J}}|)\|\bp\|_2^2\ge\kappa\|\bp\|_2^2/2$,
	as long as $|\bar{\mathcal{J}}|\ge cs\log d/\kappa$. Together with Lemma \ref{pi_constraint} and Theorem 7.13 of \cite{wainwright2019high}, as $n,d\to\infty$, $
	\|\widehat{\bbeta}_{\rm SR}-\bbeta^*\|_2=O_p\left\{(s\log d /N)^{1/2}\right\}$,  $\|\widehat{\bbeta}_{\rm SR}-\bbeta^*\|_1=O_p\left\{s(\log d /N)^{1/2}\right\}$.
	\hfill\BlackBox
	\\ \hspace*{\fill} \\
	{\bf Proof of Lemma \ref{T_11}}.
	By Lemma \ref{best_linear_slope}, 
	\begin{equation}\label{t_i}
		\begin{split}
			{E}\left[v_i\{\bW_i^\T(\theta\bbeta^*+\balpha^*)+\varepsilon_i\}\right]={E}(v_i\bW_i)^\T(\theta\bbeta^*+\balpha^*)+{E}(\bX_i\varepsilon_i)^\T(1,-{\bbeta^*}^\T)^\T=0.
		\end{split}
	\end{equation}
	By Lemma \ref{psi}, $\|\bW_i^\T(\theta\bbeta^*+\balpha^*)+\varepsilon_i\|_{\psi_2}\le \sigma_u M(C_{\bSigma}^2+1)+\sigma_u\sigma_{\varepsilon}$ and hence
	\begin{equation}\label{psi_ti}
		\|v_i\{\bW_i^\T(\theta\bbeta^*+\balpha^*)+\varepsilon_i\}\|_{\psi_1}\le \sigma_u^2\{M(C_{\bSigma}^2+1)+\sigma_\varepsilon\}(1+C_{\bSigma}^4)^{1/2}.
	\end{equation}
	The conclusion then follows by Lemma \ref{Bernstein_inequality} and $|\mathcal{I}_1|\asymp n$.
	\hfill\BlackBox
	\\ \hspace*{\fill} \\
	{\bf Proof of Lemma \ref{xi_infty}}.
	By Lemma \ref{psi}, for any $i\in\mathcal I$ and $j\in[d]$, $\|W_{ij}Y_i\|_{\psi_1}\le \sigma_u^2(M+\sigma_\varepsilon)$. By Lemma \ref{Bernstein_inequality} and Boole's inequality, 
	\begin{align*}
		{P}\bigg\{\bigg\||\mathcal{I}_2|^{-1}\sum_{i\in \mathcal{I}_2}(\bW_{i}Y_i-\bxi)\bigg\|_\infty\ge2\sigma_u^2(M+\sigma_\varepsilon)t(2+t)\bigg\}\le 2d\exp(-|\mathcal{I}_2|t^2).
	\end{align*}
	For any arbitrary constant $\alpha\in(0,1)$, take $t=6\{\log (2d/\alpha)/|\mathcal{I}_2|\}^{1/2}$. When $|\mathcal{I}_2|\ge 36\log(2d/\alpha)$, $t\le1$ and hence
	\begin{equation}\label{xi1}
		{P}\left(\left\|\widehat{\bxi}-\bxi\right\|_\infty\ge6\sigma_u^2(M+\sigma_\varepsilon)\{\log (2d/\alpha)/|\mathcal{I}_2|\}^{1/2}\right)\le \alpha.
	\end{equation}
	Since $|\mathcal{I}_2| \asymp n$, $\|\widehat{\bxi}-\bxi\|_\infty=O_p\left\{(M+\sigma_{\varepsilon})(\log d /n)^{1/2}\right\}$. Similarly, $\||\mathcal{I}_1|^{-1}\sum_{i\in \mathcal{I}_1}\bW_iY_i-\bxi\|_\infty=O_p\{(M+\sigma_{\varepsilon})(\log d /n)^{1/2}\}$.
	Hence, $\||\mathcal{I}_1|^{-1}\sum_{i\in \mathcal{I}_1}\bW_iY_i-\widehat{\bxi}\|_\infty\le  \||\mathcal{I}_1|^{-1}\sum_{i\in \mathcal{I}_1}\bW_iY_i-\bxi\|_\infty+\|\widehat{\bxi}-\bxi\|_\infty=O_p\{(M+\sigma_{\varepsilon})(\log d /n)^{1/2}\}$.
	\hfill\BlackBox
	\\ \hspace*{\fill} \\
	{\bf Proof of Lemma \ref{OmegaW_xi_constraint_cov_shift}}.
	We first prove that $\bOmega_{\bW}^{\rm (mix)}\bxi$ satisfies the constraint $\|\widehat{\bxi}-\widehat{\bSigma}_{\bW} \bu\|_\infty\le \lambda_{\bu}$ with high probability. For any arbitrary constant $\alpha\in(0,1)$, notice that
	\begin{align*}
		\big\|\widehat{\bxi}-\widehat{\bSigma}_{\bW} \bOmega_{\bW}^{\rm (mix)}\bxi\big\|_\infty\le\big\|\widehat{\bxi}-\bxi\big\|_\infty+\big\|\bxi-\widehat{\bSigma}_{\bW} \bOmega_{\bW}^{\rm (mix)}\bxi\big\|_\infty=A_1+A_2.
	\end{align*}
	For $A_{1}$, we have \eqref{xi1} when $|\mathcal{I}_2|\ge 36\log(2d/\alpha)$. For $A_{2}$, we have 
	\begin{align*}
		{E}(\bxi-\widehat{\bSigma}_{\bW} \bOmega_{\bW}^{\rm (mix)}\bxi)={E}\bigg\{\bxi-\bigg(\dfrac{r}{|\mathcal{I}_1|}\sum_{i\in\mathcal{I}_1}\bW_i\bW_i^\T+\dfrac{1-r}{m}\sum_{j\in\mathcal{J}}\bW_j\bW_j^\T\bigg)\bOmega_{\bW}^{\rm (mix)}\bxi\bigg\}=0.
	\end{align*}
	By Lemma \ref{psi}, $\|W_{ij}\bW_{i}^\T\bOmega_{\bW}^{\rm (mix)}\bxi\|_{\psi_1}\le \sigma_u^2MC_{\bSigma}^2(C_{\bSigma}^2+1)$ for each $i\in\mathcal I\cup\mathcal J$ and $j\in[d]$.
	By Lemma \ref{Bernstein_inequality} and Boole's inequality,
	\begin{align*}
		&{P}\bigg\{\bigg\||\bar{\mathcal{J}}|^{-1}\sum_{i\in \bar{\mathcal{J}}}(\bW_i\bW_i^\T\bOmega_{\bW}^{\rm (mix)}\bxi-\bxi)\bigg\|_\infty\ge 2\sigma_u^2MC_{\bSigma}^2(C_{\bSigma}^2+1)t(2+t)\bigg\}\le 2d\exp(-|\bar{\mathcal{J}}|t^2).
	\end{align*}
	Take $t=\{\log(2d/\alpha)/|\bar{\mathcal{J}}|\}^{1/2}$, when $|\bar{\mathcal{J}}|\ge \log(2d/\alpha)$, $t\le 1$ and hence
	\begin{equation}\label{lambda_u_T_2_}
		{P}\Big[\Big\|\widehat{\bSigma}_{\bW} \bOmega_{\bW}^{\rm (mix)}\bxi-\bxi\Big\|_\infty\ge 6\sigma_u^2MC_{\bSigma}^2(C_{\bSigma}^2+1)\{\log (2d/\alpha)/|\bar{\mathcal{J}}|\}^{1/2}\Big]\le \alpha.
	\end{equation}
	Combining \eqref{xi1} and \eqref{lambda_u_T_2_},
	\begin{align*}
		{P}\Big[\Big\|\widehat{\bxi}-\widehat{\bSigma}_{\bW} \bOmega_{\bW}^{\rm (mix)}\bxi\Big\|_\infty \le c_2\{\log (2d/\alpha)/|\mathcal{I}_2|\}^{1/2}+c_3\{\log (2d/\alpha)/|\bar{\mathcal{J}}|\}^{1/2}\Big]\ge 1-2\alpha,
	\end{align*}
	where $c_2=6\sigma_u^2(M+\sigma_\varepsilon)$ and $c_3=6\sigma_u^2MC_{\bSigma}^2(C_{\bSigma}^2+1)$.
	Now we prove \eqref{OmegaW_xi_2}. Note that ${E}(\bxi^\T\bOmega_{\bW}^{\rm (mix)}\widehat{\bSigma}_{\bW} \bOmega_{\bW}^{\rm (mix)}\bxi)=\bxi^\T\bOmega_{\bW}^{\rm (mix)}\bxi$. By Lemma \ref{psi}, we have $\|\bxi^\T\bOmega_{\bW}^{\rm (mix)}\bW_i\bW_i^\T\bOmega_{\bW}^{\rm (mix)}\bxi\|_{\psi_1}\le \sigma_u^2 M^2C_{\bSigma}^4(C_{\bSigma}^2+1)^2$.
	By Lemma \ref{Bernstein_inequality} with $t=1$,
	\begin{align*}
		{P}\bigg\{\bigg||\bar{\mathcal{J}}|^{-1}\sum_{i\in \bar{\mathcal{J}}}(\bxi^\T\bOmega_{\bW}^{\rm (mix)} \bW_i)^2-\bxi^\T\bOmega_{\bW}^{\rm (mix)}\bxi\bigg|> 6\sigma_u^2 M^2C_{\bSigma}^4(C_{\bSigma}^2+1)^2\bigg\}\le 2\exp(-|\bar{\mathcal{J}}|).
	\end{align*}
	Together with $\bxi^\T\bOmega_{\bW}^{\rm (mix)}\bxi\le \lambda_{\rm max}(\bOmega_{\bW}^{\rm (mix)})\|\bxi\|_2^2\le M^2C_{\bSigma}^3(C_{\bSigma}^2+1)^2$, when $|\bar{\mathcal{J}}|\ge \log(2/\alpha)$, 
	\begin{align*}
		{P}\left\{\bxi^\T\bOmega_{\bW}^{\rm (mix)} \widehat{\bSigma}_{\bW} \bOmega_{\bW}^{\rm (mix)}\bxi> M^2(6\sigma_u^2 C_{\bSigma}^4+C_{\bSigma}^3)(C_{\bSigma}^2+1)^2\right\}\le \alpha.
	\end{align*}
	Since $\widehat{\bu}_{\rm SR}$ is the global optimal of the convex problem, the proof is completed.
	\hfill\BlackBox
	\\ \hspace*{\fill} \\
	{\bf Proof of Lemma \ref{T_2_2_cov_shift}}.
	Denote $\mathcal{D}=\{(Y_i,\bW_i)_{i\in \mathcal{I}_2},(\bW_i)_{i\in \bar{\mathcal{J}}}\}$, then ${E}(\widehat{\bu}_{\rm SR}^\T \bW_iv_i\mid \mathcal{D})=0$ and
	$
	\|\widehat{\bu}_{\rm SR}^\T \bW_iv_i\|_{\psi_{2,\mathcal D}}\le |\widehat{\bu}_{\rm SR}^\T \bW_i|\cdot\|v_i\|_{\psi_2}\le \sigma_u|\widehat{\bu}_{\rm SR}^\T \bW_i|(1+ C_{\bSigma}^4)^{1/2},
	$
	where $\|\cdot\|_{\psi_{2,\mathcal D}}$ denotes the $\psi_2$-Orlicz norm defined through the probability measure conditional on $\mathcal D$. For any arbitrary constant $\alpha\in(0,1)$, by Lemma \ref{sub_gaussian_bound},
	\begin{align*}
		{P}\bigg(\bigg||\bar{\mathcal{J}}|^{-1}\sum_{i\in \bar{\mathcal{J}}}\widehat{\bu}_{\rm SR}^\T \bW_iv_i\bigg|>t_N \mid \mathcal{D}\bigg)&\le 2\exp\Bigg\{\dfrac{-t_N^2}{8|\bar{\mathcal{J}}|^{-2}\sum_{i\in  \bar{\mathcal{J}}}\sigma_u^2(\widehat{\bu}_{\rm SR}^\T \bW_i)^2(1+ C_{\bSigma}^4)}\Bigg\}=\alpha,
	\end{align*}
	where $t_N=\{8\sigma_u^2(1+ C_{\bSigma}^4)\widehat{\bu}_{\rm SR}^\T\widehat{\bSigma}_{\bW} \widehat{\bu}_{\rm SR}\log(2/\alpha)/|\bar{\mathcal{J}}\}^{1/2}$.
	Since $|\bar{\mathcal{J}}|\asymp N$, conditional on $\mathcal{D}$,
	$
	|\bar{\mathcal{J}}|^{-1}\sum_{i\in \bar{\mathcal{J}}}\widehat{\bu}_{\rm SR}^\T \bW_iv_i=O_p\{(\widehat{\bu}_{\rm SR}^\T\widehat{\bSigma}_{\bW} \widehat{\bu}_{\rm SR}/N)^{1/2}\}.
	$
	By Lemma \ref{OmegaW_xi_constraint_cov_shift}, $\widehat{\bu}_{\rm SR}^\T\widehat{\bSigma}_{\bW} \widehat{\bu}_{\rm SR}=O_p(M^2)$.
	\hfill\BlackBox
	\\ \hspace*{\fill} \\
	{\bf Proof of Lemma \ref{B_6}}.
	Let $\mathcal{D}(J)=\{\bdelta\in\mathbb{R}^d\,|\, {\rm supp}(\bdelta)\subseteq J,\|\bdelta\|_2=1\}$, where $J\subseteq [d]$. We first prove that
	\begin{equation}\label{h1h2}
		\max_{|J|\le s}\max_{\bdelta_1,\bdelta_2\in\mathcal{D}(J)}\left|\bdelta_1 \left(|\mathcal{I}_1|^{-1}\sum_{i\in\mathcal{I}_1}\bW_i\bW_i^\T-\bSigma_{\bW}\right)\bdelta_2\right|=O_p\left[\max\left\{\sqrt{ \dfrac{s\log d}{n}},\dfrac{s\log d}{n}\right\}\right].
	\end{equation}
	For any $J$ satisfy $|J|\le s$, denote $\bU_i=\bW_{i,J}$ as the sub-vector of $\bW_i$ consisting of those element $W_{ij}$ for $j\in J$ and let $\bSigma_{\bU}={E}(\bU_i\bU_i^\T)$. For any $\bdelta_1, \bdelta_2\in\mathcal{D}(J)$, 
	\begin{align*}
		\left|\bdelta_1 \left(|\mathcal{I}_1|^{-1}\sum_{i\in\mathcal{I}_1}\bW_i\bW_i^\T-\bSigma_{\bW}\right)\bdelta_2\right|
		&\le \|\bdelta_1\|_2\|\bdelta_2\|_2\left\||\mathcal{I}_1|^{-1}\sum_{i\in\mathcal{I}_1}\bU_i\bU_i^\T-\bSigma_{\bU}\right\|_2\\
		&=\left\||\mathcal{I}_1|^{-1}\sum_{i\in\mathcal{I}_1}\bU_i\bU_i^\T-\bSigma_{\bU}\right\|_2,
	\end{align*}
	where $\|\cdot\|_2$ denotes spectral norm of a matrix. Notice that the dimension of $U_i$ is at most $s$, and the number of $J\subseteq[d]$ satisfying $|J|\le s$ is
	\begin{align*}
		{d\choose1}+{d\choose2}+\dots+{d\choose s}\le d+\dfrac{d^2}{2!}+\cdots \dfrac{d^s}{s!}\le \left(\sum_{k=1}^{\infty}\dfrac{1}{k!}\right)d^s=(e-1)d^s\le 2d^s.
	\end{align*}
	By Theorem 6.5 of \cite{wainwright2019high} and Boole's inequality, there are constants $c_1,c_2,c_3>0$ such that 
	\begin{align*}
		&{P}\left[\max_{|J|\le s}\max_{\bdelta_1,\bdelta_2\in\mathcal{D}(J)}\left| \bdelta_1 \left(|\mathcal{I}_1|^{-1}\sum_{i\in\mathcal{I}_1}\bW_i\bW_i^\T-\bSigma_{\bW}\right)\bdelta_2\right|\ge c_1\sigma_u^2\left\{\sqrt{\dfrac{s}{|\mathcal{I}_1|}}+\dfrac{s}{|\mathcal{I}_1|}\right\}+\sigma_u^2\eta  \right]\\
		&\qquad\le 2c_2d^s\exp\{-c_3|\mathcal{I}_1|\min(\eta,\eta^2)\},\ \ \ {\rm for\ all}\ \eta\ge 0.
	\end{align*}
	For any arbitrary constant $\alpha\in(0,1)$, choose
	$$
	\eta=\eta_\alpha=\max\left[\dfrac{\log(2c_2d^s/\alpha)}{c_3|\mathcal{I}_1|},\left\{\dfrac{\log(2c_2d^s/\alpha)}{c_3|\mathcal{I}_1|}\right\}^{1/2}\right].
	$$ 
	Then $\eta_\alpha$ satisfies $\min(\eta_\alpha,\eta_\alpha^2)=\log(2c_2d^s/\alpha)/(c_3|\mathcal{I}_1|)$. Hence,
	\begin{align*}
		&{P}\left[\max_{|J|\le s}\max_{\bdelta_1,\bdelta_2\in\mathcal{D}(J)}\left| \bdelta_1 \left(|\mathcal{I}_1|^{-1}\sum_{i\in\mathcal{I}_1}\bW_i\bW_i^\T-\bSigma_{\bW}\right)\bdelta_2\right|\ge c_1\sigma_u^2\left\{\sqrt{\dfrac{s}{|\mathcal{I}_1|}}+\dfrac{s}{|\mathcal{I}_1|}\right\}+\sigma_u^2\eta_\alpha  \right]\le \alpha,
	\end{align*}
	and \eqref{h1h2} follows by $|\mathcal{I}_1|\asymp n$.
	
	Let $S_0={\rm supp}(\bbeta^*)$, $\mathcal{C}(S_0,3)=\{\bdelta\in\mathbb{R}^d\,|\,\|\bdelta_{S_0^c}\|_1\le 3\|\bdelta_{S_0}\|_1\}$, and $\mathbb{S}^{d-1}=\{\bdelta\,|\,\|\bdelta\|_2=1\}$ we next prove that
	\begin{align}
		\!\!\!\max_{\bdelta_1,\bdelta_2\in\mathcal{C}(S_0,3)\,\cap\, \mathbb{S}^{d-1}}\left|\bdelta_1^\T \left(|\mathcal{I}_1|^{-1}\!\sum_{i\in\mathcal{I}_1}\bW_i\bW_i^\T-\bSigma_{\bW}\right)\bdelta_2\right|=O_p\left[\max\left\{\sqrt{\dfrac{s\log d }{n}},\dfrac{s\log d }{n}\right\}\right].\label{h1h2C}
	\end{align}
	By Lemma 14 of \cite{rudelson2013reconstruction}, for any fixed $\delta\in(0,1)$,
	$\bigcup_{|J|\le s}\mathcal{C}(J,3)\cap\mathbb{S}^{d-1}\subseteq \{1/(1-\delta)\}{\rm conv}(\bigcup_{|J|\le s^\prime}V_J\cap\mathbb{S}^{d-1})$
	with some $s^\prime\asymp s$, where $V_J$ is the space spanned by $\{\be_j\}_{j\in J}$ and ${\rm conv}(\cdot)$ denotes the convex hull of a set. Let $\widehat{\bSigma}_{\rm diff}=|\mathcal{I}_1|^{-1}\sum_{i\in\mathcal{I}_1}\bW_i\bW_i^\T-\bSigma_{\bW}$. Fix any $\bdelta_2\in\mathcal{C}(s_0,3)\cap \mathbb{S}^{d-1}$,
	\begin{align}\label{fd1}
		&\max_{\bdelta_1\in \mathcal{C}(S_0,3)\,\cap\,\mathbb{S}^{d-1}}\left|\bdelta_1^\T \widehat{\bSigma}_{\rm diff}\bdelta_2\right|\le 
		\max_{\bdelta_1\in \cup_{|J|\le s}\mathcal{C}(J,3)\,\cap\,\mathbb{S}^{d-1}}\left|\bdelta_1^\T \widehat{\bSigma}_{\rm diff}\bdelta_2\right|\nonumber\\
		&\qquad\le \dfrac{1}{1-\delta}\max_{\bdelta_1\in{\rm conv}(\bigcup_{|J|\le s^\prime}V_J\,\cap\,\mathbb{S}^{d-1})}\left|\bdelta_1^\T \widehat{\bSigma}_{\rm diff}\bdelta_2\right|\le \dfrac{1}{1-\delta}\max_{\bdelta_1\in\bigcup_{|J|\le s^\prime}V_J\,\cap\,\mathbb{S}^{d-1}}\left|\bdelta_1^\T \widehat{\bSigma}_{\rm diff}\bdelta_2\right|,
	\end{align}
	where the last inequality holds since the maximum of convex function $f_1(\bdelta_1)=|\bdelta_1^\T\widehat{\bSigma}_{\rm diff}\bdelta_2|$ occurs at an extreme point of the set ${\rm conv}(\bigcup_{|J|\le s^\prime}V_J\,\cap\,\mathbb{S}^{d-1})$. We next define 
	$f_2(\bdelta_2)=\max_{\bdelta_1\in\mathcal{C}(S_0,3)\,\cap\,\mathbb{S}^{d-1}}|\bdelta_1^\T \bSigma_{\rm diff}\bdelta_2|$,
	which is a convex function since it is the maximum of convex functions. We have
	\begin{align}\label{fd2}
		&\max_{\bdelta_1,\bdelta_2\in\mathcal{C}(S_0,3)\,\cap\, \mathbb{S}^{d-1}}\left|\bdelta_1^\T \widehat{\bSigma}_{\rm diff}\bdelta_2\right|=\max_{\bdelta_2\in\mathcal{C}(S_0,3)\,\cap\, \mathbb{S}^{d-1}}f_2(\bdelta_2)\\
		&\qquad\le \max_{\bdelta_2\in \cup_{|J|\le s}\mathcal{C}(J,3)\,\cap\,\mathbb{S}^{d-1}}f_2(\bdelta_2)\le \dfrac{1}{1-\delta}\max_{\bdelta_2\in{\rm conv}(\bigcup_{|J|\le s^\prime}V_J\,\cap\,\mathbb{S}^{d-1})}f_2(\bdelta_2)\nonumber\\
		&\qquad\le \dfrac{1}{1-\delta}\max_{\bdelta_2\in\bigcup_{|J|\le s^\prime}V_J\,\cap\,\mathbb{S}^{d-1}}f_2(\bdelta_2),
	\end{align}
	where the last inequality holds as the maximum of convex function $f_2(\bdelta_2)$ occurs at an extreme point of the set ${\rm conv}(\bigcup_{|J|\le s^\prime}V_J\,\cap\,\mathbb{S}^{d-1})$. Combining \eqref{fd1} and \eqref{fd2},
	\begin{align*}
		&\max_{\bdelta_1,\bdelta_2\in\mathcal{C}(S_0,3)\,\cap\, \mathbb{S}^{d-1}}\left|\bdelta_1^\T \bSigma_{\rm diff}\bdelta_2\right|\le \dfrac{1}{(1-\delta)^2}\max_{\bdelta_1,\bdelta_2\in\bigcup_{|J|\le s^\prime}V_J\,\cap\,\mathbb{S}^{d-1}}\left|\bdelta_1^\T \bSigma_{\rm diff}\bdelta_2\right|\\
		&\qquad=\dfrac{1}{(1-\delta)^2}\max_{|J|\le s^\prime}\max_{\bdelta_1,\bdelta_2\in\mathcal{D}(J)}\left|\bdelta_1^\T \bSigma_{\rm diff}\bdelta_2\right|=O_p\left[\max\left\{\sqrt{\dfrac{s\log d }{n}},\dfrac{s\log d }{n}\right\}\right],
	\end{align*}
	where the last equality is by \eqref{h1h2}. Therefore, we conclude that \eqref{h1h2C} holds.
	
	It suffices to obtain the remaining results when $\widehat{\bbeta}_{\rm SR}\neq\bbeta^*$. Note that
	\begin{equation}\label{B_6_1}
		(\widehat{\bbeta}_{\rm SR}-\bbeta^*)^\T\bSigma_{\bW} (\widehat{\bbeta}_{\rm SR}-\bbeta^*)/\|\widehat{\bbeta}_{\rm SR}-\bbeta^*\|_2^2\le C_{\bSigma}=O(1).
	\end{equation}
	For the Lasso estimator $\widehat{\bbeta}_{\rm SR}$, Theorem 7.13 of \cite{wainwright2019high} and Lemma \ref{pi_constraint} imply that $\widehat{\bbeta}_{\rm SR}-\bbeta^*\in\mathcal{C}(S_0,3)$. By \eqref{h1h2C} and \eqref{B_6_1}, 
	\begin{align*}
		&\dfrac{|\mathcal{I}_1|^{-1}\sum_{i\in\mathcal{I}_1}\{\bW_i^\T(\widehat{\bbeta}_{\rm SR}-\bbeta^*)\}^2}{\|\widehat{\bbeta}_{\rm SR}-\bbeta^*\|_2^2}\\
		&\qquad \le \sup_{\bdelta_1,\bdelta_2\in\mathcal{C}(S_0,3)\,\cap\,\mathbb{S}^{d-1}}\left|\bdelta_1^\T \left(|\mathcal{I}_1|^{-1}\sum_{i\in\mathcal{I}_1}\bW_i\bW_i^\T-\bSigma_{\bW}\right)\bdelta_2\right|\\
		&\qquad\quad+\dfrac{(\widehat{\bbeta}_{\rm SR}-\bbeta^*)^\T\bSigma_{\bW} (\widehat{\bbeta}_{\rm SR}-\bbeta^*)}{\|\widehat{\bbeta}_{\rm SR}-\bbeta^*\|_2^2}
		=O_p\left[1+\max\left\{\sqrt{\dfrac{s\log d }{n}},\dfrac{s\log d }{n}\right\}\right].
	\end{align*}
	By Lemma \ref{delta}, $\|\widehat{\bbeta}_{\rm SR}-\bbeta^*\|_2^2=O_p(s\log d /N)$. Hence,
	\begin{align*}
		&|\mathcal{I}_1|^{-1}\sum_{i\in\mathcal{I}_1}(\bW_i^\T(\widehat{\bbeta}_{\rm SR}-\bbeta^*))^2=O_p\left[\dfrac{s\log d }{N}+\dfrac{s\log d }{N}\max\left\{\sqrt{\dfrac{s\log d }{n}},\dfrac{s\log d }{n}\right\}\right]\\
		&\qquad=O_p\left[\max\left\{\dfrac{s\log d }{(Nn)^{1/2}}\left(\dfrac{s\log d }{N}\right)^{1/2},\dfrac{s^2(\log d)^2 }{Nn}\right\}+\dfrac{s\log d }{N}\right]=O_p\left\{\dfrac{s\log d }{(Nn)^{1/2}}\right\},
	\end{align*}
	where the last equality is by $s\ll N/\log d$ under the assumption $s\ll (Nn)^{1/2}/\log d$.
	\hfill\BlackBox
	\\ \hspace*{\fill} \\
	{\bf Proof of Lemma \ref{numerator}}.
	Notice that
	\begin{align*}
		|\mathcal{I}_1|^{-1}\sum_{i\in \mathcal{I}_1}\hat{v}_i^2-\sigma_{v,1}^2&=|\mathcal{I}_1|^{-1}\sum_{i\in \mathcal{I}_1}(Z_i-\bW_i^\T\widehat{\bbeta}_{\rm SR})^2-\sigma_{v,1}^2\\
		&=\Big(|\mathcal{I}_1|^{-1}\sum_{i\in \mathcal{I}_1}v_i^2-\sigma_{v,1}^2\Big)+|\mathcal{I}_1|^{-1}\sum_{i\in \mathcal{I}_1}\Big(\bW_i^\T\bbeta^*-\bW_i^\T\widehat{\bbeta}_{\rm SR}\Big)^2\\
		&\quad+2|\mathcal{I}_1|^{-1}\sum_{i\in \mathcal{I}_1}v_i\Big(\bW_i^\T\bbeta^*-\bW_i^\T\widehat{\bbeta}_{\rm SR}\Big)=A_3+A_4+A_5.
	\end{align*}
	For $A_3$, we have ${E}(v_i^2)=\sigma_{v,1}^2$ for $i\in\mathcal{I}_1$. By Lemma \ref{psi}, $\|v_i\|_{\psi_1}=O(1)$. By Lemma \ref{sub_gaussian_bound}, ${\rm var}(|\mathcal{I}_1|^{-1}\sum_{i\in \mathcal{I}_1}v_i^2)=O(n^{-1})$. By Chebyshev's inequality, $A_3=O_p(n^{-1/2})$. In addition, by Lemma \ref{B_6}, $A_4=O_p\{s\log d /(Nn)^{1/2}\}$.
	For $A_5$, reperating the proof of Lemma \ref{delta}, $\||\mathcal{I}_1|^{-1}\sum_{i\in\mathcal{I}_1}\bW_iv_i\|_\infty=O_p\{(\log d/n)^{1/2}\}$. Together with Lemma \ref{pi_constraint},
	\begin{equation}\label{T_3_3}
		\begin{split}
			|A_5|&\le 2\|\widehat{\bbeta}_{\rm SR}-\bbeta^*\|_1\bigg\||\mathcal{I}_1|^{-1}\sum_{i\in \mathcal{I}_1}\bW_iv_i\bigg\|_\infty
			=O_p\left\{s\log d /(Nn)^{1/2}\right\}.
		\end{split}
	\end{equation}
	Therefore, $|\mathcal{I}_1|^{-1}\sum_{i\in\mathcal{I}_1}\hat{v}_i^2-\sigma_{v,1}^2=O_p\{n^{-1/2}+s\log d /(Nn)^{1/2}\}$. By $\sigma_{v,1}^2\ge 1/C_{\bSigma}$ and the assumption that $s\ll (Nn)^{1/2}/\log d$, $|\mathcal{I}_1|^{-1}\sum_{i\in \mathcal{I}}\hat{v}_i^2/\sigma_{v,1}^2=1+o_p(1)$.
	\hfill\BlackBox
	\\ \hspace*{\fill} \\
	{\bf Proof of Lemma \ref{v_i}}.
	We first prove that ${E}(\|\bW_i\|_\infty^4)=O\{(\log d )^2\}$. For any $j\in [d]$, $\|W_{ij}\|_{\psi_2}\le \sigma_u$. By Boole's inequality, for any $t>0$,
	${P}(\|\bW_{i}\|_\infty>t)\le 2d\exp(-t^2/\sigma_u^2)$. Since we also have ${P}(\|\bW_i\|_\infty>t)\le 1$,
	\begin{align*}
		{E}(\|\bW_i\|_\infty^{4})&=\int_{0}^\infty 4t^{3}{P}(\|\bW_i\|_\infty>t)dt\\
		&\le \int_{0}^{\sigma_u(\log d)^{1/2}} 4t^{3}dt+\int_{\sigma_u(\log d)^{1/2}}^{\infty}8t^{3}d\exp(-t^2/\sigma_u^2)dt=a_1+a_2,
	\end{align*}
	where $a_1=(\sigma_u^2\log d )^2=O\{(\log d )^2\}$ and
	\begin{align*}
		a_2&=8d\int_{\sigma_u(\log d)^{1/2}}^{\infty}t^3\exp\left(-\dfrac{t^2}{\sigma_u^2}\right)dt=8\sigma_u^{4}d\int_{\log d}^{\infty}\dfrac{x}{\exp(x)}dx=\dfrac{8\sigma_u^{4}d(1+\log d)}{2d}=O(\log d ).
	\end{align*}
	Hence ${E}(\|\bW_i\|_\infty^4)=O\{(\log d )^2\}$. By Markov's inequality, we have $|\mathcal{I}_1|^{-1}\sum_{i\in \mathcal{I}_1}\|\bW_i\|_\infty^4=O_p\{(\log d )^2\}$. By Lemma \ref{delta}, $\|\widehat{\bbeta}_{\rm SR}-\bbeta^*\|_1=O_p\{s(\log d /N)^{1/2}\}$. When $s\ll N^{1/2}/\log d$, 
	\begin{align*}
		&|\mathcal{I}_1|^{-1}\sum_{i\in \mathcal{I}_1}(v_i-\hat{v}_i)^{4}
		\le |\mathcal{I}_1|^{-1}\sum_{i\in \mathcal{I}_1}\|\bW_i\|_\infty^{4}\|\widehat{\bbeta}_{\rm SR}-\bbeta^*\|_1^{4}=O_p\left\{\left(s\log d /N^{1/2}\right)^4\right\}=o_p(1).
	\end{align*}
	For the second part, since $v_i$ is sub-Gaussian, ${E}(v_i^4)=O(1)$. By Markov's inequality, $|\mathcal{I}_1|^{-1}\sum_{i\in \mathcal{I}_1}v_i^4=O_p(1)$, and hence
	\begin{align*}
		|\mathcal{I}_1|^{-1}\sum_{i\in \mathcal{I}_1}\hat{v}_i^4&=|\mathcal{I}_1|^{-1}\sum_{i\in \mathcal{I}_1}\{v_i+(\hat{v}_i-v_i)\}^4\le 8|\mathcal{I}_1|^{-1}\sum_{i\in \mathcal{I}_1}v_i^4+8|\mathcal{I}_1|^{-1}\sum_{i\in \mathcal{I}_1}(v_i-\hat{v}_i)^4=O_p(1).
	\end{align*}
	\hfill\BlackBox
	\\ \hspace*{\fill} \\
	{\bf Proof of Lemma \ref{df_delta}}.
	Repeating the proof of Lemma \ref{pi_constraint}, we have
	\begin{align}
		\bigg\|n^{-1}\sum_{i\in \mathcal{I}}\bW_i(Z_i-\bW_i^\T\bbeta^*)\bigg\|_\infty&=\bigg\|n^{-1}\sum_{i\in \mathcal{I}}\bW_iv_i\bigg\|_\infty=O_p\left\{(\log d /n)^{1/2}\right\},\label{bound:Wv-I}\\
		\bigg\|N^{-1}\sum_{i\in \mathcal{I}\cup\mathcal{J}}\bW_i(Z_i-\bW_i^\T\bbeta^*)\bigg\|_\infty&=\bigg\|N^{-1}\sum_{i\in \mathcal{I}\cup\mathcal{J}}\bW_iv_i\bigg\|_\infty=O_p\left\{(\log d /N)^{1/2}\right\}.\nonumber
	\end{align}
	Hence, for any arbitrary constant $\alpha\in (0,1)$, there exists a constant $C_{\alpha}$ such that $ \lambda_{\bbeta}=C_\alpha(\log d /N)^{1/2}$ satisfies both of the constraints $\|n^{-1}\sum_{i\in \mathcal{I}\cup \mathcal{J}}\bW_i(Z_i-\bW_i^\T\bbeta^*)\|_\infty\le  \lambda_{\bbeta}$ and $\|n^{-1}\sum_{i\in \mathcal{I}}\bW_i(Z_i-\bW_i^\T\bbeta^*)\|_\infty\le (N/n)^{1/2}\cdot  \lambda_{\bbeta}$ with probability at least $1-\alpha$. Thus, the solution set of the optimization problem is non-empty and $\widehat{\bbeta}_{\rm DFA}$ exists. Repeating the proof of Lemma \ref{delta}, we have the following event holds with probability at least $1-2\alpha$ and some constant $\kappa>0$,
	\begin{align*}
		\mathcal{M}_3=\Big\{
		&\Big\|N^{-1}\sum_{i\in \mathcal{I}\cup\mathcal{J}}\bW_iZ_i-\widetilde{\bSigma}_{\bW}\bbeta^*\Big\|_\infty\le  \lambda_{\bbeta},\Big\|n^{-1}\sum_{i\in \mathcal{I}}\bW_i(Z_i-\bW_i^\T\bbeta^*)\Big\|_\infty\le (N/n)^{1/2}\cdot \lambda_{\bbeta},\\
		&\bp^\T\widetilde{\bSigma}_{\bW}\bp\ge \kappa\|\bp\|_2^2,\ \ \forall \bp\in\mathbb{R}^{d}\ \ \mbox{with}\ \ \|\bp_{S^c}\|_1\le 3\|\bp_S\|_1 \Big\},
	\end{align*}
	where $\widetilde{\bSigma}_{\bW}=N^{-1}\sum_{i\in \mathcal{I}\cup\mathcal{J}}\bW_i\bW_i^\T$. Condition on the event $\mathcal{M}_3$. Let $S_0={\rm supp}(\bbeta^*)$ and $\bdelta=\widehat{\bbeta}_{\rm DFA}-\bbeta^*$, then
	$$
	\|\bbeta^*_{S_0}\|_1-\|\bdelta_{S_0}\|_1+\|\bdelta_{S_0^{c}}\|_1\le \|\bbeta^*_{S_0}+\bdelta_{S_0}\|_1+\|\bdelta_{S_0^{c}}\|_1=\|\widehat{\bbeta}_{\rm DFA}\|_1 \le \|\bbeta^*\|_1,
	$$
	and hence $\|\bdelta_{S_0^c}\|_1\le \|\bdelta_{S_0}\|_1\le 3\|\bdelta_{S_0}\|_1$. By the construction of $\widehat{\bbeta}_{\rm DFA}$, $\|N^{-1}\sum_{i\in \mathcal{I}\cup\mathcal{J}}\bW_iZ_i-\widetilde{\bSigma}_{\bW}\widehat{\bbeta}_{\rm DFA}\|_\infty\le  \lambda_{\bbeta}$.
	Hence, 
	$$\|\widetilde{\bSigma}_{\bW}\bdelta\|_\infty\le\Big\|N^{-1}\sum_{i\in \mathcal{I}\cup\mathcal{J}}\bW_iZ_i-\widetilde{\bSigma}_{\bW}\bbeta^*\Big\|_\infty+\Big\|N^{-1}\sum_{i\in \mathcal{I}\cup\mathcal{J}}\bW_iZ_i-\widetilde{\bSigma}_{\bW}\widehat{\bbeta}_{\rm DFA}\Big\|_\infty\le2  \lambda_{\bbeta}.$$
	Moreover,
	\begin{equation*}
		\begin{split}
			\kappa\|\bdelta\|_2^2&\le\delta^\T\widetilde{\bSigma}_{\bW}\bdelta
			\le \|\bdelta\|_1\|\widetilde{\bSigma}_{\bW}\bdelta\|_\infty\le 2\|\bdelta\|_1 \lambda_{\bbeta}=2(\|\bdelta_{S_0}\|_1+\|\bdelta_{S_0^c}\|_1) \lambda_{\bbeta}\\
			&\le 4 \lambda_{\bbeta} \|\bdelta_{S_0}\|_1\le 4 \lambda_{\bbeta}s^{1/2}\|\bdelta_{S_0}\|_2\le 4 \lambda_{\bbeta}s^{1/2}\|\bdelta\|_2.
		\end{split}
	\end{equation*}
	Hence, $
	\|\bdelta\|_2\le 4 \lambda_{\bbeta} s^{1/2}/\kappa$ and 
	$\|\bdelta\|_1\le 2\|\bdelta_{S_0}\|_1\le 2s^{1/2}\|\bdelta_{S_0}\|_2\le 2s^{1/2}\|\bdelta\|_2=8s \lambda_{\bbeta}/\kappa$
	which gives that $\|\bdelta\|_2=O_p\{(s\log d /N)^{1/2}\}$ and $\|\bdelta\|_1=O_p\{s(\log d /N)^{1/2}\}$.
	\hfill\BlackBox
	\\ \hspace*{\fill} \\
	{\bf Proof of Lemma \ref{vizi}}.
	On the event $\mathcal{M}_3$, we have
	\begin{equation}
		\Big\|n^{-1}\sum_{i\in \mathcal{I}}\bW_i(Z_i-\bW_i^\T\bbeta^*)\Big\|_\infty\le (N/n)^{1/2}\cdot \lambda_{\bbeta},\label{Wv}
	\end{equation}
	In addition, by the construction of $\widehat{\bbeta}_{\rm DFA}$,
	\begin{equation}
		\Big\|n^{-1}\sum_{i\in \mathcal{I}}\bW_i(Z_i-\bW_i^\T\widehat{\bbeta}_{\rm DFA})\Big\|_\infty\le (N/n)^{1/2}\cdot \lambda_{\bbeta}.\label{Whatv}
	\end{equation}
	Taking the difference between \eqref{Wv} and \eqref{Whatv}, we have $\|n^{-1}\sum_{i\in \mathcal{I}}\bW_i\bW_i^\T(\widehat{\bbeta}_{\rm DFA}-\bbeta^*)\|_\infty\le 2(N/n)^{1/2}\cdot \lambda_{\bbeta}$.
	Therefore,
	\begin{equation}\label{hatvmv}
		\begin{split}
			n^{-1}\sum_{i\in\mathcal{I}}(\hat{v}_i-v_i)^2&\le \|\widehat{\bbeta}_{\rm DFA}-\bbeta^*\|_1\Big\|n^{-1}\sum_{i\in \mathcal{I}}\bW_i\bW_i^\T(\widehat{\bbeta}_{\rm DFA}-\bbeta^*)\Big\|_\infty\\
			&\le 2\|\widehat{\bbeta}_{\rm DFA}-\bbeta^*\|_1(N/n)^{1/2}\cdot \lambda_{\bbeta}=O_p\left\{s\log d/(Nn)^{1/2}\right\}.
		\end{split}
	\end{equation}
	Besides, since $v_i$ is Gaussian with ${E}(v_i^2)=\sigma_{v,1}^2=O(1)$, we have ${\rm var}(n^{-1}\sum_{i\in\mathcal{I}}v_i^2)=O(1/n)$. By Chebyshev's inequality, 
	\begin{equation}\label{bound:v-v}
		n^{-1}\sum_{i\in \mathcal{I}}v_i^2-\sigma_{v,1}^2=O_p(n^{-1/2}).
	\end{equation}
	Additionally,
	\begin{equation}\label{bound:v-dv}
		\begin{split}
			\Big|n^{-1}\sum_{i\in\mathcal{I}}v_i(\hat{v}_i-v_i)\Big|
			&\le\Big\|n^{-1}\sum_{i\in\mathcal{I}}v_i\bW_i\Big\|_\infty\|\widehat{\bbeta}_{\rm DFA}-\bbeta^*\|_1=O_p\left\{s\log d/(Nn)^{1/2}\right\},
		\end{split}
	\end{equation}
	where the last equality holds by \eqref{bound:Wv-I} and Lemma \ref{df_delta}. By \eqref{bound:v-v}, \eqref{hatvmv}, and \eqref{bound:v-dv},
	\begin{align*}
		\left|n^{-1}\sum_{i\in \mathcal{I}}\hat{v}_i^2-\sigma_{v,1}^2\right|&\leq\left|n^{-1}\sum_{i\in \mathcal{I}}v_i^2-\sigma_{v,1}^2\right|+n^{-1}\sum_{i\in\mathcal{I}}(\hat{v}_i-v_i)^2+2\left|n^{-1}\sum_{i\in\mathcal{I}}v_i(\hat{v}_i-v_i)\right|\\
		&=O_p\left\{n^{-1/2}+s\log d/(Nn)^{1/2}\right\}.
	\end{align*}
	Note that ${E}(v_iZ_i)={E}(v_i^2)=\sigma_{v,1}^2$ for $i\in\mathcal{I}$. Since $v_i$ and $Z_i$ are Gaussian random variables, $v_iZ_i$ is sub-exponential with $\|v_iZ_i\|_{\psi_1}=O(1)$. By Lemma \ref{Bernstein_inequality}, ${\rm var}(n^{-1}\sum_{i\in\mathcal{I}}v_iZ_i)=O(1/n)$. By Chebyshev's inequality, $n^{-1}\sum_{i\in \mathcal{I}}v_iZ_i-\sigma_{v,1}^2=O_p(n^{-1/2})$. Besides, $Z_i^2$ is sub-exponential with $\|Z_i\|_{\psi_1}=O(1)$. By Lemma \ref{Bernstein_inequality}, $E(Z_i^2)=O(1)$ and ${\rm var}(n^{-1}\sum_{i\in\mathcal{I}}Z_i^2)=O(1/n)$. By Chebyshev's inequality, $n^{-1}\sum_{i\in \mathcal{I}}Z_i^2={E}(Z_i^2)+O_p(n^{-1/2})=O_p(1)$. By \eqref{hatvmv},
	\begin{align*}
		n^{-1}\sum_{i\in \mathcal{I}}\hat{v}_iZ_i-\sigma_{v,1}^2
		&=(n^{-1}\sum_{i\in \mathcal{I}}v_iZ_i-\sigma_{v,1}^2)+n^{-1}\sum_{i\in \mathcal{I}}(\hat{v}_i-v_i)Z_i\\
		&\le(n^{-1}\sum_{i\in \mathcal{I}}v_iZ_i-\sigma_{v,1}^2)+\left\{n^{-1}\sum_{i\in \mathcal{I}}(\hat{v}_i-v_i)^2\right\}^{1/2}\left\{n^{-1}\sum_{i\in \mathcal{I}}Z_i^2\right\}^{1/2}\\
		&=O_p\left[{n^{-1/2}}+\left\{\dfrac{s\log d }{(Nn)^{1/2}}\right\}^{1/2}\right]=o_p(1).
	\end{align*}
	\hfill\BlackBox
	\\ \hspace*{\fill} \\
	{\bf Proof of Lemma \ref{Sigma_W_inf}}.
	Notice that $W_{ij}\sim\mathcal{N}(0,\be_j^\T\bSigma_{\bW}\be_j)$ with $\be_j^\T\bSigma_{\bW}\be_j\le C_{\bSigma}$ and $\bW_i^\T \bbeta^*\sim \mathcal{N}(0,{\bbeta^*}^\T\bSigma_{\bW}\bbeta^*)$ with ${\bbeta^*}^\T\bSigma_{\bW}\bbeta^*\le \lambda_{\rm max}(\bSigma_{\bW})\|\bbeta^*\|_2^2\le C_{\bSigma}^3$. Hence, there exists a constant $C>0$ such that $\|W_{ij}\bW_i^\T\bbeta^*\|_{\psi_1}\le C$ with ${E}(\bW_i\bW_i^\T\bbeta^*)=\bSigma_{\bW}\bbeta^*$. For any arbitrary constant $\alpha\in (0,1)$, by Lemma \ref{Bernstein_inequality} and Boole's inequality, let $t=\{\log(2d/\alpha)/n\}^{1/2}$. When $n\geq\log(2d/\alpha)$, $t\leq1$ and hence
	\begin{align*}
		{P}\bigg(\bigg\|n^{-1}\sum_{i\in \mathcal{I}} \bW_i\bW_i^\T\bbeta^*-\bSigma_{\bW}\bbeta^*\bigg\|_\infty\ge 6C\{\log(2d/\alpha)/n\}^{1/2} \bigg)\le 2d\exp(-nt^2)=\alpha.
	\end{align*} 
	It follows that $\|(n^{-1}\sum_{i\in\mathcal{I}} \bW_i\bW_i^\T-\bSigma_{\bW})\bbeta^*\|_\infty = O_p\{(\log d /n)^{1/2}\}$.
	\hfill\BlackBox
	\\ \hspace*{\fill} \\
	{\bf Proof of Lemma \ref{xeps}}.
	By Lemma \ref{best_linear_slope}, we have ${E}(\bX_i\varepsilon_i)=0$ for any $i\in\mathcal{I}$. For any $j \in [d]$, $\|X_{ij}\|_{\psi_2}\le c_1$, $\|\varepsilon_i\|_{\psi_2}\le c_2\sigma_{\varepsilon}$ for some constants $c_1,c_2>0$. Hence, $\|X_{ij}\varepsilon_i\|_{\psi_1} \le c_1c_2\sigma_{\varepsilon}$. By Lemma \ref{Bernstein_inequality} and Boole's inequality, 
	$$
	{P}\bigg(\bigg\|n^{-1}\sum_{i\in \mathcal{I}}\bX_i\varepsilon_i\bigg\|_\infty\ge 2c_1c_2\sigma_{\varepsilon}t(2+t)\bigg)\le 2(d+1)\exp(-nt^2).
	$$
	Let $t=\{\log(2(d+1)/\alpha)/n\}^{1/2}$. When $n\ge \log(2(d+1)/\alpha)$, $t\le 1$ and hence
	$$
	{P}\bigg[\bigg\|n^{-1}\sum_{i\in \mathcal{I}}\bX_i\varepsilon_i\bigg\|_\infty\ge 6c_1c_2\sigma_{\varepsilon}\{\log(2(d+1)/\alpha)/n\}^{1/2}\bigg]\le \alpha.
	$$
	\hfill\BlackBox

	\section{Proof of the main results}\label{sec:proof_main}
	{\bf Proof of Theorem \ref{thm:SR_consist_rate_cs} and Theorem 1}.
	In the following, we prove Theorem 1 and Theorem \ref{thm:SR_consist_rate_cs} under more general sub-Gaussian designs and allow for the presence of covariate shift, assuming Assumptions \ref{ass:MAR}-\ref{ass:sub_gau_design} and $E(v_i\mid \bW_i)=0$ holds. Note that Assumption \ref{ass:sub_gau_design} and $E(v_i\mid \bW_i)=0$ hold under the stronger Gaussian assumptions, Assumption \ref{ass:MAR} holds when there is no covariates shift. Denote $\hat{R}=|\bar{\mathcal{J}}|^{-1}\sum_{i\in \bar{\mathcal{J}}}\widehat{\bu}_{\rm SR}^{\top}\bW_i(Z_i-\bW_i^\T\widehat{\bbeta}_{\rm SR})$, we have
	\begin{align}
		\hat{\theta}_{\rm SR}-\theta
		&=\dfrac{|\mathcal{I}_1|^{-1}\sum_{i\in \mathcal{I}_1}\hat{v}_iY_i-\hat{R}-\theta\cdot |\mathcal{I}_1|^{-1}\sum_{i\in \mathcal{I}_1}\hat{v}_i^2}{|\mathcal{I}_1|^{-1}\sum_{i\in \mathcal{I}_1} \hat{v}_i^2}=\dfrac{|\mathcal{I}_1|^{-1}\sum_{i\in \mathcal{I}_1}\hat{v}_i(Y_i-\theta\hat{v}_i)-\hat{R}}{|\mathcal{I}_1|^{-1}\sum_{i\in \mathcal{I}_1} \hat{v}_i^2}\nonumber\\
		&=\dfrac{|\mathcal{I}_1|^{-1}\sum_{i\in \mathcal{I}_1}v_i(Y_i-\theta\hat{v}_i)+\big\{|\mathcal{I}_1|^{-1}\sum_{i\in \mathcal{I}_1}(\hat{v}_i-v_i)(Y_i-\theta\hat{v}_i)-\hat{R}\big\}}{|\mathcal{I}_1|^{-1}\sum_{i\in \mathcal{I}_1} \hat{v}_i^2}\nonumber\\
		&=\dfrac{T_1+T_2}{T_3}.\label{theta_dec}
	\end{align} 
	For $T_1$, notice that $
	Y_i=Z_i\theta+\bW_i^\T\balpha^*+\varepsilon_i=\bW_i^\T(\theta\bbeta^*+\balpha^*)+\theta v_i+\varepsilon_i$,
	and hence $Y_i-\hat{v}_i\theta=\bW_i^\T(\theta\bbeta^*+\balpha^*)+\varepsilon_i+\theta\cdot(\widehat{\bbeta}_{\rm SR}-\bbeta^*)^\T \bW_i$. Observe that
	\begin{align*}
		T_1&=|\mathcal{I}_1|^{-1}\sum_{i\in \mathcal{I}_1}v_i(Y_i-\hat{v}_i\theta)\\
		&=|\mathcal{I}_1|^{-1}\sum_{i\in \mathcal{I}_1}v_i\{\bW_i^\T(\theta\bbeta^*+\balpha^*)+\varepsilon_i\}+\theta\cdot |\mathcal{I}_1|^{-1}\sum_{i\in \mathcal{I}_1}(\widehat{\bbeta}_{\rm SR}-\bbeta^*)^\T \bW_iv_i=T_{1,1}+T_{1,2}.
	\end{align*}
	By Lemma \ref{T_11}, $T_{1,1}=O_p\{(M+\sigma_{\varepsilon})n^{-1/2}\}$. By Lemmas \ref{pi_constraint} and \ref{delta},
	\begin{align}\label{T_12}
		|T_{1,2}|\le |\theta|\cdot\|\widehat{\bbeta}_{\rm SR}-\bbeta^*\|_1\bigg\||\mathcal{I}_1|^{-1}\sum_{i\in \mathcal{I}_1}\bW_iv_i\bigg\|_\infty=O_p\left\{\dfrac{Ms\log d }{(Nn)^{1/2}}\right\}.
	\end{align}
	Thus,
	\begin{equation}\label{T_1}
		T_1=O_p\left\{\dfrac{M+\sigma_{\varepsilon}}{n^{1/2}}+\dfrac{Ms\log d }{(Nn)^{1/2}}\right\}.
	\end{equation}
	For $T_2$, notice that
	\begin{align*}
		T_2&=\theta\cdot |\mathcal{I}_1|^{-1}\sum_{i\in \mathcal{I}_1}(\widehat{\bbeta}_{\rm SR}-\bbeta^*)^\T \bW_i\hat{v}_i+\bigg\{|\mathcal{I}_1|^{-1}\sum_{i\in \mathcal{I}_1}(\bbeta^*-\widehat{\bbeta}_{\rm SR})^\T \bW_iY_i-\hat{R}\bigg\}=T_{2,1}+T_{2,2}.
	\end{align*}
	By Lemmas \ref{pi_constraint}, \ref{delta} and \ref{B_6}, 
	\begin{equation}\label{T_21}
		\begin{split}
			\!\!\!|T_{2,1}|&=\left|\theta\cdot |\mathcal{I}_1|^{-1}\!\sum_{i\in \mathcal{I}_1}(\widehat{\bbeta}_{\rm SR}-\bbeta^*)^\T \bW_iv_i-\theta\cdot |\mathcal{I}_1|^{-1}\!\sum_{i\in \mathcal{I}_1}\{\bW_i^\T(\widehat{\bbeta}_{\rm SR}-\bbeta^*)\}^2\right|\\
			&\le |\theta|\|\widehat{\bbeta}_{\rm SR}-\bbeta^*\|_1\bigg\||\mathcal{I}_1|^{-1}\!\sum_{i\in \mathcal{I}_1}\bW_iv_i\bigg\|_\infty\!\!+|\theta| |\mathcal{I}_1|^{-1}\!\sum_{i\in \mathcal{I}_1}\{\bW_i^\T(\widehat{\bbeta}_{\rm SR}-\bbeta^*)\}^2=O_p\left\{\dfrac{Ms\log d }{(Nn)^{1/2}}\right\}.
		\end{split}
	\end{equation}
	For $T_{2,2}$, 
	\begin{align*}
		T_{2,2}&=|\mathcal{I}_1|^{-1}\sum_{i\in \mathcal{I}_1}(\bbeta^*-\widehat{\bbeta}_{\rm SR})^\T \bW_iY_i-|\bar{\mathcal{J}}|^{-1}\sum_{i\in \bar{\mathcal{J}}}\widehat{\bu}_{\rm SR}^\T \bW_i\{v_i+\bW_i^\T(\bbeta^*-\widehat{\bbeta}_{\rm SR})\}\\
		&=\vphantom{b_N^{-1}\sum_{i\in \bar{\mathcal{J}}}\widehat{\bu}_{\rm SR}^\T \bW_iv_i}\bigg\langle|\mathcal{I}_1|^{-1}\sum_{i\in \mathcal{I}_1}\bW_iY_i-\widehat{\bSigma}_{\bW} \widehat{\bu}_{\rm SR},\bbeta^*-\widehat{\bbeta}_{\rm SR}\bigg\rangle-|\bar{\mathcal{J}}|^{-1}\sum_{i\in \bar{\mathcal{J}}}\widehat{\bu}_{\rm SR}^\T \bW_iv_i=T_{2,2,1}+T_{2,2,2},
	\end{align*}
	where $\widehat{\bxi}=|\mathcal{I}_2|^{-1}\sum_{i\in \mathcal{I}_2}\bW_iY_i$ and $\widehat{\bSigma}_{\bW} =|\bar{\mathcal{J}}|^{-1}\sum_{i\in\bar{\mathcal{J}}}\bW_i\bW_i^\T$. By the construction of $\widehat{\bu}_{\rm SR}$, $\|\widehat{\bxi}-\widehat{\bSigma}_{\bW} \widehat{\bu}_{\rm SR}\|_\infty\le \lambda_{\bu}$. Together with Lemmas \ref{delta} and \ref{xi_infty},
	\begin{equation}\label{T_221}
		\begin{split}
			|T_{2,2,1}|&\le \|\widehat{\bbeta}_{\rm SR}-\bbeta^*\|_1\bigg(\bigg\||\mathcal{I}_1|^{-1}\sum_{i\in \mathcal{I}_1}\bW_iY_i-\widehat{\bxi}\,\bigg\|_\infty+\|\widehat{\bxi}-\widehat{\bSigma}_{\bW} \widehat{\bu}_{\rm SR}\|_\infty\bigg)\\
			&=O_p\left\{\dfrac{(M+\sigma_{\varepsilon})s\log d }{(Nn)^{1/2}}\right\}.
		\end{split}
	\end{equation}
	By Lemma \ref{T_2_2_cov_shift}, $T_{2,2,2}=O_p(M/N^{1/2})$. Therefore,
	\begin{equation}\label{T_2}
		T_2=O_p\left\{\dfrac{M}{N^{1/2}}+\dfrac{(M+\sigma_{\varepsilon})s\log d }{(Nn)^{1/2}}\right\}.
	\end{equation} 
	Additionally, by Lemma \ref{numerator}, 
	\begin{equation}\label{T_3}
		1/T_3=O_p(1).
	\end{equation}
	Combining \eqref{T_1}, \eqref{T_2}, and \eqref{T_3}, 
	$
	\hat{\theta}_{\rm SR}-\theta=O_p[(M+\sigma_{\varepsilon})\{n^{-1/2}+s\log d /(Nn)^{1/2}\}].
	$
	\hfill\BlackBox
	\\ \hspace*{\fill} \\
	{\bf Proof of Theorem \ref{thm:SR_asy_normal_cs} and Theorem 2}.
	Now we proceed to prove Theorem \ref{thm:SR_asy_normal_cs}, noting that Theorem 2 serves as a corollary of Theorem \ref{thm:SR_asy_normal_cs}. Consider the representation \eqref{theta_dec}. By Lemma \ref{T_11}, $T_{1,1}=O_p\{(M+\sigma_{\varepsilon})/n^{1/2}\}$. By Lemma \ref{T_2_2_cov_shift}, $T_{2,2,2}=O_p(M/N^{1/2})$. As shown in \eqref{T_12}, \eqref{T_21}, and \eqref{T_221}, we have $T_{1,2}=O_p\{Ms\log d /(Nn)^{1/2}\}$, $T_{2,1}=O_p\{Ms\log d /(Nn)^{1/2}\}$, and $T_{2,2,1}=O_p\{(M+\sigma_{\varepsilon})s\log d /(Nn)^{1/2}\}$. Let $\mathcal{D}=\{(\bW_i,\varepsilon_i)_{i\in \mathcal{I}_1},(Y_i,\bW_i)_{i\in \mathcal{I}_2},(\bW_i)_{i\in \mathcal{J}}\}$, under Gaussian designs, 
	\begin{align*}
		&(T_{1,1}+T_{2,2,1})\mid \mathcal{D}=
		\bigg(\dfrac{1}{|\mathcal{I}_1|}\sum_{i\in \mathcal{I}_1}v_i\{\bW_i^\T(\theta\bbeta^*+\balpha^*)+\varepsilon_i\}-\dfrac{1}{|\bar{\mathcal{J}}|}\sum_{i\in \bar{\mathcal{J}}}\widehat{\bu}_{\rm SR}^\T \bW_iv_i\bigg)\mid \mathcal{D}\sim\mathcal{N}(0,\sigma_T^2),
	\end{align*}
	where
	\begin{align*}
		\sigma_T^2&=\sigma_{v,1}^2\sum_{i\in\mathcal{I}_1}\left(\dfrac{Y_i-\theta v_i}{|\mathcal{I}_1|}-\dfrac{\widehat{\bu}_{\rm SR}^\T \bW_i}{|\bar{\mathcal{J}}|}\right)^2+\sigma_{v,2}^2|\bar{\mathcal{J}}|^{-2}\sum_{i\in\mathcal{J} }(\widehat{\bu}_{\rm SR}^\T \bW_i)^2=\sigma_{T,1}^2+\sigma_{T,2}^2.
	\end{align*}
	It holds that
	\begin{equation}\label{normal:T11+T221}
		\dfrac{(T_{1,1}+T_{2,2,1})}{\sigma_T}\sim\mathcal{N}(0,1).
	\end{equation}
	We provide a lower bound for $\sigma_{T}^2$. Observe that
	\begin{align*}
		\sigma_{T,1}^2&\ge |\mathcal{I}_1|^{-1}(1/C_{\bSigma})|\mathcal{I}_1|^{-1}\sum_{i\in \mathcal{I}_1}[\bW_i^\T\{\theta\bbeta^*+\balpha^*-(|\mathcal{I}_1|/|\bar{\mathcal{J}}|)\widehat{\bu}_{\rm SR}\}+\varepsilon_i]^2\\
		&\ge \dfrac{1}{|\mathcal{I}_1|C_{\bSigma}}\Big(|\mathcal{I}_1|^{-1}\sum_{i\in \mathcal{I}_1}\varepsilon_i^2+2|\mathcal{I}_1|^{-1}\sum_{i\in \mathcal{I}_1}\bW_i^\T(\theta\bbeta^*+\balpha^*)\varepsilon_i-2|\bar{\mathcal{J}}|^{-1}\sum_{i\in \mathcal{I}_1}\bW_i^\T\widehat{\bu}_{\rm SR}\varepsilon_i\Big)\\
		&=\dfrac{1}{|\mathcal{I}_1|C_{\bSigma}}(B_1+B_2+B_3)
	\end{align*}
	For $B_1$, we first consider the term $|\mathcal{I}_1|^{-1}\sum_{i\in \mathcal{I}_1}\varepsilon_i^2/\sigma_{\varepsilon}^2$. Notice that $E(\varepsilon_i^2/\sigma_{\varepsilon}^2)=1$ and $\|\varepsilon\|_{\psi_2}=O(\sigma_{\varepsilon})$. By Lemma \ref{sub_gaussian_bound}, ${\rm var}(\varepsilon^2)=O(\sigma_{\varepsilon}^4)$ and ${\rm var}(|\mathcal{I}_1|^{-1}\sum_{i\in \mathcal{I}_1}\varepsilon_i^2/\sigma_{\varepsilon}^2)=O(1/n)$. By Chebyshev's inequality, $|\mathcal{I}_1|^{-1}\sum_{i\in \mathcal{I}_1}\varepsilon_i^2/\sigma_{\varepsilon}^2=1+O_p(n^{-1/2})=1+o_p(1)$, thus $B_1=\{1+o_p(1)\}\sigma_{\varepsilon}^2$. Define $\mathcal{D}_2=\{(\bW_i)_{i\in \mathcal{I}_1\cup\mathcal{J}},(Y_i,\bW_i)_{i\in \mathcal{I}_2}\}$. Conditional on $\mathcal{D}_2$, 
	$$
	B_2\mid \mathcal{D}_2\sim\mathcal{N}[0,4\sigma_\varepsilon^2|\mathcal{I}_1|^{-2}\sum_{i\in\mathcal{I}_1}\{\bW_i^\T(\theta\bbeta^*+\balpha^*)\}^2].
	$$
	By Lemmas \ref{sub_gaussian_bound} and \ref{psi}, we have ${E}[\{\bW_i^\T(\theta\bbeta^*+\balpha^*)\}^2]=O(M^2)$. By Markov's inequality, $|\mathcal{I}_1|^{-1}\sum_{i\in\mathcal{I}_1}\{\bW_i^\T(\theta\bbeta^*+\balpha^*)\}^2=O_p(M^2)$. Hence, $B_2=O_p(M\sigma_{\varepsilon}/n^{1/2})$. Conditional on $\mathcal{D}_2$,
	$
	B_3\mid\mathcal{D}_2\sim \mathcal{N}\left\{0,4\sigma_{\varepsilon}^2|\bar{\mathcal{J}}|^{-2}\sum_{i\in\mathcal{I}_1}(\bW_i^\T\widehat{\bu}_{\rm SR})^2\right\}$.
	By Lemma \ref{OmegaW_xi_constraint_cov_shift}, $|\bar{\mathcal{J}}|^{-1}\sum_{i\in\mathcal{I}_1}(\bW_i^\T\widehat{\bu}_{\rm SR})^2\le \widehat{\bu}_{\rm SR}^\T\widehat{\bSigma}_{\bW} \widehat{\bu}_{\rm SR}=O_p(M^2)$. Therefore, we have $B_3=O_p(M\sigma_{\varepsilon}/N^{1/2})$. We have 
	$B_2+B_3=O_p(M\sigma_{\varepsilon}/n^{1/2})=\sigma_{\varepsilon}^2O_p\{(M/\sigma_{\varepsilon})/n^{1/2}\}=\sigma_{\varepsilon}^2o_p(1)$ since $M/\sigma_{\varepsilon}= O(1)$,
	and hence
	\begin{align}\label{sigma_T}
		1/\sigma_{T}^2\le 1/\sigma_{T,1}^2&\le \dfrac{C_{\bSigma}|\mathcal{I}_1|}{\{1-o_p(1)\}\sigma_{\varepsilon}^2}=O_p(n/\sigma_{\varepsilon}^2).
	\end{align}
	When $s\ll N^{1/2}/\log d$, we have $(T_{1,2}+T_{2,1}+T_{2,2,2})/\sigma_T=O_p\{(M/\sigma_{\varepsilon}+1)s\log d /N^{1/2}\}=o_p(1)$. By Lemma \ref{numerator}, $T_3/\sigma_{v,1}^2=1+o_p(1)$. Together with \eqref{normal:T11+T221} and by Slutsky's theorem,
	$$\dfrac{(\hat{\theta}_{\rm SR}-\theta)}{\sigma_{\rm SR}}=\dfrac{T_1+T_2}{\sigma_T}\dfrac{\sigma_{v,1}^2}{T_3}=\left(\dfrac{T_{1,2}+T_{2,1}+T_{2,2,2}}{\sigma_T}+\dfrac{T_{1,1}+T_{2,2,1}}{\sigma_T}\right)\dfrac{\sigma_{v,1}^2}{T_3}\xrightarrow{\rm d}\mathcal{N}(0,1),$$ 
	where
	\begin{align*}
		\sigma_{\rm SR}^2=(\sigma_{v,1}^2)^{-1}\sum_{i\in\mathcal{I}_1}\left(\dfrac{Y_i-\theta v_i}{|\mathcal{I}_1|}-\dfrac{\widehat{\bu}_{\rm SR}^\T \bW_i}{|\bar{\mathcal{J}}|}\right)^2+\dfrac{\sigma_{v,2}^2}{(\sigma_{v,1}^2)^2}|\bar{\mathcal{J}}|^{-2}\sum_{i\in\mathcal{J} }(\widehat{\bu}_{\rm SR}^\T \bW_i)^2=\sigma_{{\rm SR},1}^2+\sigma_{{\rm SR},2}^2.
	\end{align*}
	
	We next consider the variance estimation. Denote $\hat{\sigma}_{v,1}^2=|\mathcal{I}_1|^{-1}\sum_{i\in\mathcal{I}_1}\hat{v}_i^2$, $\hat{\sigma}_{v,2}^2=m^{-1}\sum_{i\in\mathcal{J}}\hat{v}_i^2$, we will show that
	\begin{align}
		\dfrac{\hat\sigma_{{\rm SR},1}^2}{\sigma_{{\rm SR},1}^2}&= 1+o_p(1)\ \ \mbox{as}\ \ n,d\to\infty,\ \ \hat\sigma_{{\rm SR},1}^2=(\hat\sigma_{v,1}^2)^{-1}\sum_{i\in\mathcal{I}_1}\left(\dfrac{Y_i-\hat{\theta}_{\rm SR} \hat{v}_i}{|\mathcal{I}_1|}-\dfrac{\widehat{\bu}_{\rm SR}^\T \bW_i}{|\bar{\mathcal{J}}|}\right)^2,\label{hat_sigma_SR1}\\
		\dfrac{\hat\sigma_{{\rm SR},2}'^2}{\sigma_{{\rm SR},2}^2}&=1+o_p(1)\ \ \mbox{as}\ \ m,d\to\infty, \ \ \hat{\sigma}_{{\rm SR},2}'^2=\dfrac{\hat\sigma_{v,2}^2}{(\hat\sigma_{v,1}^2)^2}|\bar{\mathcal{J}}|^{-2}\sum_{i\in\mathcal{J} }(\widehat{\bu}_{\rm SR}^\T \bW_i)^2.\label{hat_sigma_SR2}
	\end{align}
	For \eqref{hat_sigma_SR2}, repeating the proof of Lemma \ref{numerator}, $\hat{\sigma}_{v,1}^2/\sigma_{v,1}^2=1+o_p(1)$ and $\hat{\sigma}_{v,2}^2/\sigma_{v,2}^2=1+o_p(1)$ when $s\ll N^{1/2}/\log d$ and $n,m\to \infty$. Hence, \eqref{hat_sigma_SR2} holds by Slutsky's theorem. To see \eqref{hat_sigma_SR1}, as $\hat{\sigma}_{v,1}^2/\sigma_{v,1}^2=1+o_p(1)$, we have
	$
	\dfrac{\breve{\sigma}_{{\rm SR},1}^2}{\sigma_{{\rm SR},1}^2}=1+o_p(1)$, where $\breve{\sigma}_{{\rm SR},1}^2=(\hat\sigma_{v,1}^2)^{-1}\sum_{i\in\mathcal{I}_1}\{(Y_i-\theta v_i)/|\mathcal{I}_1|-(\widehat{\bu}_{\rm SR}^\T \bW_i)/|\bar{\mathcal{J}}|\}^{2}.
	$
	Then it suffices to show $(\hat{\sigma}_{{\rm SR},1}^2-\breve{\sigma}_{{\rm SR},1}^2)/\sigma_{{\rm SR},1}^2=o_p(1)$.
	Note that
	\begin{align*}
		&\dfrac{\hat{\sigma}_{{\rm SR},1}^2-\breve{\sigma}_{{\rm SR},1}^2}{\sigma_{{\rm SR},1}^2}
		=\dfrac{1}{\sigma_{{\rm SR},1}^2\hat\sigma_{v,1}^2}\sum_{i\in\mathcal{I}_1}\bigg\{\bigg(\dfrac{Y_i-\hat{\theta}_{\rm SR} \hat{v}_i}{|\mathcal{I}_1|}-\dfrac{\widehat{\bu}_{\rm SR}^\T \bW_i}{|\bar{\mathcal{J}}|}\bigg)^{2}-\bigg(\dfrac{Y_i-\theta v_i}{|\mathcal{I}_1|}-\dfrac{\widehat{\bu}_{\rm SR}^\T \bW_i}{|\bar{\mathcal{J}}|}\bigg)^{2}\bigg\}\\
		&\qquad=\dfrac{1}{\sigma_{{\rm SR},1}^2\hat\sigma_{v,1}^2}\sum_{i\in\mathcal{I}_1}\bigg(\dfrac{\theta v_i-\hat{\theta}_{\rm SR}\hat{v}_i}{|\mathcal{I}_1|}\bigg)^{2}+\dfrac{2}{\sigma_{{\rm SR},1}^2\hat\sigma_{v,1}^2}\sum_{i\in\mathcal{I}_1}\bigg(\dfrac{Y_i-\theta v_i}{|\mathcal{I}_1|}-\dfrac{\widehat{\bu}_{\rm SR}^\T \bW_i}{|\bar{\mathcal{J}}|}\bigg)\bigg(\dfrac{\theta v_i-\hat{\theta}_{\rm SR}\hat{v}_i}{|\mathcal{I}_1|}\bigg)\\
		&\qquad=B_4+B_5.
	\end{align*}
	For $B_4$,
	\begin{align*}
		B_4&=\dfrac{1}{\sigma_{{\rm SR},1}^2\hat\sigma_{v,1}^2}|\mathcal{I}_1|^{-2}\sum_{i\in \mathcal{I}_1}(\theta v_i-\theta \hat{v}_i+\theta\hat{v}_i-\hat{\theta}_{\rm SR}\hat{v}_i)^2\\
		&\leq\dfrac{2}{|\mathcal{I}_1|\sigma_{{\rm SR},1}^2\hat\sigma_{v,1}^2}\bigg\{|\mathcal{I}_1|^{-1}\sum_{i\in \mathcal{I}_1}(v_i-\hat{v}_i)^2\theta^2+|\mathcal{I}_1|^{-1}\sum_{i\in \mathcal{I}_1}\hat{v}_i^2(\theta-\hat{\theta}_{\rm SR})^2\bigg\}=B_{4,1}+B_{4,2}.
	\end{align*}
	We next show that $B_{4,1}$ and $B_{4,2}$ are both $o_p(1)$ terms. 
	By Lemma \ref{numerator}, \eqref{sigma_T}, and the fact that $1/C_{\bSigma}\le\sigma_{v,1}^2\le C_{\bSigma}$,
	\begin{equation}\label{B1}
		\dfrac{1}{|\mathcal{I}_1|\sigma_{{\rm SR},1}^2\hat{\sigma}_v^2}=\dfrac{(\sigma_{v,1}^2)^2}{|\mathcal{I}_1|\sigma_{T,1}^2\hat{\sigma}_v^2}=O_p\left(\dfrac{1}{\sigma_{\varepsilon}^2}\right).
	\end{equation} 
	For $B_{4,1}$, by $M/\sigma_{\varepsilon}=O(1)$ and \eqref{B1},
	$\theta^2/(|\mathcal{I}_1|\sigma_{{\rm SR},1}^2\hat{\sigma}_{v,1}^2)=O_p(M^2/\sigma_{\varepsilon}^2)=O_p(1)$. 
	Furthermore, by Lemma \ref{B_6}, $|\mathcal{I}_1|^{-1}\sum_{i\in \mathcal{I}_1}(v_i-\hat{v}_i)^2=o_p(1)$, thus $B_{4,1}=o_p(1)$. For $B_{4,2}$, by Theorem 2.3 and \eqref{B1},
	\begin{equation}\label{B12}
		\dfrac{(\theta-\hat{\theta}_{\rm SR})^2}{|\mathcal{I}_1|\sigma_{{\rm SR},1}^2\hat{\sigma}_{v,1}^2}=O_p\left\{\left(\dfrac{M}{\sigma_{\varepsilon}}+1\right)^2\left(\dfrac{1}{n^{1/2}}+\dfrac{s\log d }{(Nn)^{1/2}}\right)^2\right\}=o_p(1).
	\end{equation}
	Combining \eqref{B12} and $|\mathcal{I}_1|^{-1}\sum_{i\in\mathcal{I}_1}\hat{v}_i^2\le 2|\mathcal{I}_1|^{-1}\sum_{i\in\mathcal{I}_1}v_i^2+2|\mathcal{I}_1|^{-1}\sum_{i\in\mathcal{I}_1}(\hat{v}_i-v_i)^2=O_p(1)$, $B_{4,2}=o_p(1)$. For $B_2$,
	\begin{align*}
		B_5&\le 2\left\{\dfrac{1}{\sigma_{{\rm SR},1}^2\hat\sigma_{v,1}^2}\sum_{i\in\mathcal{I}_1}\bigg(\dfrac{Y_i-\theta v_i}{|\mathcal{I}_1|}-\dfrac{\widehat{\bu}_{\rm SR}^\T \bW_i}{|\bar{\mathcal{J}}|}\bigg)^2\right\}^{1/2}\left\{\dfrac{1}{\sigma_{{\rm SR},1}^2\hat\sigma_{v,1}^2}\sum_{i\in\mathcal{I}_1}\bigg(\dfrac{\theta v_i-\hat{\theta}_{\rm SR}\hat{v}_i}{|\mathcal{I}_1|}\bigg)^2\right\}^{1/2}\\
		&\le 2\{B_1\breve{\sigma}_{{\rm SR},1}^2/\sigma_{{\rm SR},1}^2\}^{1/2}=o_p(1).
	\end{align*}
	Therefore, we conclude that \eqref{hat_sigma_SR1} holds. 
	
	(a) When there is no covariate shift, notice that at this point, $\sigma_{v,1}^2=\sigma_{v,2}^2$. Let
	$\hat\sigma_{{\rm SR},2}^2= (\hat\sigma_{v,1}^2)^{-1}|\bar{\mathcal{J}}|^{-2}\sum_{i\in\mathcal{J} }(\widehat{\bu}_{\rm SR}^\T \bW_i)^2$. Since $\hat{\sigma}_{v,1}^2/\sigma_{v,1}^2=1+o_p(1)$, where $\sigma_{v,1}^2=(\sigma_{v,1}^2)^2/\sigma_{v,2}^2$, we have $\hat\sigma_{{\rm SR},2}^2/\sigma_{{\rm SR},2}^2=1+o_p(1)$. Together with \eqref{hat_sigma_SR1}, $\hat\sigma_{\rm SR}^2/\sigma_{\rm SR}^2=1+o_p(1)$, where $\hat\sigma_{\rm SR}^2=\hat\sigma_{{\rm SR},1}^2+\hat\sigma_{{\rm SR},2}^2$. By Slutsky's theorem, $(\hat{\theta}_{\rm SR}-\theta)/\hat\sigma_{\rm SR}\xrightarrow{\rm d}\mathcal{N}(0,1)$.
	
	(b) When there exists potential covariate shift, let $\hat\sigma_{\rm SR}'^2=\hat\sigma_{{\rm SR},1}^2+\hat\sigma_{{\rm SR},2}'^2$. Combining \eqref{hat_sigma_SR1} and \eqref{hat_sigma_SR2}, $\hat\sigma_{\rm SR}'^2/\sigma_{\rm SR}^2=1+o_p(1)$ as $n,m,d\to\infty$ since it lies between $\hat\sigma_{{\rm SR},1}^2/\sigma_{{\rm SR},1}^2$ and $\hat\sigma_{{\rm SR},2}^2/\sigma_{{\rm SR},2}^2$. By Slutsky's theorem, $(\hat{\theta}_{\rm SR}-\theta)/\hat\sigma_{\rm SR}'\xrightarrow{\rm d}\mathcal{N}(0,1)$.
	\hfill\BlackBox
	\\ \hspace*{\fill} \\
	{\bf Proof of Theorem \ref{thm:DF_consist_rate_cs} and the consistency results of Theorem 3}.
	In the following, we provide the proof for Theorem \ref{thm:DF_consist_rate_cs}. The consistency results outlined in Theorem 3 follow directly from Theorem \ref{thm:DF_consist_rate_cs}. As shown in the proof of Theorem 2.5 of \cite{bellec2022biasing}, the supervised nuisance estimator $\widehat{\bgamma}_{\rm DFA}$ satisfies
	\begin{equation}
		\|\widehat{\bgamma}_{\rm DFA}-\bgamma^*\|_2=O_p\left\{\sigma_{\varepsilon}(k\log d /n)^{1/2}\right\},\ \|\widehat{\bgamma}_{\rm DFA}-\bgamma^*\|_1=O_p\left\{\sigma_{\varepsilon}k(\log d /n)^{1/2}\right\}.\label{gamma_DF}
	\end{equation}
	By Lemma \ref{df_delta},
	\begin{equation}
		\|\widehat{\bbeta}_{\rm DFA}-\bbeta^*\|_2=O_p\left\{(s\log d /N)^{1/2}\right\},\ \|\widehat{\bbeta}_{\rm DFA}-\bbeta^*\|_1=O_p\left\{s(\log d /N)^{1/2}\right\}.\label{beta_DF}
	\end{equation}
	
	\textbf{Case (a)}. Let $\widehat{\gamma}_{\rm DFA}=(\be_1^\T\widehat{\bgamma}_{\rm DFA},\widehat{\balpha}_{\rm DFA}^\T)^\T$. We first have
	\begin{equation}\label{casea}
		\begin{split}
			&\!\!\!n^{1/2}(1-\hat{q}/n)(\hat{\theta}_{\rm DFA}-\theta)\\
			&\!\!\!\qquad=n^{1/2}\bigg[(1-\hat{q}/n)(\be_1^\T\widehat{\gamma}_{\rm DFA}-\theta)+\dfrac{n^{-1}\sum_{i\in\mathcal{I}} \hat{v}_i\{\varepsilon_i-\bX_i^\T(\widehat{\bgamma}_{\rm DFA}-\bgamma^*)\}}{n^{-1}\sum_{i\in\mathcal{I}} \hat{v}_iZ_i}\bigg]\\
			&\!\!\!\qquad=n^{1/2}\bigg\{-(\hat{q}/n)(\be_1^\T\widehat{\bgamma}_{\rm DFA}-\theta)+\dfrac{n^{-1}\sum_{i\in\mathcal{I}}\hat{v}_i\bW_i^\T(\balpha^*-\widehat{\balpha}_{\rm DFA})}{n^{-1}\sum_{i\in\mathcal{I}}\hat{v}_iZ_i}+\dfrac{n^{-1}\sum_{i\in\mathcal{I}}\hat{v}_i\varepsilon_i}{n^{-1}\sum_{i\in\mathcal{I}}\hat{v}_iZ_i}\bigg\}\\
			&\!\!\!\qquad=F_1+F_2+F_3.
		\end{split}
	\end{equation}
	Repeating the proof of Lemma \ref{psi}, when 
	$1/C_{\bSigma}\le \lambda_{\rm min}(\bSigma)\le\lambda_{\rm max}(\bSigma)\le C_{\bSigma}$,
	\begin{equation}\label{C_sigma}
		1/C_{\bSigma}\le \sigma_{v,1}^2\le C_{\bSigma}\ \ {\rm and}\ \
		\|\bbeta^*\|_2\le C_{\bSigma}^2.
	\end{equation} 
	By Lemma \ref{vizi}, when $s\ll(Nn)^{1/2}/\log d$, $1/(n^{-1}\sum_{i\in \mathcal{I}}\hat{v}_iZ_i)=O_p(1)$. For $F_{1}$, by Lemma 8.1 of \cite{bellec2022biasing}, $\hat{q}=O_p(k)$. Additionally, $|\be_1^\T\widehat{\bgamma}_{\rm DFA}-\theta|\le \|\widehat{\bgamma}_{\rm DFA}-\bgamma^*\|_2=O_p\{\sigma_{\varepsilon}(k\log d /n)^{1/2}\}$. Since $k\ll n/\log d$, 
	$$
	F_1=O_p\left\{\dfrac{\sigma_{\varepsilon}k^{3/2}(\log d)^{1/2} }{n}\right\}=O_p\left\{\dfrac{\sigma_{\varepsilon}k\log d }{n^{1/2}}\left(\dfrac{k}{n\log d }\right)^{1/2}\right\}=O_p\left(\dfrac{\sigma_{\varepsilon}k\log d }{n^{1/2}}\right).
	$$ 
	For $F_{2}$, we first have
	\begin{align*}
		n^{1/2}\cdot n^{-1}\sum_{i\in \mathcal{I}}\hat{v}_i\bW_i^\T(\balpha^*-\widehat{\balpha}_{\rm DFA})&\le n^{1/2}\, \|\widehat{\balpha}_{\rm DFA}-\balpha^*\|_1\Big\|n^{-1}\sum_{i\in\mathcal{I}} \bW_i\hat{v}_i\Big\|_\infty\\
		&\le n^{1/2}\,\|\widehat{\bgamma}_{\rm DFA}-\bgamma^*\|_1\Big\|n^{-1}\sum_{i\in\mathcal{I}} \bW_i\hat{v}_i\Big\|_\infty=O_p\left(\dfrac{\sigma_{\varepsilon}k\log d }{n^{1/2}}\right)
	\end{align*}
	since \eqref{gamma_DF} holds and by the construction of $\widehat{\bbeta}_{\rm DFA}$. Together with $1/(n^{-1}\sum_{i\in \mathcal{I}}\hat{v}_iZ_i)=O_p(1)$, $F_{2}=O_p(\sigma_{\varepsilon}k\log d /n^{1/2})$. For $F_{3}$, let $\mathcal{D}_4=(Z_i,\bW_i)_{i\in \mathcal{I}\cup\mathcal{J}}$. Conditional on $\mathcal{D}_4$, 
	\begin{equation}\label{A3}
		F_3\mid \mathcal{D}_4\sim \mathcal{N}\left\{0,\dfrac{n^{-1}\sum_{i\in \mathcal{I}}\hat{v}_i^2\sigma_\varepsilon^2}{(n^{-1}\sum_{i\in \mathcal{I}}\hat{v}_iZ_i)^2}\right\}.
	\end{equation}
	By Lemma \ref{vizi}, when $s\ll(Nn)^{1/2}/\log d$, the conditional variance of $F_3$
	\begin{equation}\label{sigma_DFA}
		\dfrac{n^{-1}\sum_{i\in \mathcal{I}}\hat{v}_i^2\sigma_\varepsilon^2}{(n^{-1}\sum_{i\in \mathcal{I}}\hat{v}_iZ_i)^2}\xrightarrow{\rm p} \dfrac{\sigma_{\varepsilon}^2}{\sigma_{v,1}^2},
	\end{equation}
	which is independent of $\mathcal{D}_4$. Together with \eqref{C_sigma}, $F_3=O_p(\sigma_\varepsilon)$. As $k\ll n/\log d$,  $\hat{q}/n=O_p(k/n)=o_p(1)$. Combining the upper bounds for $F_1$, $F_2$ and $F_3$, $
	\hat{\theta}_{\rm DFA}-\theta=O_p\{\sigma_\varepsilon n^{-1/2}+\sigma_{\varepsilon}k\log d /n\}$.
	
	\textbf{Case (b)}. Define the oracle degree-of-freedom adjusted estimator $\hat{\theta}_{\rm ODF}$ as follows
	\begin{equation}\label{breve_beta_1}
		\hat{\theta}_{\rm ODF}=\be_1^\T\widehat{\bgamma}_{\rm DFA}+\dfrac{n^{-1}\sum_{i\in\mathcal{I}} v_i(Y_i-\bX_i^\T \widehat{\gamma}_{\rm DFA})}{(1-\hat{q}/n)n^{-1}\sum_{i\in\mathcal{I}} v_i^2}.
	\end{equation}
	By Theorem 2.1 of \cite{bellec2022biasing}, 
	\begin{align}\label{odf}
		n^{1/2}(1-\hat{q}/n)(\hat{\theta}_{\rm ODF}-\theta)\xrightarrow{\rm d}\mathcal{N}(0,\sigma_\varepsilon^2/\sigma_{v,1}^2)
	\end{align}
	We next control the term $n^{1/2}(1-\hat{q}/n)(\hat{\theta}_{\rm DFA}-\hat{\theta}_{\rm ODF})$:
	\begin{equation}
		\begin{split}\label{caseb}
			&\!n^{1/2}(1-\hat{q}/n)(\hat{\theta}_{\rm DFA}-\hat{\theta}_{\rm ODF})=n^{1/2}(1-\hat{q}/n)(\hat{\theta}_{\rm DFA}-\theta)-n^{1/2}(1-\hat{q}/n)(\hat{\theta}_{\rm ODF}-\theta)\\
			&\!\quad=n^{1/2}\bigg(n^{-1}\sum_{i\in\mathcal{I}} \bigg(\dfrac{\hat{v}_i}{n^{-1}\sum_{i\in\mathcal{I}} \hat{v}_iZ_i}-\dfrac{v_i}{n^{-1}\sum_{i\in\mathcal{I}} v_i^2}\bigg)(Y_i-\bX_i^\T\widehat{\bgamma}_{\rm DFA}) \bigg)\\
			&\!\quad=\dfrac{n^{1/2}\cdot n^{-1}\sum_{i\in\mathcal{I}}(\hat{v}_i-v_i)(Y_i-\bX_i^\T\widehat{\bgamma}_{\rm DFA})}{n^{-1}\sum_{i\in\mathcal{I}}\hat{v}_iZ_i}\\
			&\!\quad\quad + n^{1/2}\bigg(\dfrac{1}{n^{-1}\sum_{i\in\mathcal{I}}\hat{v}_iZ_i}-\dfrac{1}{n^{-1}\sum_{i\in\mathcal{I}}v_i^2}\bigg)\bigg\{n^{-1}\sum_{i\in\mathcal{I}}v_i(Y_i-\bX_i^\T\widehat{\bgamma}_{\rm DFA})\bigg\}=F_{4}+F_{5}.
		\end{split}
	\end{equation}
	For $F_4$, we have
	\begin{align*}
		F_{4}&\le n^{1/2}\cdot\|\widehat{\bbeta}_{\rm DFA}-\bbeta^*\|_1\Big\|n^{-1}\sum_{i\in\mathcal{I}}\bW_i(Y_i-\bX_i^\T\widehat{\bgamma}_{\rm DFA})\Big\|_\infty\\
		&\le n^{1/2}\cdot\|\widehat{\bbeta}_{\rm DFA}-\bbeta^*\|_1 \Big\|n^{-1}\sum_{i\in\mathcal{I}}\bX_i(Y_i-\bX_i^\T\widehat{\bgamma}_{\rm DFA})\Big\|_\infty\\
		&\le n^{1/2}(\lambda_{\gamma}/2)\|\widehat{\bbeta}_{\rm DFA}-\bbeta^*\|_1=O_p\left(\dfrac{\sigma_{\varepsilon}s\log d }{N^{1/2}}\right),
	\end{align*}
	where the third inequality is by the KKT condition of $\widehat{\bgamma}_{\rm DFA}$, the last equality is by \eqref{beta_DF}. Denote $F_{5,1}=1/\{(n^{-1}\sum_{i\in\mathcal{I}}\hat{v}_iZ_i)(n^{-1}\sum_{i\in\mathcal{I}}v_i^2)\}$ and
	$\widehat{\bSigma}_{\bW} =n^{-1}\sum_{i\in \mathcal{I}}\bW_i\bW_i^\T$, then
	\begin{align*}
		F_5&=n^{1/2}F_{5,1}\Big\{n^{-1}\sum_{i\in\mathcal{I}}v_i(Y_i-\bX_i^\T\widehat{\bgamma}_{\rm DFA})\Big\}
		\Big\{n^{-1}\sum_{i\in\mathcal{I}}v_i(v_i-\hat{v}_i)+n^{-1}\sum_{i\in\mathcal{I}}\hat{v}_i(v_i-Z_i)\Big\}\\
		&=n^{1/2}F_{5,1}\Big\{n^{-1}\sum_{i\in\mathcal{I}}v_i(Y_i-\bX_i^\T\widehat{\bgamma}_{\rm DFA})\Big\}
		\Big\{(\widehat{\bbeta}_{\rm DFA}-\bbeta^*)^\T\bSigma_{\bW}\bbeta^*\Big\}\\
		&\quad+n^{1/2}F_{5,1}\Big\{n^{-1}\sum_{i\in\mathcal{I}}v_i(Y_i-\bX_i^\T\widehat{\bgamma}_{\rm DFA})\Big\}\\
		&\quad\quad
		\cdot\Big\{n^{-1}\sum_{i\in\mathcal{I}}v_i\bW_i^\T(\widehat{\bbeta}_{\rm DFA}-\bbeta^*)+(\widehat{\bbeta}_{\rm DFA}-\bbeta^*)^\T(\widehat{\bSigma}_{\bW} -\bSigma_{\bW})\bbeta^*-n^{-1}\sum_{i\in\mathcal{I}}v_i\bW_i^\T\bbeta^*\Big\}\\
		&=F_{5,2}+F_{5,3}.
	\end{align*}
	When $s\ll(Nn)^{1/2}/\log d$, by Lemma \ref{vizi} and \eqref{C_sigma}, 
	\begin{equation}\label{B21}
		F_{5,1}=O_p(1).
	\end{equation}
	We now consider the term $F_{5,2}$. Denote $S_0={\rm supp}(\bbeta^*)$, by \eqref{C_sigma}, $\|(1,-{\bbeta^*}^\T)^\T\|_1=1+\|\bbeta^*\|_1= 1+\|\bbeta^*_{S_0}\|_1\le 1+s^{1/2}\|\bbeta^*\|_2\le 1+s^{1/2}C_{\bSigma}^2\asymp s^{1/2}$. Together with the KKT condition of $\widehat{\bgamma}_{\rm DFA}$,
	\begin{equation}\label{B221}
		\begin{split}
			&n^{-1}\sum_{i\in\mathcal{I}}v_i(Y_i-\bX_i^\T\widehat{\bgamma}_{\rm DFA})
			=n^{-1}\sum_{i\in\mathcal{I}}(1,-\bbeta^*)^\T \bX_i(Y_i-\bX_i^\T\widehat{\bgamma}_{\rm DFA})\\
			&\qquad\le\|(1,{-\bbeta^*}^\T)^\T\|_1 \Big\|n^{-1}\sum_{i\in\mathcal{I}}\bX_i(Y_i-\bX_i^\T\widehat{\bgamma}_{\rm DFA})\Big\|_\infty=O_p\left\{\sigma_{\varepsilon}(s\log d /n)^{1/2}\right\}.
		\end{split}
	\end{equation}
	By \eqref{C_sigma}, $\|\bSigma_{\bW}\bbeta^*\|_2\le C_{\bSigma}\|\bbeta^*\|_2=O(1)$. Together with \eqref{beta_DF},
	\begin{equation}\label{B222}
		(\widehat{\bbeta}_{\rm DFA}-\bbeta^*)^\T\bSigma_{\bW}\bbeta^*\le \|\widehat{\bbeta}_{\rm DFA}-\bbeta^*\|_2\|\bSigma_{\bW}\bbeta^*\|_2=O_p\left\{(s\log d /N)^{1/2}\right\}.
	\end{equation}
	Combining \eqref{B21}, \eqref{B221}, and \eqref{B222},
	\begin{equation}\label{B_2}
		F_{5,2}=O_p\left(\dfrac{\sigma_{\varepsilon}s\log d }{N^{1/2}}\right).
	\end{equation} 
	For $F_{5,3}$, we have ${E}(v_i\varepsilon_i)=0$ and ${\rm var}(n^{-1}\sum_{i\in\mathcal{I}}v_i\varepsilon_i)=n^{-1}\sigma_v^2\sigma_{\varepsilon}^2=O(\sigma_\varepsilon^2/n)$. By Chebyshev's inequality, $|n^{-1}\sum_{i\in\mathcal{I}}v_i\varepsilon_i|=O_p(\sigma_\varepsilon/n^{1/2})$. By Lemma \ref{vizi}, $|n^{-1}\sum_{i\in\mathcal{I}}v_iZ_i|=|\sigma_{v,1}^2+o_p(1)|=O_p(1)$. Therefore,
	\begin{equation}\label{B231}
		\begin{split}
			&n^{-1}\sum_{i\in\mathcal{I}}v_i(Y_i-\bX_i^\T\widehat{\bgamma}_{\rm DFA})\\
			&\qquad=n^{-1}\sum_{i\in\mathcal{I}} v_i\varepsilon_i+n^{-1}\sum_{i\in\mathcal{I}}v_iZ_i(\theta-\be_1^\T\widehat{\bgamma}_{\rm DFA})+n^{-1}\sum_{i\in\mathcal{I}} v_i\bW_i^\T(\balpha^*-\widehat{\balpha}_{\rm DFA})\\
			&\qquad\le \Big|n^{-1}\sum_{i\in \mathcal{I}} v_i\varepsilon_i\Big|+|\be_1^\T\widehat{\bgamma}_{\rm DFA}-\theta| \Big|n^{-1}\sum_{i\in \mathcal{I}} v_iZ_i\Big|+\|\widehat{\balpha}_{\rm DFA}-\balpha^*\|_1\Big\|n^{-1}\sum_{i\in \mathcal{I}} \bW_iv_i\Big\|_\infty\\
			&\qquad\le \Big|n^{-1}\sum_{i\in \mathcal{I}} v_i\varepsilon_i\Big|+\|\widehat{\bgamma}_{\rm DFA}-\bgamma^*\|_2 \Big|n^{-1}\sum_{i\in \mathcal{I}} v_iZ_i\Big|+\|\widehat{\bgamma}_{\rm DFA}-\bgamma^*\|_1\Big\|n^{-1}\sum_{i\in \mathcal{I}} \bW_iv_i\Big\|_\infty\\
			&\qquad=O_p\left\{\dfrac{\sigma_{\varepsilon}}{n^{1/2}}+\sigma_{\varepsilon}\left(\dfrac{k\log d }{n}\right)^{1/2}+\dfrac{\sigma_{\varepsilon}k\log d }{n}\right\}.
		\end{split}
	\end{equation}
	Repeating the proof of Lemma \ref{pi_constraint}, $\|n^{-1}\sum_{i\in\mathcal{I}}\bW_iv_i\|_\infty=O_p\{(\log d /n)^{1/2}\}$. Together with \eqref{beta_DF},
	\begin{equation}\label{B2321}
		n^{-1}\sum_{i\in\mathcal{I}}v_i\bW_i^\T(\widehat{\bbeta}_{\rm DFA}-\bbeta^*)\le \|\widehat{\bbeta}_{\rm DFA}-\bbeta^*\|_1\Big\|n^{-1}\sum_{i\in\mathcal{I}}v_i\bW_i\Big\|_\infty=O_p\left\{\dfrac{s\log d }{(Nn)^{1/2}}\right\}.
	\end{equation}
	By \eqref{beta_DF} and Lemma \ref{Sigma_W_inf}, 
	\begin{equation}\label{B2322}
		(\widehat{\bbeta}_{\rm DFA}-\bbeta^*)^\T(\widehat{\bSigma}_{\bW} -\bSigma_{\bW})\bbeta^*\le \|\widehat{\bbeta}_{\rm DFA}-\bbeta^*\|_1\|(\widehat{\bSigma}_{\bW} -\bSigma_{\bW})\bbeta^*\|_\infty=O_p\left\{\dfrac{s\log d }{(Nn)^{1/2}}\right\}.
	\end{equation}
	By Lemma \ref{best_linear_slope}, ${E}(v_i\bW_i^\T\bbeta^*)=0$. Repeating the proof of Lemma \ref{psi}, $v_i\bW_i^\T\bbeta^*$ is sub-exponential with $\|v_i\bW_i^\T\bbeta^*\|_{\psi_1}=O(1)$. By Lemma \ref{Bernstein_inequality}, ${\rm var}(n^{-1}\sum_{i\in\mathcal{I}}v_i\bW_i^\T\bbeta^*)=O(1/n)$. By Chebyshev's inequality,
	\begin{align}
		n^{-1}\sum_{i\in\mathcal{I}}v_i\bW_i^\T\bbeta^*&=O_p\left(n^{-1/2}\right).\label{B2323}
	\end{align}
	Combing \eqref{B21}, \eqref{B231}-\eqref{B2323}, when $k\ll n/\log d$, 
	\begin{equation}\label{k22}
		\begin{split}
			F_{5,3}&=O_p\left[n^{1/2}\left\{\dfrac{\sigma_{\varepsilon}}{n^{1/2}}+\sigma_{\varepsilon}\left(\dfrac{k\log d }{n}\right)^{1/2}+\dfrac{\sigma_{\varepsilon}k\log d }{n}\right\}\Bigg\{\dfrac{1}{n^{1/2}}+\dfrac{s\log d }{(Nn)^{1/2}}\Bigg\}\right]\\
			&=o_p\left(\sigma_{\varepsilon}+\dfrac{\sigma_{\varepsilon}s\log d}{N^{1/2}}\right).
		\end{split}
	\end{equation}
	Since $k\ll n/\log d$, $\hat{q}/n=O_p(k/n)=o_p(1)$, combining \eqref{odf}, \eqref{B_2} and \eqref{k22}, $
	\hat\theta_{\rm DFA}-\theta=O_p\{\sigma_{\varepsilon}/n^{1/2}+\sigma_{\varepsilon}s\log d /(Nn)^{1/2}\}$,
	which gives the proof of Theorem \ref{thm:DF_consist_rate_cs} and the first part of Theorem 3. In the next we will prove the second part of Theorem 3.
	\hfill\BlackBox
	\\ \hspace*{\fill} \\
	{\bf Proof of Theorem \ref{thm:DF_asy_normal_cs} and the asymptotic normal results in Theorem 3}.
	Now we prove Theorem \ref{thm:DF_asy_normal_cs}, and the asymptotic normality result of Theorem 3 follows by Theorem \ref{thm:DF_asy_normal_cs}. When $k\ll n^{1/2}/\log d$ and $s\ll(Nn)^{1/2}/\log d$. Consider the decomposition $n^{1/2}(1-\hat{q}/n)(\hat{\theta}_{\rm DFA}-\theta)=F_1+F_2+F_3$, where $F_1$, $F_2$, and $F_3$ are defined in \eqref{casea}. By \eqref{A3} and \eqref{sigma_DFA}, let $\sigma=\sigma_\varepsilon/\sigma_{v,1}$, we have $F_3/\sigma\xrightarrow{\rm d} \mathcal{N}(0,1)$. Moreover, as shown in the proof Theorem 3.1, $F_1+F_2=O_p\{\sigma_{\varepsilon}k\log(d)/n^{1/2}\}$, $(F_1+F_2)/\sigma=O_p(k\log(d)/n^{1/2})=o_p(1)$. Hence, $n^{1/2}(1-q/n)(\hat{\theta}_{\rm DFA}-\theta)/\sigma\xrightarrow{\rm d} \mathcal{N}(0,1)$,
	where $\sigma  = \sigma_\varepsilon/\sigma_{v,1}$. When $k\ll n/\log d$ and $s\ll N^{1/2}/\log d$. As shown in the proof Theorem 3.1, $n^{1/2}(1-\hat{q}/n)(\hat{\theta}_{\rm DFA}-\hat{\theta}_{\rm ODF})/\sigma=O_p\{s\log d/N^{1/2}\}+o_p(1)=o_p(1)$. Moreover, by \eqref{odf}, $n^{1/2}(1-q/n)(\hat{\theta}_{\rm DFA}-\theta)/\sigma\xrightarrow{\rm d} \mathcal{N}(0,1)$,
	where $\sigma =\sigma_\varepsilon/\sigma_{v,1}$. 
	
	Now we consider the variance estimation. We first show that 
	\begin{equation}\label{hatDFA1}
		n^{-1}\sum_{i\in \mathcal{I}}(Y_i-\bX_i^\T\widehat{\bgamma}_{\rm DFA})^2/\sigma_\varepsilon^2=1+o_p(1).
	\end{equation}
	We have
	\begin{align*}
		n^{-1}\sum_{i\in \mathcal{I}}(Y_i-\bX_i^\T\widehat{\bgamma}_{\rm DFA})^2/\sigma_\varepsilon^2
		&\le n^{-1}\sum_{i\in \mathcal{I}}\varepsilon_i^2/\sigma_\varepsilon^2+n^{-1}\sum_{i\in \mathcal{I}}(\bX_i^\T(\bgamma^*-\widehat{\bgamma}_{\rm DFA}))^2/\sigma_\varepsilon^2\\
		&\quad+2\left(n^{-1}\sum_{i\in \mathcal{I}}\varepsilon_i^2/\sigma_\varepsilon^2\right)^{1/2}\left[n^{-1}\sum_{i\in \mathcal{I}}\{\bX_i^\T(\bgamma^*-\widehat{\bgamma}_{\rm DFA})\}^2/\sigma_\varepsilon^2\right]^{1/2}\\
		&=F_6+F_7+F_8.
	\end{align*}
	For $F_6$, we have ${E}(\varepsilon_i^2/\sigma_\varepsilon^2)=1$. Since $\|\varepsilon\|_{\psi_2}=O(\sigma_{\varepsilon})$, by Lemma \ref{sub_gaussian_bound}, ${\rm var}(\epsilon_i^2)=O(\sigma_{\varepsilon}^4)$. Moreover, ${\rm var}(n^{-1}\sum_{i\in \mathcal{I}}\varepsilon_i^2/\sigma_\varepsilon^2)=n^{-1}{\rm var}(\varepsilon_i^2)/\sigma_{\varepsilon}^4=O(1/n)$. By Chebyshev's inequality, $F_6=1+O_p(n^{-1/2})=1+o_p(1)$. Then it suffices to show that both $F_7$ and $F_8$ are $o_p(1)$ terms. By the KKT condition of $\widehat{\bgamma}_{\rm DFA}$,
	\begin{align}
		\Big\|n^{-1}\sum_{i\in\mathcal{I}}\bX_i(Y_i-\bX_i^\T\widehat{\bgamma}_{\rm DFA})\Big\|_\infty\le \lambda_{\gamma}/2=O_p\left\{\sigma_\varepsilon(\log d /n)^{1/2}\right\}.\label{KKT1}
	\end{align}
	By Lemma \ref{xeps},
	\begin{align}
		\Big\|n^{-1}\sum_{i\in\mathcal{I}}\bX_i(Y_i-\bX_i^\T\bgamma^*)\Big\|_\infty=\Big\|n^{-1}\sum_{i\in\mathcal{I}}\bX_i\varepsilon_i\Big\|_\infty=O_p\left\{\sigma_\varepsilon(\log d /n)^{1/2}\right\}.\label{KKT2}
	\end{align}
	Taking the difference between \eqref{KKT1} and \eqref{KKT2}, we have $\|n^{-1}\sum_{i\in \mathcal{I}}\bX_i\bX_i^\T(\widehat{\bgamma}_{\rm DFA}-\bgamma^*)\|_\infty=O_p\{\sigma_\varepsilon(\log d /n)^{1/2}\}$. Together with \eqref{gamma_DF}, 
	\begin{align*}
		F_7\le \dfrac{1}{\sigma_\varepsilon^2}\|\widehat{\bgamma}_{\rm DFA}-\bgamma^*\|_1\bigg\|n^{-1}\sum_{i\in \mathcal{I}}\bX_i\bX_i^\T(\widehat{\bgamma}_{\rm DFA}-\bgamma^*)\bigg\|_\infty=O_p\left(\dfrac{k\log d }{n}\right)=o_p(1).
	\end{align*} 
	Moreover, $F_8\le 2(F_6F_7)^{1/2}=o_p(1)$. By Lemma \ref{B_6} and $\sigma_v^2\ge 1/C_{\bSigma}$,
	\begin{equation}\label{hatDFA2}
		n^{-1}\sum_{i\in \mathcal{I}}(Z_i-\bW_i^\T\widehat{\bbeta}_{\rm DFA})^2/\sigma_{v,1}^2=1+o_p(1).
	\end{equation}
	Let $\hat\sigma_{\rm DFA}^2={\sum_{i\in \mathcal{I}}(Y_i-\bX_i^\T\widehat{\gamma}_{\rm DFA})^2}/{\sum_{i\in \mathcal{I}}(Z_i-\bW_i^\T\widehat{\bbeta}_{\rm DFA})^2}$. Combining \eqref{hatDFA1} and $\eqref{hatDFA2}$, we have $\hat\sigma_{\rm DFA}^2/\sigma^2=1+o_p(1)$. 
	\hfill\BlackBox
	\\ \hspace*{\fill} \\
	{\bf Proof of Theorem \ref{thm:SSR_consist_rate}}.
	We first show that with high probability, $\bOmega_{\bW}^{\rm (mix)}\bxi$ satisfies the constraint $\max_{i\in \mathcal{J}_2} | \bW_i^\T u|\le \hat{V}_Y^{1/2}|\mathcal{J}_2|^{q}$. Note that $|\mathcal J_1|\asymp|\mathcal J_2|\asymp N$. Since $\bX_i$ has mean zero, $E(\bW_i^\T\bOmega_{\bW}^{\rm (mix)}\bxi)=0$. By Lemma \ref{psi}, $\|\bW_i^\T\bOmega_{\bW}^{\rm (mix)}\bxi\|_{\psi_2}\le \sigma_u MC_{\bSigma}^2(C_{\bSigma}^2+1)$ for $i\in \mathcal{J}_2$. By Lemma \ref{sub_gaussian_bound} and Boole's inequality, with $t={\rm var}(Y_i)^{1/2}|\mathcal{J}_2|^q/2$ and some constant $c>0$,
	\begin{equation}\label{sigma_Y}
		\begin{split}
			{P}\left\{\max_{i\in \mathcal{J}_2}|\bW_{i}^\T\bOmega_{\bW}^{\rm (mix)}\bxi|\le {\rm var}(Y_i)^{1/2}|\mathcal{J}_2|^q/2\right\}
			&\ge1- 2|\mathcal{J}_2|\exp\left[\dfrac{-{\rm var}(Y_i)|\mathcal{J}_2|^{2q}}{32\{\sigma_u MC_{\bSigma}^2(C_{\bSigma}^2+1)\}^2}\right]\\
			&\ge 1-|\mathcal{J}_2|\exp\left\{\dfrac{-c(M^2+\sigma_{\varepsilon}^2)|\mathcal{J}_2|^{2q}}{M^2}\right\}\\[4pt]
			& \ge 1-|\mathcal{J}_2|\exp(-c|\mathcal{J}_2|^{2q})=1-o(1)
		\end{split}
	\end{equation}
	as $|\mathcal{J}_2|\to \infty$,
	where the second inequality holds by ${\rm var}(Y_i)={\bgamma^*}^\T\bSigma\bgamma^*+\sigma_{\varepsilon}^2$ and $ {\bgamma^*}^\T\bSigma\bgamma^*\asymp M^2$ under Assumption 3. We next prove that $\hat{V}_Y/{\rm var}(Y_i)=1+o_p(1)$, notice that
	\begin{align*}
		&\dfrac{\hat{V}_Y}{{\rm var}(Y_i)}=\dfrac{|\mathcal{I}_3|^{-1}\sum_{i\in\mathcal{I}_3}\{Y_i-{E}(Y_i)+{E}(Y_i)-\bar{Y}\}^2}{{\rm var}(Y_i)}\\
		&\quad=|\mathcal{I}_3|^{-1}\sum_{i\in\mathcal{I}_3}\dfrac{\{Y_i-{E}(Y_i)\}^2}{{\rm var}(Y_i)}+\dfrac{\{{E}(Y_i)-\bar{Y}\}^2}{{\rm var}(Y_i)}+\dfrac{2|\mathcal{I}_3|^{-1}\sum_{i\in\mathcal{I}_3}\{Y_i-{E}(Y_i)\}\{{E}(Y_i)-\bar{Y}\}}{{\rm var}(Y_i)}\\
		&\quad=B_6+B_7+B_8.
	\end{align*}
	For $B_6$, since ${E}[\{Y_i-{E}(Y_i)^2\}/{\rm var}(Y_i)]=1<\infty$, by the law of large numbers, $B_6=1+o_p(1)$. For $B_7$, we have ${E}[\{Y_i-{E}(Y_i)\}/\{{\rm var}(Y_i)\}^{1/2}]=0$ and ${E}[\{Y_i-{E}(Y_i)\}^2/{\rm var}(Y_i)]=1<\infty$, thus ${E}[|\{Y_i-{E}(Y_i)\}/\{{\rm var}(Y_i)\}^{1/2}|]<\infty$. By the law of large numbers,
	$$
	B_7=\left[|\mathcal{I}_3|^{-1}\sum_{i\in\mathcal{I}_3}\dfrac{Y_i-{E}(Y_i)}{\{{\rm var}(Y_i)\}^{1/2}}\right]^2={E}\left[\dfrac{Y_i-{E}(Y_i)}{\{{\rm var}(Y_i)\}^{1/2}}\right]^2+o_p(1)=o_p(1).
	$$
	Additionally, $B_8\le 2(B_6B_7)^{1/2}=o_p(1)$. Hence, $\hat{V}_Y^{1/2}/\{{\rm var}(Y_i)\}^{1/2}=1-o_p(1)$ and it follows that
	\begin{align}\label{sigma_Y_hat}
		{P}\left\{{\rm var}(Y_i)^{1/2}b_N^q/2\le \hat{V}_Y^{1/2}b_N^q\right\}={P}\left[\hat{V}_Y^{1/2}/\{{\rm var}(Y_i)\}^{1/2}\ge 1/2\right]=1-o(1).
	\end{align}
	Combining \eqref{sigma_Y} and \eqref{sigma_Y_hat}, $P(\max_{i\in\mathcal{J}_2}|\bW_i^\T \bOmega_{\bW}\bxi|\le \hat{V}_Y^{1/2}b_N^q)=1-o(1)$. Together with Lemma \ref{OmegaW_xi_constraint_cov_shift}, for any arbitrary constant $\alpha>0$, with some appropriately chosen $\lambda_{\bu}\asymp(M+\sigma_\varepsilon)(\log d/n)^{1/2}$, $u=\bOmega_{\bW}^{\rm (mix)}\bxi$ satisfies the constraints in (2.1) and hence $\widetilde{\bu}_{\rm SR}$ exists, with probability at least $1-\alpha-o(1)$. Denote $\bdelta=\widetilde{\bbeta}_{\rm SR}-\bbeta^*$. Similar to Lemma \ref{delta}, we also have 
	\begin{equation}\label{rate:delta}
		\|\bdelta\|_2=O_p\left\{(s\log d /N)^{1/2}\right\},\ \  \|\bdelta\|_1=O_p\left\{s(\log d /N)^{1/2}\right\}.\end{equation}
	Recall that $\widetilde{\bxi}=|\mathcal{I}_3|^{-1}\sum_{i\in \mathcal{I}_3}\bW_iY_i$ and $\widetilde{\bSigma}_{\bW}=|\mathcal{J}_2|^{-1}\sum_{i\in\mathcal{J}_2}\bW_i\bW_i^\T$. Consider the following decomposition
	\begin{equation}\label{rep:thetatil}
		\tilde{\theta}_{\rm SR}-\theta=\dfrac{R_1+R_2+R_3}{|\mathcal{I}_1|^{-1}\sum_{i\in \mathcal{I}_1}\tilde{v}_i^2},
	\end{equation}
	where 
	\begin{align}
		R_1&=|\mathcal{I}_1|^{-1}\sum_{i\in \mathcal{I}_1}v_i\{\bW_i^\T(\theta\bbeta^*+\balpha^*)+\varepsilon_i\},\quad R_2=-|\mathcal{J}_2|^{-1}\sum_{i\in \mathcal{J}_2}\widetilde{\bu}_{\rm SR}^\T \bW_iv_i,\label{R_2}\\
		R_3&=\theta\cdot |\mathcal{I}_1|^{-1}\sum_{i\in \mathcal{I}_1}\bdelta^\T \bW_iv_i+\theta\cdot |\mathcal{I}_1|^{-1}\sum_{i\in \mathcal{I}_1}\bdelta^\T \bW_i\tilde{v}_i\\
		&\quad-\vphantom{\sum_{i\in\mathcal{I}_1}}\bigg\langle|\mathcal{I}_1|^{-1}\sum_{i\in\mathcal{I}_1}\bW_iY_i-\widetilde{\bSigma}_{\bW}\widetilde{\bu}_{\rm SR},\bdelta\bigg\rangle=R_{3,1}+R_{3,2}+R_{3,3}\nonumber.
	\end{align}
	As in \eqref{T_12} and \eqref{T_21}, $R_{3,1}=O_p\{Ms\log d /(Nn)^{1/2}\}$ and $R_{3,2}=O_p\{Ms\log d /(Nn)^{1/2}\}$. By the construction of $\widetilde{\bu}_{\rm SR}$, $\|\widetilde{\bxi}-\widetilde{\bSigma}_{\bW}\widetilde{\bu}_{\rm SR}\|_\infty\leq\lambda_{\bu}$.
	Repeating the proof of Lemma \ref{xi_infty} and together with \eqref{rate:delta}, we have
	$$
	R_{3,3}
	\le \|\bdelta\|_1\bigg(\bigg\||\mathcal{I}_1|^{-1}\sum_{i\in\mathcal{I}_1}\bW_iY_i-\widetilde{\bxi}\,\bigg\|_\infty+\|\widetilde{\bxi}-\widetilde{\bSigma}_{\bW}\widetilde{\bu}_{\rm SR}\|_\infty\bigg)=O_p\left\{\dfrac{(M+\sigma_{\varepsilon})s\log d }{(Nn)^{1/2}}\right\}.
	$$
	Therefore, 
	\begin{equation}\label{rate:R3}
		R_3=R_{3,1}+R_{3,2}+R_{3,3}=O_p\left\{\dfrac{(M+\sigma_{\varepsilon})s\log d }{(Nn)^{1/2}}\right\}.
	\end{equation}
	As shown in Lemmas \ref{T_11}, \ref{T_2_2_cov_shift}, and \ref{numerator}, $
	R_1=O_p\{(M+\sigma_{\varepsilon})/n^{1/2}\}$, $R_2=O_p(MN^{-1/2})$, and $1/(|\mathcal{I}_1|^{-1}\sum_{i\in \mathcal{I}_1}\tilde{v}_i^2)=O_p(1)$.
	Thus, $\tilde{\theta}_{\rm SR}-\theta=O_p[(M+\sigma_{\varepsilon})\{n^{-1/2}+s\log d /(Nn)^{1/2}\}]$.
	\hfill\BlackBox
	\\ \hspace*{\fill} \\
	{\bf Proof of Theorem \ref{thm:SSR_asy_normal}}.
	Consider the representation \eqref{rep:thetatil}. We first show the asymptotic normality of $R_1+R_2$, where $R_1$ and $R_2$ are defined in \eqref{R_2}, respectively. Denote $t_i=v_i\{\bW_i^\T(\theta\bbeta^*+\balpha^*)+\varepsilon_i\}$. Since ${E}(|v_i|^{4+c}\mid \bW_i)\le C_v$, with some constant $C\ge C_v$,
	\begin{align}\label{C_v}
		{E}(|v_i|^{4+c}\mid \bW_i)\le C,\ \ 
		{\rm var}(v_i^2\mid \bW_i)\le C.
	\end{align}
	By \eqref{t_i}, ${E}(t_i)=0$. By \eqref{psi_ti} and Lemma \ref{Bernstein_inequality}, there exists a constant $C_t>0$ such that
	\begin{align}\label{C_t}
		{E}(|t_i|^{4+c})\le C_t(M+\sigma_{\varepsilon})^{4+c},\ \ 
		{\rm var}(t_i^2)\le C_t(M+\sigma_{\varepsilon})^{4}.
	\end{align}
	By assumption,
	\begin{align}\label{c_t}
		{\rm var}(t_i)\ge c_t(M+\sigma_{\varepsilon})^2,\ \ {\rm var}(v_i\mid \bW_i)\ge c_v.
	\end{align} 
	Let $\mathcal{D}_3=\{(Y_i,\bW_i)_{i\in \mathcal{I}_3},(\bW_i)_{i\in \mathcal{J}_2}\}$. By \eqref{C_v} and \eqref{C_t},
	\begin{align*}
		&\sum_{i\in \mathcal{I}_1}{E}(||\mathcal{I}_1|^{-1}t_i|^{4+c}\mid \mathcal{D}_3)+\sum_{i\in \mathcal{J}_2}{E}\{||\mathcal{J}_2|^{-1}(-\widetilde{\bu}_{\rm SR}^\T \bW_i)v_i|^{4+c}|\mathcal{D}_3\}\\
		&\qquad\le C_t(M+\sigma_{\varepsilon})^{4+c}|\mathcal{I}_1|^{-3-c}+C |\mathcal{J}_2|^{-3-c}\max_{i\in\mathcal{J}_2}|\widetilde{\bu}_{\rm SR}^\T \bW_i|^{4+c}\\
		&\qquad\le  C_t(M+\sigma_{\varepsilon})^{4+c}|\mathcal{I}_1|^{-3-c}+C \hat{V}_Y^{(4+c)/2}|\mathcal{J}_2|^{-3-c+(4+c)q}.
	\end{align*}
	On the other hand, by \eqref{c_t},
	\begin{align*}
		&\sum_{i\in \mathcal{I}_1}{\rm var}(|\mathcal{I}_1|^{-1}t_i\mid \mathcal{D}_3)+\sum_{i\in \mathcal{J}_2}{\rm var}\{|\mathcal{J}_2|^{-1}(-\widetilde{\bu}_{\rm SR}^\T \bW_i)v_i\mid \mathcal{D}_3\}
		\ge |\mathcal{I}_1|^{-1}c_t(M+\sigma_{\varepsilon})^2.
	\end{align*}
	We check the Lyapunov's condition below. Since $q< (2+c)/(8+2c)$,
	\begin{align*}
		&\dfrac{\sum_{i\in \mathcal{I}_1}{E}(||\mathcal{I}_1|^{-1}t_i|^{4+c}\mid \mathcal{D}_3)+\sum_{i\in \mathcal{J}_2}{E}\{||\mathcal{J}_2|^{-1}(-\widetilde{\bu}_{\rm SR}^\T \bW_i)v_i|^{4+c}\mid \mathcal{D}_3\}}{\left[\sum_{i\in \mathcal{I}_1}{\rm var}(|\mathcal{I}_1|^{-1}t_i\mid \mathcal{D}_3)+\sum_{i\in \mathcal{J}_2}{\rm var}\{|\mathcal{J}_2|^{-1}(-\widetilde{\bu}_{\rm SR}^\T \bW_i)v_i\mid \mathcal{D}_3\}\right]^{(4+c)/2}}\\[1mm]
		&\qquad\le \dfrac{C_t(M+\sigma_{\varepsilon})^{4+c}|\mathcal{I}_1|^{-3-c}+C \hat{V}_Y^{(4+c)/2}|\mathcal{J}_2|^{-3-c+(4+c)q}}{c_t^{(4+c)/2}(M+\sigma_{\varepsilon})^{4+c}|\mathcal{I}_1|^{-(4+c)/2}}\\
		&\qquad=O\left\{|\mathcal{I}_1|^{-(2+c)/2}+\dfrac{\hat{V}_Y^{(4+c)/2}}{(M+\sigma_{\varepsilon})^{4+c}}\cdot |\mathcal{J}_2|^{-(2+c)/2+(4+c)q}\right\}=o_p(1),
	\end{align*}
	where the last equality holds as $-(2+c)/2+(4+c)q<0$ and
	$$
	\dfrac{\hat{V}_Y^{1/2}}{M+\sigma_{\varepsilon}}=\dfrac{\hat{V}_Y^{1/2}}{\{{\rm var}(Y_i)\}^{1/2}}\dfrac{\{{\rm var}(Y_i)\}^{1/2}}{M+\sigma_{\varepsilon}}\le\dfrac{\hat{V}_Y^{1/2}}{\{{\rm var}(Y_i)\}^{1/2}}\dfrac{\{C_{\bSigma}M^2+\sigma_{\varepsilon}^2\}^{1/2}}{M+\sigma_{\varepsilon}}=O_p(1).
	$$ 
	Thus $(R_1+R_2)/\sigma_R=\{|\mathcal{I}_1|^{-1}\sum_{i\in \mathcal{I}_1}t_i+|\mathcal{J}_2|^{-1}\sum_{i\in \mathcal{J}_2}(-\widetilde{\bu}_{\rm SR}^\T \bW_i)v_i\}/\sigma_R\xrightarrow{\rm d} \mathcal{N}(0,1)$ by Lyapunov's central limit theorem,
	where $\sigma_R^2=|\mathcal{I}_1|^{-1}{E}(t_i^2)+|\mathcal{J}_2|^{-2}\sum_{i\in \mathcal{J}_2}(\widetilde{\bu}_{\rm SR}^\T \bW_i)^2{\rm var}(v_i\mid \bW_i).$ By \eqref{c_t}, we have
	$1/\sigma_R\le 1/\{|\mathcal{I}_1|^{-1}{E}(t_i^2)\}^{1/2}=O\{n^{1/2}/(M+\sigma_{\varepsilon})\}$.
	By \eqref{rate:R3}, $R_3/\sigma_R=O_p(s\log d /N^{1/2})=o_p(1)$ when $s\ll N^{1/2}/\log d$. As in Lemma \ref{numerator}, $|\mathcal{I}_1|^{-1}\sum_{i\in \mathcal{I}_1}\tilde{v}_i^2/\sigma_{v,1}^2=1+o_p(1)$, by Slutsky's theorem, $(\tilde{\theta}_{\rm SR}-\theta)/\sigma_{\rm SR}^2\xrightarrow{\rm d} \mathcal{N}(0,1)$, where $\sigma_{\rm SR}^2=\sigma_R^2/(\sigma_{v,1}^2)^2$.
	
	We next consider the estimation of $\sigma_R^2$. In the following, we will show that
	\begin{align}
		V_1/\sigma_R^2&=1+o_p(1),\ \ V_1=|\mathcal{I}_1|^{-2}\sum_{i\in\mathcal{I}_1}t_i^2+|\mathcal{J}_2|^{-2}\sum_{i\in \mathcal{J}_2}(\widetilde{\bu}_{\rm SR}^\T \bW_i)^2v_i^2,\label{sigma_sd_1}\\
		(V_2-V_1)/\sigma_R^2&=o_p(1),\ \ V_2=|\mathcal{I}_1|^{-2}\sum_{i\in\mathcal{I}_1}\tilde{v}_i^2(Y_i-\tilde{\theta}_{\rm SR}\tilde{v}_i)^2+|\mathcal{J}_2|^{-2}\sum_{i\in \mathcal{J}_2}(\widetilde{\bu}_{\rm SR}^\T \bW_i)^2v_i^2\label{sigma_sd_2},\\
		(V_3-V_2)/\sigma_R^2&=o_p(1),\ \ 
		V_3=|\mathcal{I}_1|^{-2}\sum_{i\in\mathcal{I}_1}\tilde{v}_i^2(Y_i-\tilde{\theta}_{\rm SR}\tilde{v}_i)^2+|\mathcal{J}_2|^{-2}\sum_{i\in \mathcal{J}_2}(\widetilde{\bu}_{\rm SR}^\T \bW_i)^2\tilde{v}_i^2,\label{sigma_sd_3}
	\end{align}
	thus $V_3/\sigma_R^2=1+o_p(1)$. We have ${E}(V_1/\sigma_R^2\mid \mathcal{D}_3)=1$. Let $C_1=\max(C_t,C)/\min(c_t^2,c_v^2)$,
	\begin{align*}
		&{\rm var}(V_1/\sigma_R^2|\mathcal{D}_3)=\dfrac{|\mathcal{I}_1|^{-3}{\rm var}(t_i^2)+|\mathcal{J}_2|^{-4}\sum_{i\in \mathcal{J}_2}(\widetilde{\bu}_{\rm SR}^\T \bW_i)^4{\rm var}(v_i^2\mid \bW_i)}{\left\{|\mathcal{I}_1|^{-1}{E}(t_i^2)+|\mathcal{J}_2|^{-2}\sum_{i\in \mathcal{J}_2}(\widetilde{\bu}_{\rm SR}^\T \bW_i)^2{\rm var}(v_i\mid \bW_i)\right\}^2}\\
		&\qquad\le C_1\dfrac{|\mathcal{I}_1|^{-3}(M+\sigma_{\varepsilon})^4+|\mathcal{J}_2|^{-4}\sum_{i\in \mathcal{J}_2}(\widetilde{\bu}_{\rm SR}^\T \bW_i)^4}{\left\{|\mathcal{I}_1|^{-1}(M+\sigma_{\varepsilon})^2+|\mathcal{J}_2|^{-2}\sum_{i\in \mathcal{J}_2}(\widetilde{\bu}_{\rm SR}^\T \bW_i)^2\right\}^2}\\
		&\qquad\le \dfrac{C_1 \max\big\{|\mathcal{I}_1|^{-2}(M+\sigma_{\varepsilon})^2,\hat{V}_Y|\mathcal{J}_2|^{2q-2}\big\}}{\{|\mathcal{I}_1|^{-1}(M+\sigma_{\varepsilon})^2+|\mathcal{J}_2|^{-2}\sum_{i\in\mathcal{J}_2}(\widetilde{\bu}_{\rm SR}^\T \bW_i)^2\}^2}\bigg\{\dfrac{(M+\sigma_{\varepsilon})^2}{|\mathcal{I}_1|}+|\mathcal{J}_2|^{-2}\sum_{i\in\mathcal{J}_2}(\widetilde{\bu}_{\rm SR}^\T \bW_i)^2\bigg\}\\
		&\qquad=\dfrac{C_1\max\big\{|\mathcal{I}_1|^{-2}(M+\sigma_{\varepsilon})^2,\hat{V}_Y|\mathcal{J}_2|^{2q-2}\big\}}{|\mathcal{I}_1|^{-1}(M+\sigma_{\varepsilon})^2+|\mathcal{J}_2|^{-2}\sum_{i\in\mathcal{J}_2}(\widetilde{\bu}_{\rm SR}^\T \bW_i)^2}\le C_1\max\left\{|\mathcal{I}_1|^{-1},\dfrac{\hat{V}_Y|\mathcal{J}_2|^{2q-1}}{(M+\sigma_{\varepsilon})^2}\right\}=o_p(1),
	\end{align*}
	where the first inequality is by \eqref{C_v}, \eqref{C_t}, and \eqref{c_t}, the last equality is by $\hat{V}_Y/(M+\sigma_{\varepsilon})^2=O_p(1)$ and we choose $q<(2+c)/(8+2c)\le1/2$.
	Thus by Chebyshev's inequality, \eqref{sigma_sd_1} holds. For \eqref{sigma_sd_2},
	\begin{align*}
		(V_2-V_1)/\sigma_R^2&\le\dfrac{|\mathcal{I}_1|^{-1}\sum_{i\in\mathcal{I}_1}\tilde{v}_i^2(Y_i-\tilde{\theta}_{\rm SR}\tilde{v}_i)^2-|\mathcal{I}_1|^{-1}\sum_{i\in\mathcal{I}_1}t_i^2}{{E}(t_i^2)}\\
		&=\dfrac{|\mathcal{I}_1|^{-1}\sum_{i\in\mathcal{I}_1}\tilde{v}_i^2(Y_i-\tilde{\theta}_{\rm SR} \tilde{v}_i)^2-|\mathcal{I}_1|^{-1}\sum_{i\in\mathcal{I}_1}\tilde{v}_i^2(Y_i-\theta v_i)^2}{{E}(t_i^2)}\\
		&\quad+\dfrac{|\mathcal{I}_1|^{-1}\sum_{i\in\mathcal{I}_1}\tilde{v}_i^2(Y_i-\theta v_i)^2-|\mathcal{I}_1|^{-1}\sum_{i\in\mathcal{I}_1}t_i^2}{{E}(t_i^2)}=B_9+B_{10}.
	\end{align*}
	We next show that $B_9$ and $B_{10}$ are both $o_p(1)$ terms. For $B_{10}$, 
	\begin{align*}
		B_{10}&=\dfrac{\sum_{i\in\mathcal{I}_1}(\tilde{v}_i-v_i)^2(Y_i-\theta v_i)^2}{|\mathcal{I}_1|{E}(t_i^2)}+\dfrac{2\sum_{i\in\mathcal{I}_1}v_i(\tilde{v}_i-v_i)(Y_i-\theta v_i)^2}{|\mathcal{I}_1|{E}(t_i^2)}=B_{10,1}+B_{10,2}.
	\end{align*}
	For $B_{10,1}$, by Lemma \ref{v_i},  $|\mathcal{I}_1|^{-1}\sum_{i\in \mathcal{I}_1}(v_i-\tilde{v}_i)^4=o_p(1)$. By Lemma \ref{psi}, $\|Y_i-\theta v_i\|_{\psi_2}=\|\bW_i^\T(\theta\bbeta^*+\balpha^*)+\varepsilon_i\|_{\psi_2}=O(M+\sigma_{\varepsilon})$. By Lemma \ref{sub_gaussian_bound}, ${E}\{(Y_i-\theta v_i)^4\}=O\{(M+\sigma_{\varepsilon})^4\}$. By Markov's inequality, $|\mathcal{I}_1|^{-1}\sum_{i\in\mathcal{I}_1}(Y_i-\theta v_i)^4=O_p\{(M+\sigma_{\varepsilon})^4\}$. Together with \eqref{c_t},
	\begin{align*}
		B_{10,1}&\le\{|\mathcal{I}_1|^{-1}\sum_{i\in\mathcal{I}_1}(\tilde{v}_i-v_i)^4\}^{1/2}\{|\mathcal{I}_1|^{-1}\sum_{i\in\mathcal{I}_1}(Y_i-\theta v_i)^4\}^{1/2}/{E}(t_i^2)=o_p(1).
	\end{align*}
	For $B_{10,2}$, we have $E\{t_i^2/{E}(t_i^2)\}=1$. Combining \eqref{C_t} and \eqref{c_t}, 
	${\rm var}\{|\mathcal{I}_1|^{-1}\sum_{i\in\mathcal{I}_1}t_i^2/{E}(t_i^2)\}=|\mathcal{I}_1|^{-1}{\rm var}(t_i^2)/\{{E}(t_i^2)\}^2=O(1/n)$. 
	By Chebyshev's inequality, 
	\begin{equation}\label{sumt_i}
		|\mathcal{I}_1|^{-1}\sum_{i\in\mathcal{I}_1}t_i^2/{E}(t_i^2)=1+O_p(n^{-1/2})=O_p(1).
	\end{equation} 
	Therefore, $B_{10,2}\le 2\{B_{10,1}\cdot |\mathcal{I}_1|^{-1}\sum_{i\in\mathcal{I}_1}t_i^2/{E}(t_i^2)\}^{1/2}=o_p(1)$.
	For $B_9$, we have
	\begin{align*}
		B_9&=\dfrac{\sum_{i\in\mathcal{I}_1}\tilde{v}_i^2(\theta v_i-\tilde{\theta}_{\rm SR}\tilde{v}_i)^2}{|\mathcal{I}_1|{E}(t_i^2)}+\dfrac{2\sum_{i\in\mathcal{I}_1}\tilde{v}_i^2(Y_i-\theta v_i)(\theta v_i-\tilde{\theta}_{\rm SR}\tilde{v}_i)}{|\mathcal{I}_1|{E}(t_i^2)}=B_{9,1}+B_{9,2}.
	\end{align*}
	It suffices to show that both $B_{9,1}$ and $B_{9,2}$ are $o_p(1)$ terms. For $B_{9,1}$, by Lemma \ref{v_i}, $|\mathcal{I}_1|^{-1}\sum_{i\in \mathcal{I}_1}(v_i-\hat{v}_i)^4=o_p(1)$, $|\mathcal{I}_1|^{-1}\sum_{i\in \mathcal{I}_1}\hat{v}_i^4=O_p(1)$. By \eqref{c_t}, $\theta^2/{E}(t_i^2)=O\{M^2/(M+\sigma_{\varepsilon})^2\}=O(1)$. Therefore,
	\begin{align}\label{A111}
		\!\!\!\dfrac{|\mathcal{I}_1|^{-1}\sum_{i\in \mathcal{I}_1}(v_i-\tilde{v}_i)^2\tilde{v}_i^2\theta^2}{{E}(t_i^2)}&\le \dfrac{\theta^2}{{E}(t_i^2)}\{|\mathcal{I}_1|^{-1}\!\sum_{i\in \mathcal{I}_1}(v_i-\tilde{v}_i)^4\}^{1/2}\{|\mathcal{I}_1|^{-1}\!\sum_{i\in \mathcal{I}_1}\tilde{v}_i^4\}^{1/2}=o_p(1).
	\end{align}
	Furthermore, by Theorem 2.6, $\tilde{\theta}_{\rm SR}-\theta=O_p[(M+\sigma_{\varepsilon})\{n^{-1/2}+s\log d /(Nn)^{1/2}\}]$. Together with \eqref{c_t},
	\begin{align}\label{A112}
		\dfrac{|\mathcal{I}_1|^{-1}\sum_{i\in \mathcal{I}_1}\tilde{v}_i^4(\tilde{\theta}_{\rm SR}-\theta)^2}{{E}(t_i^2)}&=O_p\left[\left\{\dfrac{1}{n^{1/2}}+\dfrac{s\log d }{(Nn)^{1/2}}\right\}^{2}\right]=o_p(1).
	\end{align}
	Combining \eqref{A111} and \eqref{A112},
	\begin{align*}
		B_{9,1}&\le \dfrac{|\mathcal{I}_1|^{-1}\sum_{i\in \mathcal{I}_1}(v_i-\tilde{v}_i)^2\tilde{v}_i^2\theta^2}{{E}(t_i^2)}+\dfrac{|\mathcal{I}_1|^{-1}\sum_{i\in \mathcal{I}_1}\tilde{v}_i^4(\tilde{\theta}_{\rm SR}-\theta)^2}{{E}(t_i^2)}\\
		&\quad+2\left\{\dfrac{|\mathcal{I}_1|^{-1}\sum_{i\in \mathcal{I}_1}(v_i-\tilde{v}_i)^2\tilde{v}_i^2\theta^2}{{E}(t_i^2)}\right\}^{1/2}\left\{\dfrac{|\mathcal{I}_1|^{-1}\sum_{i\in \mathcal{I}_1}\tilde{v}_i^4(\tilde{\theta}_{\rm SR}-\theta)^2}{{E}(t_i^2)}\right\}^{1/2}=o_p(1).
	\end{align*}
	For $B_{9,2}$, combining \eqref{sumt_i} and the upper bound for $A_{1,1}$ and $A_2$, $B_{9,2}\le 2[B_{9,1}\{B_{10}+|\mathcal{I}_1|^{-1}\sum_{i\in\mathcal{I}_1}t_i^2/{E}(t_i^2)\}]=o_p(1)$.
	For \eqref{sigma_sd_3}, 
	\begin{align*}
		\dfrac{V_3-V_2}{\sigma_R^2}&=\dfrac{\sum_{i\in \mathcal{J}_2}(\widetilde{\bu}_{\rm SR}^\T \bW_i)^2(\tilde{v}_i-v_i)^2}{|\mathcal{J}_2|^2\sigma_R^2}+\dfrac{2\sum_{i\in \mathcal{J}_2}(\widetilde{\bu}_{\rm SR}^\T \bW_i)^2v_i(\tilde{v}_i-v_i)}{|\mathcal{J}_2|^2\sigma_R^2}=B_{11}+B_{12}.
	\end{align*}
	We next show that both $B_{11}$ and $B_{12}$ are $o_p(1)$ terms. By \eqref{c_t}, we first have
	\begin{align*}
		B_{11}&\le \dfrac{\sum_{i\in \mathcal{J}_2}(\widetilde{\bu}_{\rm SR}^\T \bW_i)^2(\tilde{v}_i-v_i)^2}{\sum_{i\in \mathcal{J}_2}(\widetilde{\bu}_{\rm SR}^\T \bW_i)^2{\rm var}(v_i\mid \bW_i)}\le \dfrac{\sum_{i\in \mathcal{J}_2}(\widetilde{\bu}_{\rm SR}^\T \bW_i)^2}{c_v\sum_{i\in \mathcal{I}_1}(\widetilde{\bu}_{\rm SR}^\T \bW_i)^2}\max_{i\in \mathcal{J}_2}\{\bW_i^\T(\bbeta^*-\widetilde{\bbeta}_{\rm SR})\}^2\\
		&=\max_{i\in\mathcal{J}_2,j\in [d]}|W_{ij}|^2\|\widetilde{\bbeta}_{\rm SR}-\bbeta^*\|_1^2/=O_p\left\{\log(Nd)\dfrac{s^2\log d }{N}\right\}
		=o_p(1),
	\end{align*}
	where the second equality is by $\max_{i\in\mathcal{J}_2,j\in [d]}|W_{ij}|=O_p[\{\log(Nd)\}^{1/2}]$, the third equality is by the assumption that $\liminf \log d /\log (N)>0$ and $s\ll N^{1/2}/\log d$. For $B_{12}$, since $V_1/\sigma_R^2=1+o_p(1)$, $B_{12}\le 2\{B_{11}V_1/\sigma_R^2\}^{1/2}=o_p(1)$. Combining \eqref{sigma_sd_1}, \eqref{sigma_sd_2} and \eqref{sigma_sd_3}, $V_3/\sigma_R^2=1+o_p(1)$. By Lemma \ref{numerator}, we have $\tilde{\sigma}_{v,1}^2/\sigma_{v,1}^2=1+o_p(1)$, where $\tilde{\sigma}_{v,1}^2=|\mathcal{I}_1|^{-1}\sum_{i\in \mathcal{I}_1}\tilde{v}_i^2$. Therefore, $\tilde{\sigma}_{\rm SR}^2/\sigma_{\rm SR}^2= 1+o_p(1)$, where
	$$ 
	\tilde{\sigma}_{\rm SR}^2=\dfrac{V_3}{(\tilde{\sigma}_{v,1}^2)^2}
	=(\tilde{\sigma}_{v,1}^2)^{-2}|\mathcal{I}_1|^{-2}\sum_{i\in\mathcal{I}_1}\tilde{v}_i^2(Y_i-\hat{\theta}_{\rm SR}\tilde{v}_i)^2+(\tilde{\sigma}_{v,1}^2)^{-2}|\mathcal{J}_2|^{-2}\sum_{i\in \mathcal{J}_2}(\widehat{\bu}_{\rm SR}^\T \bW_i)^2\tilde{v}_i^2.
	$$
	By Slutsky's theorem, $(\tilde{\theta}_{\rm SR}-\theta)/\tilde\sigma_{\rm SR}\xrightarrow{\rm d} \mathcal{N}(0,1)$.
	\hfill\BlackBox
	\\ \hspace*{\fill} \\
	{\bf Proof of Theorem  \ref{thm:DR_consist_rate}}.
	Denote $\widehat{\bgamma}_{{\rm DR},-k}=(\be_1^\T\widehat{\bgamma}_{{\rm DR},-k},\widehat{\balpha}_{{\rm DR},-k}^\T)^\T$ and
	$
	\hat\theta_{{\rm DR},k}=\be_1^\T\widehat{\bgamma}_{{\rm DR},-k}+\{\sum_{i\in\mathcal{I}_{k}} (Z_i-\bW_i^\T\widehat{\bbeta}_{{\rm DR},-k})(Y_i-\bX_i^\T\widehat{\bgamma}_{{\rm DR},-k})\}/\{\sum_{i\in \mathcal{I}_{k}} (Z_i-\bW_i^\T\widehat{\bbeta}_{{\rm DR},-k})Z_i\}$.
	For each $k$, repeating the proof of Lemma \ref{delta}, we have the following upper bounds:
	\begin{align}
		\|\widehat{\balpha}_{{\rm DR},-k}-\balpha^*\|_2=O_p\left\{\sigma_{\varepsilon}\left(\dfrac{k\log d }{n}\right)^{1/2}\right\}, \|\widehat{\bbeta}_{{\rm DR},-k}-\bbeta^*\|_2&=O_p\left\{\left(\dfrac{s\log d }{N}\right)^{1/2}\right\}.\label{beta_DR}
	\end{align}
	For each $k$,
	\begin{align*}
		&\hat{\theta}_{{\rm DR},k}-\theta=\be_1^\T\widehat{\bgamma}_{{\rm DR},-k}-\theta+\dfrac{|\mathcal{I}_k|^{-1}\sum_{i\in\mathcal{I}_k} (Z_i-\bW_i^\T\widehat{\bbeta}_{{\rm DR},-k})\{\varepsilon_i+\bX_i^\T(\bgamma^*-\widehat{\gamma}_{{\rm DR},-k})\}}{|\mathcal{I}_k|^{-1}\sum_{i\in \mathcal{I}_{k}} (Z_i-\bW_i^\T\widehat{\bbeta}_{{\rm DR},-k})Z_i}\\
		&\quad=\dfrac{|\mathcal{I}_k|^{-1}\sum_{i\in\mathcal{I}_k} (Z_i-\bW_i^\T\widehat{\bbeta}_{{\rm DR},-k})\{\varepsilon_i+\bW_i^\T(\balpha^*-\widehat{\balpha}_{{\rm DR},-k})\}}{|\mathcal{I}_k|^{-1}\sum_{i\in \mathcal{I}_{k}} (Z_i-\bW_i^\T\widehat{\bbeta}_{{\rm DR},-k})Z_i}\\
		&\quad=\bigg\{|\mathcal{I}_k|^{-1}\sum_{i\in \mathcal{I}_{k}} (Z_i-\bW_i^\T\widehat{\bbeta}_{{\rm DR},-k})Z_i\bigg\}^{-1}\bigg\{|\mathcal{I}_k|^{-1}\sum_{i\in\mathcal{I}_k}v_i\varepsilon_i\\
		&\quad\quad+|\mathcal{I}_k|^{-1}\sum_{i\in\mathcal{I}_k}\bW_i^\T(\bbeta^*-\widehat{\bbeta}_{{\rm DR},-k})\varepsilon_i+|\mathcal{I}_k|^{-1}\sum_{i\in \mathcal{I}_k}v_i\bW_i^\T(\balpha^*-\widehat{\balpha}_{{\rm DR},-k})\\
		&\quad\quad+|\mathcal{I}_k|^{-1}\sum_{i\in \mathcal{I}_k}\bW_i^\T(\bbeta^*-\widehat{\bbeta}_{{\rm DR},-k})\bW_i^\T(\balpha^*-\widehat{\balpha}_{{\rm DR},-k})\bigg\}=G_5^{-1}(G_1+G_2+G_3+G_4).
	\end{align*}
	Repeating the proof of Lemma \ref{psi}, when 
	$1/C_{\bSigma}\le \lambda_{\rm min}(\bSigma)\le\lambda_{\rm max}(\bSigma)\le C_{\bSigma}$,
	\begin{equation}\label{C_sigma2}
		1/C_{\bSigma}\le \sigma_{v,1}^2\le C_{\bSigma}\ \ {\rm and}\ \
		\|\bbeta^*\|_2\le C_{\bSigma}^2.
	\end{equation} 
	Let $\mathcal{D}_{-k}=\{(Z_i,\bW_i,Y_i)_{i\in \mathcal{I}_{-k}},(Z_i,\bW_i)_{i\in \mathcal{J}_{-k}}\}$. Since $v_iZ_i$ is sub-exponential with ${E}(v_iZ_i)=\sigma_{v,1}^2$ and $\|v_iZ_i\|_{\psi_1}=O(1)$, by Lemma \ref{Bernstein_inequality}, ${\rm var}(|\mathcal{I}_k|^{-1}\sum_{i\in\mathcal{I}}v_iZ_i)=O(1/n)$. By Chebyshev's inequality, $|\mathcal{I}_k|^{-1}\sum_{i\in \mathcal{I}_{k}}v_iZ_i-\sigma_{v,1}^2=O_p(n^{-1/2})$. Conditional on $\mathcal{D}_{-k}$, notice that $|\mathcal{I}_k|^{-1}\sum_{i\in \mathcal{I}_{k}}\{\bW_i^\T(\widehat{\bbeta}_{{\rm DR},-k}-\bbeta^*)\}^2$ is the summation of i.i.d. random variable with mean
	\begin{align*}
		{E}\left[\{\bW_i^\T(\widehat{\bbeta}_{{\rm DR},-k}-\bbeta^*)\}^2\mid \mathcal{D}_{-k}\right]&=(\widehat{\bbeta}_{{\rm DR},-k}-\bbeta^*)^\T\bSigma_{\bW}(\widehat{\bbeta}_{{\rm DR},-k}-\bbeta^*)=O(\|\widehat{\bbeta}_{{\rm DR},-k}-\bbeta^*\|_2^2).
	\end{align*}
	By Markov's inequality and \eqref{beta_DR}, 
	\begin{equation}\label{beta2}
		|\mathcal{I}_k|^{-1}\sum_{i\in \mathcal{I}_{k}}\{\bW_i^\T(\widehat{\bbeta}_{{\rm DR},-k}-\bbeta^*)\}^2=O_p\left(\dfrac{s\log d }{N}\right).
	\end{equation}
	Since $Z_i^2$ is sub-exponential with $\|Z_i^2\|_{\psi_1}=O(1)$. By Lemma \ref{Bernstein_inequality}, $E(Z_i^2)=O(1)$ and ${\rm var}(|\mathcal{I}_k|^{-1}\sum_{i\in\mathcal{I}}Z_i^2)=O(1/n)$. By Chebyshev's inequality, we have $|\mathcal{I}_k|^{-1}\sum_{i\in \mathcal{I}_{k}}Z_i^2={E}(Z_i^2)+O_p(1/n^{1/2})=O_p(1)$. For $G_5$, we have
	\begin{equation}\label{R_5}
		\begin{split}
			&G_5-\sigma_{v,1}^2
			=||\mathcal{I}_k|^{-1}\sum_{i\in \mathcal{I}_{k}}v_iZ_i-\sigma_{v,1}^2|+|\mathcal{I}_k|^{-1}\sum_{i\in \mathcal{I}_{k}}\bW_i^\T(\bbeta^*-\widehat{\bbeta}_{{\rm DR},-k})Z_i\\
			&\quad\le||\mathcal{I}_k|^{-1}\sum_{i\in \mathcal{I}_{k}}v_iZ_i-\sigma_{v,1}^2|+\sqrt{|\mathcal{I}_k|^{-1}\sum_{i\in \mathcal{I}_{k}}\{\bW_i^\T(\widehat{\bbeta}_{{\rm DR},-k}-\bbeta^*)\}^2}\sqrt{|\mathcal{I}_k|^{-1}\sum_{i\in \mathcal{I}_{k}}Z_i^2}\\
			&\quad=O_p\left\{n^{-1/2}+\left(\dfrac{s\log d }{N}\right)^{1/2}\right\}.
		\end{split}
	\end{equation}
	By \eqref{C_sigma2}, when $s\ll N/\log d$, $1/G_5=O_p(1)$. For $G_1$, by Lemma \ref{best_linear_slope}, ${E}(v_i\varepsilon_i)=0$. By Lemma \ref{psi}, $v_i\varepsilon_i$ is sub-exponential with $\|v_i\varepsilon_i\|_{\psi_1}=O(\sigma_{\varepsilon})$. By Lemma \ref{Bernstein_inequality}, ${\rm var}(|\mathcal{I}_k|^{-1}\sum_{i\in\mathcal{I}_k}v_i\varepsilon_i)=O(\sigma_{\varepsilon}/n)$. By Chebyshev's inequality, 
	\begin{equation}\label{R1}
		G_1=O_p\left(\sigma_{\varepsilon}n^{-1/2}\right).
	\end{equation}
	Conditional on $\mathcal{D}_{-k}$, $G_2$ is the sum of i.i.d. random variables. By Lemma \ref{best_linear_slope}, ${E}(G_2\mid \mathcal{D}_{-k})=(\bbeta^*-\widehat{\bbeta}_{{\rm DR},-k})^\T{E}(\bW_i\varepsilon_i)=0$.
	Conditional on $\mathcal{D}_{-k}$, $\bW_i^\T(\bbeta^*-\widehat{\bbeta}_{{\rm DR},-k})$ is sub-Gaussian with $\|\bW_i^\T(\bbeta^*-\widehat{\bbeta}_{{\rm DR},-k})\|_{\psi_2}=\|\widehat{\bbeta}_{{\rm DR},-k}-\bbeta^*\|_2$, and hence ${E}\{(\bW_i^\T(\bbeta^*-\widehat{\bbeta}_{{\rm DR},-k}))^4\mid \mathcal{D}_{-k}\}=O(\|\bbeta^*-\widehat{\bbeta}_{{\rm DR},-k}\|_2^4)$. Additionally, ${E}(\varepsilon_i^4)=O(\sigma_\varepsilon^4)$. Hence,
	\begin{align*}
		&{\rm var}(G_2\mid \mathcal{D}_{-k})=|\mathcal{I}_k|^{-1}{E}\left[\{\bW_i^\T(\bbeta^*-\widehat{\bbeta}_{{\rm DR},-k})\}^2\varepsilon_i^2\mid \mathcal{D}_{-k}\right]\\
		&\qquad\le |\mathcal{I}_k|^{-1}\left[{E}\left\{(\bW_i^\T(\bbeta^*-\widehat{\bbeta}_{{\rm DR},-k}))^4|\mathcal{D}_{-k}\right\}{E}(\varepsilon_i^4)\right]^{1/2}=O\left(\sigma_\varepsilon^2\|\widehat{\bbeta}_{{\rm DR},-k}-\bbeta^*\|_2^2/n\right).
	\end{align*}
	By Chebyshev's inequality and \eqref{beta_DR}, $G_2=O_p\{\sigma_{\varepsilon}(s\log d/Nn)^{1/2}\}$. Conditional on $\mathcal{D}_{-k}$, $G_3$ is the sum of i.i.d. random variables. By Lemma \ref{best_linear_slope}, ${E}(G_3\mid \mathcal{D}_{-k})=(\balpha^*-\widehat{\balpha}_{{\rm DR},-k}){E}(\bW_iv_i)=0$.
	Conditional on $\mathcal{D}_{-k}$, $\bW_i^\T(\balpha^*-\widehat{\balpha}_{{\rm DR},-k})$ is sub-Gaussian and $\|\bW_i^\T(\balpha^*-\widehat{\balpha}_{{\rm DR},-k})\|_{\psi_2}=\|\widehat{\balpha}_{{\rm DR},-k}-\balpha^*\|_2$. Hence, ${E}[\{\bW_i^\T(\balpha^*-\widehat{\balpha}_{{\rm DR},-k})\}^4\mid \mathcal{D}_{-k}]=O(\|\balpha^*-\widehat{\balpha}_{{\rm DR},-k}\|_2^4)$. Additionally, ${E}(v_i^4)=O(1)$, and we have
	\begin{align*}
		&{\rm var}(G_3\mid \mathcal{D}_{-k})
		=|\mathcal{I}_k|^{-1}{E}\left[\{\bW_i^\T(\balpha^*-\widehat{\balpha}_{{\rm DR},-k})\}^2v_i^2|\mathcal{D}_{-k}\right]
		\\
		&\qquad\le |\mathcal{I}_k|^{-1}\left[{E}\left\{(\bW_i^\T(\balpha^*-\widehat{\balpha}_{{\rm DR},-k}))^4\mid \mathcal{D}_{-k}\right\}{E}(v_i^4)\right]^{1/2}=O_p\left(\|\widehat{\balpha}_{{\rm DR},-k}-\balpha^*\|_2^2/n\right).
	\end{align*}
	By Chebyshev's inequality and \eqref{gamma_DF}, $G_3=O_p\left\{\sigma_{\varepsilon}(k\log d)^{1/2}/n\right\}$. Conditional on $\mathcal{D}_{-k}$, $|\mathcal{I}_k|^{-1}\sum_{i\in \mathcal{I}_{k}}\{\bW_i^\T(\balpha^*-\widehat{\balpha}_{{\rm DR},-k})\}^2$ is the sum of i.i.d. random variables. Additionally,
	\begin{align*}
		{E}\left[\{\bW_i^\T(\balpha^*-\widehat{\balpha}_{{\rm DR},-k})\}^2|\mathcal{D}_{-k}\right]&=(\widehat{\balpha}_{{\rm DR},-k}-\balpha^*)\bSigma_{\bW}(\widehat{\balpha}_{{\rm DR},-k}-\balpha^*)=O(\|(\widehat{\balpha}_{{\rm DR},-k}-\balpha^*)\|_2^2).
	\end{align*}
	By Markov's inequality and \eqref{gamma_DF}, 
	\begin{align}
		|\mathcal{I}_k|^{-1}\sum_{i\in \mathcal{I}_{k}}\{\bW_i^\T(\balpha^*-\hat\alpha_{-k})\}^2
		=O_p\left(\dfrac{\sigma_{\varepsilon}^2k\log d }{n}\right).\label{gamma2}
	\end{align} 
	Combining \eqref{beta2} and \eqref{gamma2}, 
	\begin{align*}
		G_4&\le \bigg[{|\mathcal{I}_k|^{-1}\sum_{i\in \mathcal{I}_{k}}\{\bW_i^T(\bbeta^*-\widehat{\bbeta}_{{\rm DR},-k})\}^2}\bigg]^{1/2}\bigg[|\mathcal{I}_k|^{-1}\sum_{i\in \mathcal{I}_{k}}\{\bW_i^\T(\balpha^*-\widehat{\balpha}_{{\rm DR},-k})\}^2\bigg]^{1/2}\\
		&=O_p\left\{\sigma_{\varepsilon}\left(\dfrac{s\log d k\log d }{Nn}\right)^{1/2}\right\}.
	\end{align*}
	Combining the bounds for $G_1,G_2,\dots,G_5$, when $k\ll n/\log d$ and $s\ll N/\log d$,
	\begin{align*}
		\hat{\theta}_{{\rm DR,}k}-\theta&=O_p\left\{\dfrac{\sigma_{\varepsilon}}{n^{1/2}}+\sigma_{\varepsilon}\left(\dfrac{s\log d }{Nn}\right)^{1/2}+\dfrac{\sigma_{\varepsilon}(k\log d)^{1/2} }{n}+\sigma_{\varepsilon}\left(\dfrac{s\log d \cdot k\log d }{Nn}\right)^{1/2}\right\}\\
		&=O_p\left(\dfrac{\sigma_{\varepsilon}}{n^{1/2}}+\sigma_{\varepsilon}\left(\dfrac{s\log d \cdot k\log d }{Nn}\right)^{1/2}\right).
	\end{align*}
	Hence, $\hat{\theta}_{\rm DR}-\theta=K^{-1}\sum_{k=1}^{K}\hat{\theta}_{{\rm DR,}k}-\theta=O_p[\sigma_{\varepsilon}\{1/n^{1/2}+(sk)^{1/2}\log d/(Nn)^{1/2}\}]$.
	\hfill\BlackBox
	\\ \hspace*{\fill} \\
	{\bf Proof of Theorem  \ref{thm:DR_asy_normal}}.
	When $sk\log^2d\ll N$, as shown in the proof of Theorem 3.3, $G_2+G_3+G_4=o_p(\sigma_{\varepsilon}/n^{1/2})$ and $G_5\to {E}(v_i^2)$ in probabliity. Under the assumption ${E}(\varepsilon_i^2\mid \bX_i)\ge c_\varepsilon \sigma_{\varepsilon}^2$, 
	\begin{equation}\label{Eveps_lower}
		{E}(v_i^2\varepsilon_i^2)={E}\left[v_i^2{E}(\varepsilon_i^2\mid \bX_i)\right]\ge c_\varepsilon \sigma_{\varepsilon}^2/C_{\bSigma}.
	\end{equation}
	Together with the fact that ${E}(|v_i\varepsilon_i|^3)=O(\sigma_{\varepsilon}^3)$,
	$
	{E}(|v_i\varepsilon_i|^3)/\{{E}(v_i^2\varepsilon_i^2)\}^{3/2}=O(1),
	$ 
	that is, the Lyapunov condition holds. Let $\sigma ^2={E}(v_i^2\varepsilon_i^2)/(\sigma_{v,1}^2)^2$, we have
	\begin{align*}
		&\dfrac{n^{1/2}(\hat{\theta}_{{\rm DR}}-\theta)}{\sigma }
		=\dfrac{1}{K}\sum_{k=1}^{K}\dfrac{n^{1/2}\cdot |\mathcal{I}_k|^{-1}\sum_{i\in\mathcal{I}_k}v_i\varepsilon_i}{|\mathcal{I}_k|^{-1}\sum_{i\in \mathcal{I}_{k}} (Z_i-\bW_i^\T\widehat{\bbeta}_{{\rm DR},-k})Z_i}\cdot\dfrac{\sigma_{v,1}^2}{{E}(v_i^2\varepsilon_i^2)^{1/2}}+o_p(1)\\
		&\quad=\dfrac{1}{K}\sum_{k=1}^{K}\dfrac{n^{1/2}\cdot |\mathcal{I}_k|^{-1}\sum_{i\in\mathcal{I}_k}v_i\varepsilon_i}{{E}(v_i^2\varepsilon_i^2)^{1/2}}\\
		&\quad\quad+
		\dfrac{1}{K}\sum_{k=1}^{K}\dfrac{n^{1/2}\cdot |\mathcal{I}_k|^{-1}\sum_{i\in\mathcal{I}_k}v_i\varepsilon_i}{{E}(v_i^2\varepsilon_i^2)^{1/2}}
		\left(\dfrac{\sigma_{v,1}^2}{|\mathcal{I}_k|^{-1}\sum_{i\in \mathcal{I}_{k}} (Z_i-\bW_i^\T\widehat{\bbeta}_{{\rm DR},-k})Z_i}-1\right)+o_p(1)\\
		&\quad=G_6+G_7+o_p(1).
	\end{align*}
	For $G_6$, by Lyapunov CLT, $G_6=n^{1/2}|\mathcal{I}_k|^{-1}\sum_{i\in\mathcal{I}}v_i\varepsilon_i/{E}(v_i^2\varepsilon_i^2)^{1/2}\xrightarrow{\rm d}\mathcal{N}(0,1)$. For $G_7$, we have
	$n^{1/2}\cdot |\mathcal{I}_k|^{-1}\sum_{i\in\mathcal{I}_k}v_i\varepsilon_i/{E}(v_i^2\varepsilon_i^2)^{1/2}=O_p(1)$. Together with \eqref{R_5} and $\sigma_{v,1}^2\ge 1/C_{\bSigma}$,  
	$
	\sigma_{v,1}^2/\{|\mathcal{I}_1|^{-1}\sum_{i\in \mathcal{I}_{k}} (Z_i-\bW_i^\T\widehat{\bbeta}_{{\rm DR},-k})Z_i\}-1=o_p(1),
	$
	thus $G_7=o_p(1)$. It holds that $n^{1/2}(\hat{\theta}_{{\rm DR}}-\theta)/\sigma \xrightarrow{\rm d} \mathcal{N}(0,1)$. 
	
	We next consider the variance estimation. Let $\mathcal{D}_{-k}=\{(Z_i,\bW_i,Y_i)_{i\in \mathcal{I}_{-k}},(Z_i,\bW_i)_{i\in \mathcal{J}_{-k}}\}$. We first prove that for each $k$,
	\begin{equation}\label{hatDR1}
		n^{-1}\sum_{k=1}^{K}\sum_{i\in \mathcal{I}_{k}}(Z_i-\bW_i^\T\widehat{\bbeta}_{{\rm DR},-k})^2/\sigma_{v,1}^2=1+o_p(1).
	\end{equation}  
	Notice that
	\begin{align*}
		&|\mathcal{I}_k|^{-1}\sum_{i\in \mathcal{I}_{k}}(Z_i-\bW_i^\T\widehat{\bbeta}_{{\rm DR},-k})^2-\sigma_{v,1}^2
		=|\mathcal{I}_k|^{-1}\sum_{i\in \mathcal{I}_{k}}\{\bW_i^\T(\bbeta^*-\widehat{\bbeta}_{{\rm DR},-k})\}^2\\
		&\qquad+(|\mathcal{I}_k|^{-1}\sum_{i\in\mathcal{I}_k}v_i^2-\sigma_{v,1}^2)+2|\mathcal{I}_k|^{-1}\sum_{i\in\mathcal{I}_k}v_i\bW_i^\T(\bbeta^*-\widehat{\bbeta}_{{\rm DR},-k})=G_8+G_9+G_{10}.
	\end{align*}
	Since $v_i^2$ is sub-exponential with $E(v_i^2)=\sigma_{v,1}^2$ and $\|v_i^2\|_{\psi_1}=O(1)$. By Lemma \ref{Bernstein_inequality}, we have ${\rm var}(|\mathcal{I}_k|^{-1}\sum_{i\in\mathcal{I}_k}v_i^2)=O(1/n)$. By Chebyshev's inequality, $G_8=O_p(n^{-1/2})=o_p(1)$. By \eqref{beta2}, $G_9=O_p(s\log d /N)=o_p(1)$. Moreover, $G_{10}\le 2\{G_9\cdot |\mathcal{I}_k|^{-1}\sum_{i\in\mathcal{I}_k}v_i^2\}^{1/2}=o_p(1)$, and it follows that $|\mathcal{I}_k|^{-1}\sum_{i\in \mathcal{I}_{k}}(Z_i-\bW_i^\T\widehat{\bbeta}_{{\rm DR},-k})^2/\sigma_{v,1}^2= 1+o_p(1)$. Together with \eqref{C_sigma2}, we have $|\mathcal{I}_k|^{-1}\sum_{i\in \mathcal{I}_{k}}(Z_i-\bW_i^\T\widehat{\bbeta}_{{\rm DR},-k})^2/\sigma_{v,1}^2=1+o_p(1)$. Finally, taking the average over $k$ terms, \eqref{hatDR1} is proved. We next show that 
	$|\mathcal{I}_k|^{-1}\sum_{i\in \mathcal{I}_{k}}(Z_i-\bW_i^\T\widehat{\bbeta}_{{\rm DR},-k})^2(Y_i-\bX_i^\T\widehat{\bgamma}_{{\rm DR},-k})^2/{E}(v_i^2\varepsilon_i^2)=1+o_p(1)$ for each $k$.
	It suffices to show that
	\begin{align}
		&\dfrac{V_4}{{E}(v_i^2\varepsilon_i^2)}=1+o_p(1),\ \ V_4=|\mathcal{I}_k|^{-1}\sum_{i\in \mathcal{I}_{k}}v_i^2\varepsilon_i^2,\label{veps1}\\
		&\dfrac{V_5-V_4}{{E}(v_i^2\varepsilon_i^2)}=o_p(1),\ \ V_5=|\mathcal{I}_k|^{-1}\sum_{i\in \mathcal{I}_{k}}(Z_i-\bW_i^\T\widehat{\bbeta}_{{\rm DR},-k})^2\varepsilon_i^2,\label{veps2}\\
		&\dfrac{V_6-V_5}{{E}(v_i^2\varepsilon_i^2)}=o_p(1).\ \ V_6=|\mathcal{I}_k|^{-1}\sum_{i\in \mathcal{I}_{k}}(Z_i-\bW_i^\T\widehat{\bbeta}_{{\rm DR},-k})^2(Y_i-\bX_i^\T\widehat{\bgamma}_{{\rm DR},-k})^2.\label{veps3}
	\end{align}
	To establish \eqref{veps1}, we first have ${E}\{v_i^2\varepsilon_i^2/{E}(v_i^2\varepsilon_i^2)\}=1$. By Lemma \ref{psi}, $v_i\varepsilon_i$ is sub-exponential with $\|v_i\varepsilon_i\|_{\psi_1}=O(\sigma_{\varepsilon})$. By Lemma \ref{Bernstein_inequality}, ${\rm var}(v_i^2\varepsilon_i^2)=O(\sigma_{\varepsilon}^4)$. Together with \eqref{Eveps_lower}, we have ${\rm var}\{V_1/E(v_i^2\varepsilon_i^2)\}=|\mathcal{I}_k|^{-1}{\rm var}(v_i^2\varepsilon_i^2)/\{E(v_i^2\varepsilon_i^2)\}^2=O(1/n)$. By Chebyshev's inequality, we have $V_4/E(v_i^2\varepsilon^2)=1+O_p(n^{-1/2})=1+o_p(1)$, and hence \eqref{veps1} follows. For \eqref{veps2}, 
	\begin{align*}
		\dfrac{V_5-V_4}{{E}(v_i^2\varepsilon_i^2)}
		&\le \dfrac{\sum_{i\in \mathcal{I}_{k}}\{\bW_i^\T(\widehat{\bbeta}_{{\rm DR},-k}-\bbeta^*)\}^2\varepsilon_i^2}{|\mathcal{I}_k|{E}(v_i^2\varepsilon_i^2)}+\dfrac{2\sum_{i\in \mathcal{I}_{k}}v_i\{\bW_i^\T(\widehat{\bbeta}_{{\rm DR},-k}-\bbeta^*)\}\varepsilon_i^2}{|\mathcal{I}_k|{E}(v_i^2\varepsilon_i^2)}=G_{11}+G_{12}.
	\end{align*}
	We next show that $G_{11}$ and $G_{12}$ are both $o_p(1)$ terms. Notice that conditional on $\mathcal{D}_{-k}$, $|\mathcal{I}_k|^{-1}\sum_{i\in \mathcal{I}_{k}}\{\bW_i^\T(\widehat{\bbeta}_{{\rm DR},-k}-\bbeta^*)\}^4$ is the sum of i.i.d. random variables. By Lemma \ref{sub_gaussian_bound},
	$${E}\left[\{\bW_i^\T(\widehat{\bbeta}_{{\rm DR},-k}-\bbeta^*)\}^4\mid \mathcal{D}_{-k}\right]=O(\|\widehat{\bbeta}_{{\rm DR},-k}-\bbeta^*\|_2^4).$$
	By Markov inequality, conditional on $\mathcal{D}_{-k}$, 
	$|\mathcal{I}_k|^{-1}\sum_{i\in \mathcal{I}_{k}}\{\bW_i^\T(\widehat{\bbeta}_{-k}-\bbeta^*)\}^4=O_p(\|\widehat{\bbeta}_{{\rm DR},-k}-\bbeta^*\|_2^4)$.
	Together with \eqref{beta_DR},
	\begin{align}\label{beta4}
		|\mathcal{I}_k|^{-1}\sum_{i\in \mathcal{I}_{k}}\{\bW_i^\T(\widehat{\bbeta}_{{\rm DR},-k}-\bbeta^*)\}^4=O_p\left\{\bigg(\dfrac{s\log d }{N}\bigg)^{2}\right\}=o_p(1).
	\end{align}
	By Markov's inequality, $|\mathcal{I}_k|^{-1}\sum_{i\in\mathcal{I}_k}\varepsilon_i^4=O_p(\sigma_{\varepsilon}^4)$. By \eqref{Eveps_lower}, 
	\begin{align*}
		G_{11}
		&\le \dfrac{1}{{E}(v_i^2\varepsilon_i^2)}\sqrt{|\mathcal{I}_k|^{-1}\sum_{i\in \mathcal{I}_{k}}\{\bW_i^\T(\widehat{\bbeta}_{{\rm DR},-k}-\bbeta^*)\}^4}\sqrt{|\mathcal{I}_k|^{-1}\sum_{i\in \mathcal{I}_{k}}\varepsilon_i^4}=O_p\left(\dfrac{s\log d }{N}\right)=o_p(1).
	\end{align*}
	Moreover, we have $G_{12}\le 2\sqrt{G_{11}(V_4/{E}(v_i^2\varepsilon_i^2))}=o_p(1)$.
	For \eqref{veps3}, we have
	\begin{align*}
		\dfrac{V_5-V_4}{{E}(v_i^2\varepsilon_i^2)}&=\dfrac{|\mathcal{I}_k|^{-1}\sum_{i\in \mathcal{I}_{k}}(Z_i-\bW_i^\T\widehat{\bbeta}_{{\rm DR},-k})^2(\bX_i^\T\widehat{\bgamma}_{{\rm DR},-k}-\bX_i^\T\bgamma^*)^2}{{E}(v_i^2\varepsilon_i^2)}\\
		&\quad+\dfrac{|\mathcal{I}_k|^{-1}\sum_{i\in \mathcal{I}_{k}}(Z_i-\bW_i^\T\widehat{\bbeta}_{{\rm DR},-k})^2(\bX_i^\T\widehat{\bgamma}_{{\rm DR},-k}-\bX_i^\T\bgamma^*)\varepsilon_i}{{E}(v_i^2\varepsilon_i^2)}=G_{13}+G_{14}.
	\end{align*}
	Since $v_i$ is sub-Gaussian with $\|v_i\|_{\psi_2}=O(1)$, by Lemma \ref{sub_gaussian_bound}, we have $E(v_i^4)=O(1)$ and ${\rm var}(|\mathcal{I}_k|^{-1}\sum_{i\in \mathcal{I}_{k}}v_i^4)=O(1/n)$. By Chebyshev's inequality, it holds that $|\mathcal{I}_k|^{-1}\sum_{i\in \mathcal{I}_{k}}v_i^4={E}(v_i^4)+O_p(n^{-1/2})=O_p(1)$. Together with \eqref{beta4},
	\begin{align*}
		&|\mathcal{I}_k|^{-1}\sum_{i\in \mathcal{I}_{k}}(Z_i-\bW_i^\T\widehat{\bbeta}_{{\rm DR},-k})^4=8|\mathcal{I}_k|^{-1}\sum_{i\in \mathcal{I}_{k}}v_i^4+8|\mathcal{I}_k|^{-1}\sum_{i\in \mathcal{I}_{k}}\{\bW_i^\T(\widehat{\bbeta}_{{\rm DR},-k}-\bbeta^*)\}^4=O_p(1).
	\end{align*}
	Conditional on $\mathcal{D}_{-k}$, $|\mathcal{I}_k|^{-1}\sum_{i\in \mathcal{I}_{k}}(\bX_i^\T\widehat{\bgamma}_{{\rm DR},-k}-\bX_i^\T\bgamma^*)^4$ is the sum of i.i.d. random variables. In addition, ${E}\left[(\bX_i^\T\widehat{\bgamma}_{{\rm DR},-k}-\bX_i^\T\bgamma^*)^4\mid \mathcal{D}_{-k}\right]=O_p(\|\widehat{\bgamma}_{{\rm DR},-k}-\bgamma^*\|_2^4)$. By Markov's inequality, conditional on $\mathcal{D}_{-k}$, $|\mathcal{I}_k|^{-1}\sum_{i\in \mathcal{I}_{k}}(\bX_i^\T\widehat{\bgamma}_{{\rm DR},-k}-\bX_i^\T\bgamma^*)^4=O_p(\|\widehat{\bgamma}_{{\rm DR},-k}-\bgamma^*\|_2^4)$. By \eqref{beta_DR}, $|\mathcal{I}_k|^{-1}\sum_{i\in \mathcal{I}_{k}}(\bX_i^\T\widehat{\bgamma}_{{\rm DR},-k}-\bX_i^\T\bgamma^*)^4=O_p\{\sigma_\varepsilon^4(k\log d/n)^2\}$
	and
	\begin{align*}
		G_{13}&\le\dfrac{1}{{E}(v_i^2\varepsilon_i^2)}\bigg\{|\mathcal{I}_k|^{-1}\sum_{i\in \mathcal{I}_{k}}(Z_i-\bW_i^\T\widehat{\bbeta}_{{\rm DR},-k})^4\bigg\}^{1/2}\bigg\{|\mathcal{I}_k|^{-1}\sum_{i\in \mathcal{I}_{k}}(\bX_i^\T\widehat{\bgamma}_{{\rm DR},-k}-\bX_i^\T\bgamma^*)^4\bigg\}^{1/2}\\
		&=O_p\left(\dfrac{k\log d }{n}\right)=o_p(1).
	\end{align*}
	Furthermore, $G_{14}\le 2[G_{13}\{V_5/{E}(v_i^2\varepsilon_i^2)\}]^{1/2}=o_p(1)$. Combining \eqref{veps1}-\eqref{veps3}, $V_6/E(v_i^2\varepsilon_i^2)=1+o_p(1)$. 
	Taking the average over $k$ terms,  
	\begin{equation}\label{hatDR2}
		n^{-1}\sum_{k=1}^{K}\sum_{i\in \mathcal{I}_{k}}(Z_i-\bW_i^\T\widehat{\bbeta}_{{\rm DR},-k})^2(Y_i-\bX_i^\T\widehat{\bgamma}_{{\rm DR},-k})^2/{E}(v_i^2\varepsilon_i^2)=1+o_p(1).
	\end{equation}
	Let 
	$
	\hat\sigma_{\rm DR}^2=\{n^{-1}\sum_{k=1}^{K}\sum_{i\in\mathcal I_k}(Z_i-\bW_i^\T\widehat{\bbeta}_{{\rm DR},-k})^2(Y_i-\bX_i^\T\widehat{\gamma}_{{\rm DR},-k})^2\}/\{n^{-1}\sum_{k=1}^{K}\sum_{i\in \mathcal{I}_{k}}(Z_i-\bW_i^\T\widehat{\bbeta}_{{\rm DR},-k})^2\}^2.
	$ Combining \eqref{hatDR1} and \eqref{hatDR2}, we have $\hat\sigma_{\rm DR}^2/\sigma^2=1+o_p(1)$. 
	\hfill\BlackBox

	\section{Data Availability}
	The NHEFS dataset is available at https://miguelhernan.org/whatifbook.


\begin{thebibliography}{}

\bibitem[\protect\citeauthoryear{Angelopoulos, Bates, Fannjiang, Jordan, and
	Zrnic}{Angelopoulos et~al.}{2023}]{angelopoulos2023prediction}
Angelopoulos, A.~N., Bates, S., Fannjiang, C., Jordan, M.~I., and Zrnic, T.
(2023).
\newblock Prediction-powered inference.
\newblock {\em Science} {\bf 382,} 669--674.

\bibitem[\protect\citeauthoryear{Azriel, Brown, Sklar, Berk, Buja, and
	Zhao}{Azriel et~al.}{2022}]{azriel2022semi}
Azriel, D., Brown, L.~D., Sklar, M., Berk, R., Buja, A., and Zhao, L. (2022).
\newblock Semi-supervised linear regression.
\newblock {\em Journal of the American Statistical Association} {\bf 117,}
2238--2251.

\bibitem[\protect\citeauthoryear{Bellec and Zhang}{Bellec and
	Zhang}{2022}]{bellec2022biasing}
Bellec, P.~C. and Zhang, C.-H. (2022).
\newblock De-biasing the lasso with degrees-of-freedom adjustment.
\newblock {\em Bernoulli} {\bf 28,} 713--743.

\bibitem[\protect\citeauthoryear{Bradic, Fan, and Zhu}{Bradic
	et~al.}{2022}]{bradic2022testability}
Bradic, J., Fan, J., and Zhu, Y. (2022).
\newblock Testability of high-dimensional linear models with nonsparse
structures.
\newblock {\em The Annals of Statistics} {\bf 50,} 615--639.

\bibitem[\protect\citeauthoryear{Buja, Brown, Berk, George, Pitkin, Traskin,
	Zhang, and Zhao}{Buja et~al.}{2019}]{buja2019models}
Buja, A., Brown, L., Berk, R., George, E., Pitkin, E., Traskin, M., Zhang, K.,
and Zhao, L. (2019).
\newblock Models as approximations I: Consequences illustrated with linear regression.
\newblock {\em Statistical Science} {\bf 34,} 523--544.

\bibitem[\protect\citeauthoryear{Cai, Li, and Liu}{Cai
	et~al.}{2024}]{cai2024semi}
Cai, T., Li, M., and Liu, M. (2024).
\newblock Semi-supervised triply robust inductive transfer learning.
\newblock {\em Journal of the American Statistical Association} {\bf 120,} 1037--1047.

\bibitem[\protect\citeauthoryear{Cai and Guo}{Cai and
	Guo}{2017}]{cai2017confidence}
Cai, T.~T. and Guo, Z. (2017).
\newblock Confidence intervals for high-dimensional linear regression: Minimax
rates and adaptivity.
\newblock {\em The Annals of Statistics} {\bf 45,} 615--646.

\bibitem[\protect\citeauthoryear{Cai and Guo}{Cai and
	Guo}{2020}]{tony2020semisupervised}
Cai, T.~T. and Guo, Z. (2020).
\newblock Semisupervised inference for explained variance in high dimensional
linear regression and its applications.
\newblock {\em Journal of the Royal Statistical Society Series B: Statistical
	Methodology} {\bf 82,} 391--419.

\bibitem[\protect\citeauthoryear{Chakrabortty and Cai}{Chakrabortty and
	Cai}{2018}]{chakrabortty2018efficient}
Chakrabortty, A. and Cai, T. (2018).
\newblock Efficient and adaptive linear regression in semi-supervised settings.
\newblock {\em The Annals of Statistics} {\bf 46,} 1541--1572.

\bibitem[\protect\citeauthoryear{Chakrabortty and Dai}{Chakrabortty and
	Dai}{2022}]{chakrabortty2022general}
Chakrabortty, A. and Dai, G. (2022).
\newblock A general framework for treatment effect estimation in
semi-supervised and high dimensional settings.
\newblock {\em arXiv preprint arXiv:2201.00468} .

\bibitem[\protect\citeauthoryear{Chapelle, Sch{\"o}lkopf, and Zien}{Chapelle
	et~al.}{2006}]{chapelle2006semi}
Chapelle, O., Sch{\"o}lkopf, B., and Zien, A. (2006).
\newblock {\em Semi-Supervised Learning}.
\newblock MIT Press.

\bibitem[\protect\citeauthoryear{Cheng, Ananthakrishnan, and Cai}{Cheng
	et~al.}{2021}]{cheng2021robust}
Cheng, D., Ananthakrishnan, A.~N., and Cai, T. (2021).
\newblock Robust and efficient semi-supervised estimation of average treatment
effects with application to electronic health records data.
\newblock {\em Biometrics} {\bf 77,} 413--423.

\bibitem[\protect\citeauthoryear{Chernozhukov, Chetverikov, Demirer, Duflo,
	Hansen, Newey, and Robins}{Chernozhukov
	et~al.}{2018}]{chernozhukov2018double}
Chernozhukov, V., Chetverikov, D., Demirer, M., Duflo, E., Hansen, C., Newey,
W., and Robins, J. (2018).
\newblock Double/debiased machine learning for treatment and structural
parameters.
\newblock {\em The Econometrics Journal} {\bf 21,} C1--C68.

\bibitem[\protect\citeauthoryear{Cui, Kosorok, Sverdrup, Wager, and Zhu}{Cui
	et~al.}{2023}]{cui2023estimating}
Cui, Y., Kosorok, M.~R., Sverdrup, E., Wager, S., and Zhu, R. (2023).
\newblock Estimating heterogeneous treatment effects with right-censored data
via causal survival forests.
\newblock {\em Journal of the Royal Statistical Society Series B: Statistical
	Methodology} {\bf 85,} 179--211.

\bibitem[\protect\citeauthoryear{Deng, Ning, Zhao, and Zhang}{Deng
	et~al.}{2024}]{deng2023optimal}
Deng, S., Ning, Y., Zhao, J., and Zhang, H. (2024).
\newblock Optimal and safe estimation for high-dimensional semi-supervised
learning.
\newblock {\em Journal of the American Statistical Association} {\bf 119,}
2748--2759.

\bibitem[\protect\citeauthoryear{Fan, Lu, Song, and Zhou}{Fan
	et~al.}{2017}]{fan2017concordance}
Fan, C., Lu, W., Song, R., and Zhou, Y. (2017).
\newblock Concordance-assisted learning for estimating optimal individualized
treatment regimes.
\newblock {\em Journal of the Royal Statistical Society Series B: Statistical
	Methodology} {\bf 79,} 1565--1582.

\bibitem[\protect\citeauthoryear{Gao, Bonzel, Hong, Varghese, Zakir, and
	Gronsbell}{Gao et~al.}{2024}]{gao2024semi}
Gao, J., Bonzel, C.-L., Hong, C., Varghese, P., Zakir, K., and Gronsbell, J.
(2024).
\newblock Semi-supervised ROC analysis for reliable and streamlined evaluation
of phenotyping algorithms.
\newblock {\em Journal of the American Medical Informatics Association} {\bf
	31,} 640--650.

\bibitem[\protect\citeauthoryear{Hammer, Katzenstein, Hughes, Gundacker,
	Schooley, Haubrich, Henry, Lederman, Phair, Niu, et~al\mbox{.}}{Hammer
	et~al.}{1996}]{hammer1996trial}
Hammer, S.~M., Katzenstein, D.~A., Hughes, M.~D., Gundacker, H., Schooley,
R.~T., Haubrich, R.~H., Henry, W.~K., Lederman, M.~M., Phair, J.~P., Niu, M.,
et~al. (1996).
\newblock A trial comparing nucleoside monotherapy with combination therapy in
HIV-infected adults with CD4 cell counts from 200 to 500 per cubic
millimeter.
\newblock {\em New England Journal of Medicine} {\bf 335,} 1081--1090.

\bibitem[\protect\citeauthoryear{Javanmard and Montanari}{Javanmard and
	Montanari}{2014a}]{javanmard2014confidence}
Javanmard, A. and Montanari, A. (2014a).
\newblock Confidence intervals and hypothesis testing for high-dimensional
regression.
\newblock {\em The Journal of Machine Learning Research} {\bf 15,} 2869--2909.

\bibitem[\protect\citeauthoryear{Javanmard and Montanari}{Javanmard and
	Montanari}{2014b}]{javanmard2014hypothesis}
Javanmard, A. and Montanari, A. (2014b).
\newblock Hypothesis testing in high-dimensional regression under the
{G}aussian random design model: Asymptotic theory.
\newblock {\em IEEE Transactions on Information Theory} {\bf 60,} 6522--6554.

\bibitem[\protect\citeauthoryear{Javanmard and Montanari}{Javanmard and
	Montanari}{2018}]{javanmard2018debiasing}
Javanmard, A. and Montanari, A. (2018).
\newblock Debiasing the Lasso: Optimal sample size for {G}aussian designs.
\newblock {\em The Annals of Statistics} {\bf 46,} 2593--2622.

\bibitem[\protect\citeauthoryear{Li, Zheng, He, He, Li, Cui, Yang, Dong, Shen,
	and Zheng}{Li et~al.}{2022}]{li2022treatment}
Li, N., Zheng, H.-Y., He, W.-Q., He, X.-Y., Li, R., Cui, W.-B., Yang, W.-L.,
Dong, X.-Q., Shen, Z.-Q., and Zheng, Y.-T. (2022).
\newblock Treatment outcomes amongst older people with HIV infection receiving antiretroviral therapy.
\newblock {\em Aids} {\bf 38,} 803--812.

\bibitem[\protect\citeauthoryear{Li, Yang, Wei, and Liu}{Li
	et~al.}{2024}]{li2024adaptive}
Li, Y., Yang, X., Wei, Y., and Liu, M. (2024).
\newblock Adaptive and efficient learning with blockwise missing and
semi-supervised data.
\newblock {\em arXiv preprint arXiv:2405.18722} .

\bibitem[\protect\citeauthoryear{Liang and Yan}{Liang and
	Yan}{2020}]{liang2020empirical}
Liang, W. and Yan, Y. (2020).
\newblock Empirical likelihood-based estimation and inference in randomized
controlled trials with high-dimensional covariates.
\newblock {\em arXiv preprint arXiv:2010.01772} .

\bibitem[\protect\citeauthoryear{Lu, Zhang, and Zeng}{Lu
	et~al.}{2013}]{lu2013variable}
Lu, W., Zhang, H.~H., and Zeng, D. (2013).
\newblock Variable selection for optimal treatment decision.
\newblock {\em Statistical Methods in Medical Research} {\bf 22,} 493--504.

\bibitem[\protect\citeauthoryear{Lu, Gu, and Duan}{Lu
	et~al.}{2024}]{lu2024enhancing}
Lu, Y., Gu, T., and Duan, R. (2024).
\newblock Enhancing genetic risk prediction through federated semi-supervised
transfer learning with inaccurate electronic health record data.
\newblock {\em Statistics in Biosciences} pages 1--22.

\bibitem[\protect\citeauthoryear{Hern{\'a}n and Robins}{Hern{\'a}n
	and Robins}{2020}]{hernan2020causal}
Hern{\'a}n, M. A. and Robins, J. M. (2020).
\newblock {\em Causal Inference: What If}.
\newblock Boca Raton: Chapman \& Hall/CRC.

\bibitem[\protect\citeauthoryear{Negin, Martiniuk, Cumming, Naidoo,
	Phaswana-Mafuya, Madurai, Williams, and Kowal}{Negin
	et~al.}{2012}]{negin2012prevalence}
Negin, J., Martiniuk, A., Cumming, R.~G., Naidoo, N., Phaswana-Mafuya, N.,
Madurai, L., Williams, S., and Kowal, P. (2012).
\newblock Prevalence of HIV and chronic comorbidities among older adults.
\newblock {\em Aids} {\bf 26,} S55--S63.

\bibitem[\protect\citeauthoryear{Seeger}{Seeger}{2001}]{seeger2001learning}
Seeger, M. (2001).
\newblock \textit{Learning with labeled and unlabeled data}.
\newblock Technical Report, University of Edinburgh.


\bibitem[\protect\citeauthoryear{Song, Lin, and Zhou}{Song
	et~al.}{2024}]{song2023general}
Song, S., Lin, Y., and Zhou, Y. (2024).
\newblock A general M-estimation theory in semi-supervised framework.
\newblock {\em Journal of the American Statistical Association} {\bf 119,}
1065--1075.

\bibitem[\protect\citeauthoryear{Tibshirani and Taylor}{Tibshirani and
	Taylor}{2012}]{Tibshirani2012degrees}
Tibshirani, R.~J. and Taylor, J. (2012).
\newblock Degrees of freedom in lasso problems.
\newblock {\em The Annals of Statistics} {\bf 40,} 1198--1232.

\bibitem[\protect\citeauthoryear{Tsiatis, Davidian, Zhang, and Lu}{Tsiatis
	et~al.}{2008}]{tsiatis2008covariate}
Tsiatis, A.~A., Davidian, M., Zhang, M., and Lu, X. (2008).
\newblock Covariate adjustment for two-sample treatment comparisons in
randomized clinical trials: a principled yet flexible approach.
\newblock {\em Statistics in Medicine} {\bf 27,} 4658--4677.

\bibitem[\protect\citeauthoryear{Van~de Geer, B{\"u}hlmann, Ritov, and
	Dezeure}{Van~de Geer et~al.}{2014}]{van2014asymptotically}
Van~de Geer, S., B{\"u}hlmann, P., Ritov, Y., and Dezeure, R. (2014).
\newblock On asymptotically optimal confidence regions and tests for
high-dimensional models.
\newblock {\em The Annals of Statistics} {\bf 42,} 1166--1202.

\bibitem[\protect\citeauthoryear{Van~Engelen and Hoos}{Van~Engelen and
	Hoos}{2020}]{van2020survey}
Van~Engelen, J.~E. and Hoos, H.~H. (2020).
\newblock A survey on semi-supervised learning.
\newblock {\em Machine Learning} {\bf 109,} 373--440.

\bibitem[\protect\citeauthoryear{Wang, Wang, Liao, and Cai}{Wang
	et~al.}{2024}]{wang2024semisupervised}
Wang, L., Wang, X., Liao, K.~P., and Cai, T. (2024).
\newblock Semisupervised transfer learning for evaluation of model
classification performance.
\newblock {\em Biometrics} {\bf 80,} ujae002.

\bibitem[\protect\citeauthoryear{Zhang, Brown, and Cai}{Zhang
	et~al.}{2019}]{zhang2019semi}
Zhang, A., Brown, L.~D., and Cai, T.~T. (2019).
\newblock Semi-supervised inference: General theory and estimation of means.
\newblock {\em The Annals of Statistics} {\bf 47,} 2538--2566.

\bibitem[\protect\citeauthoryear{Zhang and Zhang}{Zhang and
	Zhang}{2014}]{zhang2014confidence}
Zhang, C.-H. and Zhang, S.~S. (2014).
\newblock Confidence intervals for low dimensional parameters in high
dimensional linear models.
\newblock {\em Journal of the Royal Statistical Society Series B: Statistical
	Methodology} {\bf 76,} 217--242.

\bibitem[\protect\citeauthoryear{Zhang and Oles}{Zhang and
	Oles}{2000}]{zhang2000value}
Zhang, T. and Oles, F. (2000).
\newblock The value of unlabeled data for classification problems.
\newblock In {\em Proceedings of the Seventeenth International Conference on
	Machine Learning}, pages 1191--1198.

\bibitem[\protect\citeauthoryear{Zhang and Bradic}{Zhang and
	Bradic}{2022}]{zhang2022high}
Zhang, Y. and Bradic, J. (2022).
\newblock High-dimensional semi-supervised learning: in search of optimal
inference of the mean.
\newblock {\em Biometrika} {\bf 109,} 387--403.

\bibitem[\protect\citeauthoryear{Zhu}{Zhu}{2005}]{zhu2005semi}
Zhu, X. (2005).
\newblock Semi-supervised learning literature survey.
\newblock Technical Report, Computer Sciences Department, University of
Wisconsin-Madison.

\bibitem[\protect\citeauthoryear{Zhu and Bradic}{Zhu and
	Bradic}{2018}]{zhu2018linear}
Zhu, Y. and Bradic, J. (2018).
\newblock Linear hypothesis testing in dense high-dimensional linear models.
\newblock {\em Journal of the American Statistical Association} {\bf 113,}
1583--1600.

\bibitem[\protect\citeauthoryear{Zou and Hastie}{Zou and
	Hastie}{2007}]{zou2007degrees}
Zou, H. and Hastie, T. (2007).
\newblock On the ``degrees of freedom'' of the lasso.
\newblock {\em The Annals of Statistics} {\bf 35,} 2173--2192.

\end{thebibliography}

\begin{thebibliography}{}
	
	\bibitem[\protect\citeauthoryear{Athey, Imbens, and Wager}{Athey
		et~al.}{2018}]{athey2018approximate}
	Athey, S., Imbens, G.~W., and Wager, S. (2018).
	\newblock Approximate residual balancing: debiased inference of average
	treatment effects in high dimensions.
	\newblock {\em Journal of the Royal Statistical Society Series B: Statistical Methodology} {\bf 80,}
	597--623.
	
	\bibitem[\protect\citeauthoryear{Bellec and Zhang}{Bellec and
		Zhang}{2022}]{bellec2022biasing}
	Bellec, P.~C. and Zhang, C.-H. (2022).
	\newblock De-biasing the lasso with degrees-of-freedom adjustment.
	\newblock {\em Bernoulli} {\bf 28,} 713--743.
	
	\bibitem[\protect\citeauthoryear{Bradic, Fan, and Zhu}{Bradic
		et~al.}{2022}]{bradic2022testability}
	Bradic, J., Fan, J., and Zhu, Y. (2022).
	\newblock Testability of high-dimensional linear models with nonsparse
	structures.
	\newblock {\em The Annals of Statistics} {\bf 50,} 615--639.
	
	\bibitem[\protect\citeauthoryear{Cai and Guo}{Cai and
		Guo}{2017}]{cai2017confidence}
	Cai, T.~T. and Guo, Z. (2017).
	\newblock Confidence intervals for high-dimensional linear regression: Minimax
	rates and adaptivity.
	\newblock {\em The Annals of Statistics} {\bf 45,} 615--646.
	
	\bibitem[\protect\citeauthoryear{Cai and Guo}{Cai and
		Guo}{2020}]{tony2020semisupervised}
	Cai, T.~T. and Guo, Z. (2020).
	\newblock Semisupervised inference for explained variance in high dimensional
	linear regression and its applications.
	\newblock {\em Journal of the Royal Statistical Society Series B: Statistical
		Methodology} {\bf 82,} 391--419.
	
	\bibitem[\protect\citeauthoryear{Celentano, Montanari, and Wei}{Celentano
		et~al.}{2023}]{celentano2023lasso}
	Celentano, M., Montanari, A., and Wei, Y. (2023).
	\newblock The Lasso with general Gaussian designs with applications to
	hypothesis testing.
	\newblock {\em The Annals of Statistics} {\bf 51,} 2194--2220.
	
	\bibitem[\protect\citeauthoryear{Chakrabortty, Lu, Cai, and Li}{Chakrabortty
		et~al.}{2019}]{chakrabortty2019high}
	Chakrabortty, A., Lu, J., Cai, T.~T., and Li, H. (2019).
	\newblock High dimensional M-estimation with missing outcomes: A
	semi-parametric framework.
	\newblock {\em arXiv preprint arXiv:1911.11345} .
	
	\bibitem[\protect\citeauthoryear{Chernozhukov, Chetverikov, Demirer, Duflo,
		Hansen, Newey, and Robins}{Chernozhukov
		et~al.}{2018}]{chernozhukov2018double}
	Chernozhukov, V., Chetverikov, D., Demirer, M., Duflo, E., Hansen, C., Newey,
	W., and Robins, J. (2018).
	\newblock Double/debiased machine learning for treatment and structural
	parameters.
	\newblock {\em The Econometrics Journal} {\bf 21,} C1--C68.
	
	\bibitem[\protect\citeauthoryear{Deng, Ning, Zhao, and Zhang}{Deng
		et~al.}{2024}]{deng2023optimal}
	Deng, S., Ning, Y., Zhao, J., and Zhang, H. (2024).
	\newblock Optimal and safe estimation for high-dimensional semi-supervised
	learning.
	\newblock {\em Journal of the American Statistical Association} {\bf 119,}
	2748--2759.
	
	\bibitem[\protect\citeauthoryear{Ertefaie, Hejazi, and van~der Laan}{Ertefaie
		et~al.}{2023}]{ertefaie2020nonparametric}
	Ertefaie, A., Hejazi, N.~S., and van~der Laan, M.~J. (2023).
	\newblock Nonparametric inverse-probability-weighted estimators based on the
	highly adaptive lasso.
	\newblock {\em Biometrics} {\bf 79,} 1029--1041.
	
	\bibitem[\protect\citeauthoryear{Hern{\'a}n and Robins}{Hern{\'a}n
	and Robins}{2020}]{hernan2020causal}
Hern{\'a}n, M. A. and Robins, J. M. (2020).
\newblock {\em Causal Inference: What If}.
\newblock Boca Raton: Chapman \& Hall/CRC.
	
	\bibitem[\protect\citeauthoryear{Javanmard and Montanari}{Javanmard and
		Montanari}{2014}]{javanmard2014confidence}
	Javanmard, A. and Montanari, A. (2014).
	\newblock Confidence intervals and hypothesis testing for high-dimensional
	regression.
	\newblock {\em The Journal of Machine Learning Research} {\bf 15,} 2869--2909.
	
	\bibitem[\protect\citeauthoryear{Javanmard and Montanari}{Javanmard and
		Montanari}{2018}]{javanmard2018debiasing}
	Javanmard, A. and Montanari, A. (2018).
	\newblock Debiasing the lasso: Optimal sample size for Gaussian designs.
	\newblock {\em The Annals of Statistics} {\bf 46,} 2593--2622.
	
	\bibitem[\protect\citeauthoryear{Liang and Yan}{Liang and
		Yan}{2020}]{liang2020empirical}
	Liang, W. and Yan, Y. (2020).
	\newblock Empirical likelihood-based estimation and inference in randomized
	controlled trials with high-dimensional covariates.
	\newblock {\em arXiv preprint arXiv:2010.01772} .
	
	\bibitem[\protect\citeauthoryear{Rudelson and Zhou}{Rudelson and
		Zhou}{2013}]{rudelson2013reconstruction}
	Rudelson, M. and Zhou, S. (2013).
	\newblock Reconstruction from anisotropic random measurements.
	\newblock {\em IEEE Transactions on Information Theory} {\bf 59,} 3434--3447.
	
	\bibitem[\protect\citeauthoryear{Vershynin}{Vershynin}{2010}]{vershynin2010introduction}
	Vershynin, R. (2010).
	\newblock Introduction to the non-asymptotic analysis of random matrices.
	\newblock {\em arXiv preprint arXiv:1011.3027} .
	
	\bibitem[\protect\citeauthoryear{Wainwright}{Wainwright}{2019}]{wainwright2019high}
	Wainwright, M.~J. (2019).
	\newblock {\em High-dimensional Statistics: A Non-Asymptotic Viewpoint},
	\newblock Cambridge University Press.
	
	\bibitem[\protect\citeauthoryear{Zhang, Brown, and Cai}{Zhang
		et~al.}{2019}]{zhang2019semi}
	Zhang, A., Brown, L.~D., and Cai, T.~T. (2019).
	\newblock Semi-supervised inference: General theory and estimation of means.
	\newblock {\em The Annals of Statistics} {\bf 47,} 2538--2566.
	
	\bibitem[\protect\citeauthoryear{Zhang and Bradic}{Zhang and
		Bradic}{2022}]{zhang2022high}
	Zhang, Y. and Bradic, J. (2022).
	\newblock High-dimensional semi-supervised learning: in search of optimal
	inference of the mean.
	\newblock {\em Biometrika} {\bf 109,} 387--403.
	
	\bibitem[\protect\citeauthoryear{Zhang, Chakrabortty, and Bradic}{Zhang
		et~al.}{2023a}]{zhang2023decaying}
	Zhang, Y., Chakrabortty, A., and Bradic, J. (2023a).
	\newblock The decaying missing-at-random framework: Doubly robust causal
	inference with partially labeled data.
	\newblock {\em arXiv preprint arXiv:2305.12789} .
	
	\bibitem[\protect\citeauthoryear{Zhang, Chakrabortty, and Bradic}{Zhang
		et~al.}{2023b}]{zhang2023double}
	Zhang, Y., Chakrabortty, A., and Bradic, J. (2023b).
	\newblock Double robust semi-supervised inference for the mean: Selection bias
	under mar labeling with decaying overlap.
	\newblock {\em Information and Inference: A Journal of the IMA} {\bf 12,}
	2066--2159.
	
	\bibitem[\protect\citeauthoryear{Zhang, Giessing, and Chen}{Zhang
		et~al.}{2023}]{zhang2023efficient}
	Zhang, Y., Giessing, A., and Chen, Y.-C. (2023).
	\newblock Efficient inference on high-dimensional linear models with missing
	outcomes.
	\newblock {\em arXiv preprint arXiv:2309.06429} .
	
	\bibitem[\protect\citeauthoryear{Zhu and Bradic}{Zhu and
		Bradic}{2018}]{zhu2018linear}
	Zhu, Y. and Bradic, J. (2018).
	\newblock Linear hypothesis testing in dense high-dimensional linear models.
	\newblock {\em Journal of the American Statistical Association} {\bf 113,}
	1583--1600.
	
\end{thebibliography}
\end{document}